\documentclass[letterpaper]{article} 

\usepackage[usenames,dvipsnames]{color}
\usepackage{bm}
\usepackage{graphicx,fancyhdr,lscape,dsfont,amsmath,amssymb} 
\usepackage{color, amsthm}
\usepackage{longtable}
\usepackage{arydshln}
\usepackage{psfrag}
\usepackage{subfigure} 
\usepackage{multirow}
\usepackage{stmaryrd}
\usepackage[titletoc]{appendix}
\usepackage{natbib}
\usepackage{graphicx}
\usepackage{ulem} 
\usepackage{mathrsfs}
\usepackage{bbm}
\usepackage{upgreek}
\usepackage{epstopdf}
\usepackage{booktabs}

\allowdisplaybreaks

\newcommand{\tensor}[1]{  {\bm {#1}} } 

\begin{document}

\title{Wave propagation modeling in periodic elasto-thermo-diffusive materials via multifield asymptotic homogenization
}

\author{Francesca Fantoni$^{1,*}$, Andrea Bacigalupo$^{2,}$\footnote{Corresponding authors: Tel:+39 0303711330, $\hspace{10cm}$ 
 E-mail addresses: francesca.fantoni@unibs.it; andrea.bacigalupo@unige.it}
\\
\begin{small}
$^{1}$ DICATAM, Universit\`a degli Studi di Brescia, via Branze 43, 25123, Brescia, Italy
\end{small}
\\
\begin{small}
$^{2}$ DICCA, Universit\`a degli Studi di Genova,via Montallegro 1,
16145 Genova, Italy
 \end{small}
}

\maketitle

\begin{abstract}
A multifield asymptotic homogenization technique for periodic thermo-diffusive elastic materials is provided in the present study.
Field equations for the first-order equivalent medium are derived and overall constitutive tensors are obtained in closed form.
These lasts depend upon  the micro constitutive properties of the different phases composing the composite material and upon periodic perturbation functions, which allow taking into account the effects of microstructural heterogeneities.
Perturbation functions are determined as solutions of recursive non homogeneous cell problems emanated from the substitution of asymptotic expansions of the micro fields in powers of the microstructural characteristic size into local balance equations.
Average field equations of infinite order are also provided, whose formal solution can be obtained through  asymptotic expansions of the macrofields.
With the aim of investigating dispersion properties of waves propagating inside the medium, proper integral transforms are applied to governing field equations of the homogenized medium.
A quadratic generalized eigenvalue problem is thus obtained, whose solution characterizes the complex valued frequency band structure of the first-order equivalent material.
The validity of the proposed technique has been confirmed by the very good matching obtained between dispersion curves of the homogenized medium and the lowest frequency ones relative to the heterogeneous material.
These lasts are computed from the resolution of a quadratic generalized eigenvalue problem over the periodic cell subjected to Floquet-Bloch boundary conditions.
An illustrative benchmark is conducted referring to a Solid Oxide Fuel Cell (SOFC)-like material, whose microstructure can be modeled through the spatial tessellation of the domain with a periodic cell subjected to thermo-diffusive phenomena.
\end{abstract}
\section{Introduction}
\label{Sec::Inroduction}
The increasing need of energy diversification and employment of alternative and renewable energy sources motivates the growth in the use  of fuel cells as power generating systems. 
Substitution of conventional fuel combustion with an electrochemical reaction in order to generate electricity make fuel cells  clean and  sustainable energy devices, nowadays exploited for a wide range of applications, from powering satellites to generating power for vehicles and buildings.
Fuel cells consist of two porous heat resistant electrodes, the negative one (anode) and the positive one (cathode), undergoing electrochemical reaction in order to produce an electric current. They are sandwiched around a porous electrolyte, which is the ion conductor.
Fuel cells differ according to the electrolyte employed, which influences the type of occurring electrochemical reaction, of the catalyst, and of the fuel, thus achieving distinct levels of efficiency \citep{Brandon2006}.
In this context, Solid Oxide Fuel Cells (SOFCs) are characterized by having  a doped, solid, ceramic material to form the electrolyte  and they excel for  their high electrical efficiency and low operating costs \citep{Zhu2003,Bove2008}.
The cathode of SOFCs is supplied both with oxygen, acting as the oxidant, and electrons coming from  the external electrical circuit.
Oxygen ions intercalate into the electrolyte as a consequence of the reduction process taking place at the cathode side.
Through the solid electrolyte, negative oxygen ions are therefore conducted from the cathode to the anode, where they combine with the hydrogen fuel, thus generating both water and electrons as products of the oxidation reaction.
Electrical current is hence generated by  electrons travelling along the external circuit and then reentering into the cathode material.
In addition, every single cell is characterized by flow channels for air and fuel and by a metallic or ceramic interconnect separator, which allows connecting cells in series with the aim to produce sufficient voltage for the practical use.

Macroscopic engineering response of such multiphase materials is strongly influenced by the mechanics and physics  occurring at the microscale, whose characteristic size is very small compared to the structural one.
For this reason a numerical analysis of microstructured devices like a SOFCs stack could reveal extremely challenging in terms of computational and temporal resources \citep{Hajimolana2011,Dev2014}.
When scales separation holds, homogenization techniques result to be remarkably useful in order to provide  an accurate and concise  description of the medium which properly take into account the behavior and the mechanical response of the microstructure. 
The application of homogenization methods  and multiscale modelings allows avoiding the demanding numerical computation of the whole heterogeneous medium leading to the identification of effective macroscopic properties for the equivalent continuum.
In order to study the overall properties of composite materials, numerous homogenization approaches have been provided over the last decades, which can be divided in asymptotic techniques \citep{SanchezPalencia1974,
Bensoussan1978,
Bakhvalov1984,
GambinKroner1989,
Allaire1992,
Bacigalupo2014,
fantoni2017multi,
fantoni2018design}, variational-asymptotic techniques \citep{Smyshlyaev2000,
PeerlingsFleck2004,
BacigalupoGambarotta2014b}, and numerous identification approaches including the analytical \citep{Bigoni2007,
Milton2007,
BaccaBigoniDalCorsoVeber2013a,
BaccaBigoniDalCorsoVeber2013b,
BaccaDalCorsoVeberBigoni2013,
Nassar2015,
BacigalupoBigoni2018}  and computational methods
\citep{Forest1998,
Ostoja1999,
Feyel2000,
Kouznetsova2002,
Forest2002,
Feyel2003,
KouznetsovaGeers2004,
Lew2004,
Kaczmarczyk2008,
Yuan2008,
Scarpa2009,
BacigalupoGambarotta2010b,
ForestTrinh2011,
DeBellis2011,
AdessiDeBellisSacco2013,
ZahMiehe2013,
SalvadoriEtAlJMPS2013,
Trovalusci2015}.
The present study is devoted to provide a multifield asymptotic homogenization technique for periodic thermo-diffusive materials considering as periodic cell the typical SOFC building block.
An accurate prediction of the overall response   of SOFCs is of crucial importance in order to guarantee the satisfaction of design requirements and the reliability of the entire system.
Battery devices like SOFCs, in fact, are subjected to severe stresses due to  high operating temperatures ($600^\circ-1000^\circ$)  \citep{Pitakthapanaphong2005} and intense particle diffusion, which could compromise their efficiency in terms of power generation and energy conversion, ultimately impacting on their failure behavior \citep{Atkinson2007,Kuebler2010,Delette2013}.
Previous numerical models of SOFCs focused on electrochemical aspects can be found in \citep{Kakac2007,Colpan2008}, while mechanical properties of each phase forming the composite battery device are presented in \citep{Hasanov2011}.
Latterly, different multiscale modeling of SOFCs have been provided focusing on computational homogenization \citep{Kim2009,Muramatsu2015,Molla2016},  asymptotic first-order homogenization of thermo-mechanical properties \citep{BacigalupoMorini2016}, and asymptotic non local homogenization of elastic properties \citep{BacigalupoMoriniPiccolroaz2014} where the influence of  temperature upon local and non local overall constitutive tensors has been studied. 
Furthermore, an investigation  of the complex frequency band structure  of periodic SOFCs based on a micromechanical perspective has been recently presented by one of the author in \citep{BacigalupoDeBellisGnecco2019}.
Nevertheless, to the best of authors' knowledge,  a rigorous quantitative multiscale description  of mechanical, thermal, and diffusive properties of SOFC-like material and their coupling is still missing.
In the followings, down-scaling relations are provided.
They relate the microfields, specifically the displacement, the relative temperature and the chemical potential to the macroscopic fields and their gradients by means of perturbation functions.
These lasts are regular, periodic functions derived through the resolution of recursive, non homogeneous differential problems, known as cell problems, obtained inserting an asymptotic expansion of the microfields in powers of the microstructural length scale into the local balance equations and reordering at the different orders of the micro characteristic size.
Following the rigorous approach described in \citep{Smyshlyaev2000,Bacigalupo2014}, average field equations of infinite order are obtained from the substitution of down-scaling relations into micro governing field equations.
A formal solution of the average field equations of infinite order can be attained by performing an asymptotic expansion of the macrofields in powers of the micro length scale, and truncation of resulting equations to the zeroth order allows characterizing field equations of the first-order equivalent medium for the class of periodic thermo-diffusive materials considered.
Coefficients of obtained field equations are related to the overall constitutive tensors, whose expression is provided in closed form in terms of perturbation functions and microscopic constitutive properties.

With the aim of investigating the dispersive free waves propagation within the periodic microstructured material, bilateral Laplace transform in time and Fourier transform in space are applied to field equations of the homogenized medium, thus obtaining a quadratic generalized eigenvalue problem, whose solution characterizes the complex frequency band structure of the first-order   equivalent medium.
The validity of the proposed approach is assessed by comparing the obtained complex frequency spectra with the ones relative to the heterogeneous thermo-diffusive material.
In this case, a generalization of the  Floquet-Bloch theory is employed, which allows determining dispersion properties of the heterogeneous material by solving a generalized quadratic eigenvalue problem over the periodic cell endowed with Floquet-Bloch boundary conditions.
Finally, an asymptotic approximation of the complex spectrum for the first-order equivalent medium is performed via perturbative technique.
This allows achieving a parametric approximation of the complex frequency in powers of the wave vector in terms of the overall constitutive parameters  and obtained explicit dispersion curves demonstrate to match very well with the ones relative to the homogenized medium.
The work is organized as follows: Section \ref{Sec::AsymHom} describes the governing microscopic field equations and   recursive differential problems obtained through asymptotic expansion of the microfields in powers of the microstructural length scale.
Cell problems and relative perturbation functions at the different orders of the micro characteristic size are detailed in Section \ref{Sec::CellProblems}.
Section \ref{Sec::DownScalingUpScaling} is devoted to the determination of down-scaling and up-scaling relations, while in Section \ref{Sec::FieldEquatuationsCauchyHomMedium} field equations of the first-order equivalent continuum are presented and overall constitutive tensors  are provided in closed form.
The determination of complex frequency band structure for the first-order homogenized medium is described in Section \ref{Sec::PropagazioneContinuoalPrimoOrdine}, together with its asymptotic approximation via perturbative method in Section \ref{SubSec::AsympApproxOfComplexSpectrum}. 
In order to evaluate the capabilities of the proposed method a representative example is performed in Section \ref{Sec:Benchmark}, where the complex frequency band structure and its asymptotic approximation are provided for the equivalent continuum in relation to a typical SOFC and obtained results are compared with the ones of  the relative heterogeneous periodic cell.
Final remarks are then proposed in Section \ref{Sec::Conclusions}.
\section{Periodic heterogeneous thermo-diffusive material: field equations and multi-scale description}
\label{Sec::AsymHom}
\begin{figure}[h!]
  \centering
  \includegraphics[width=11cm]{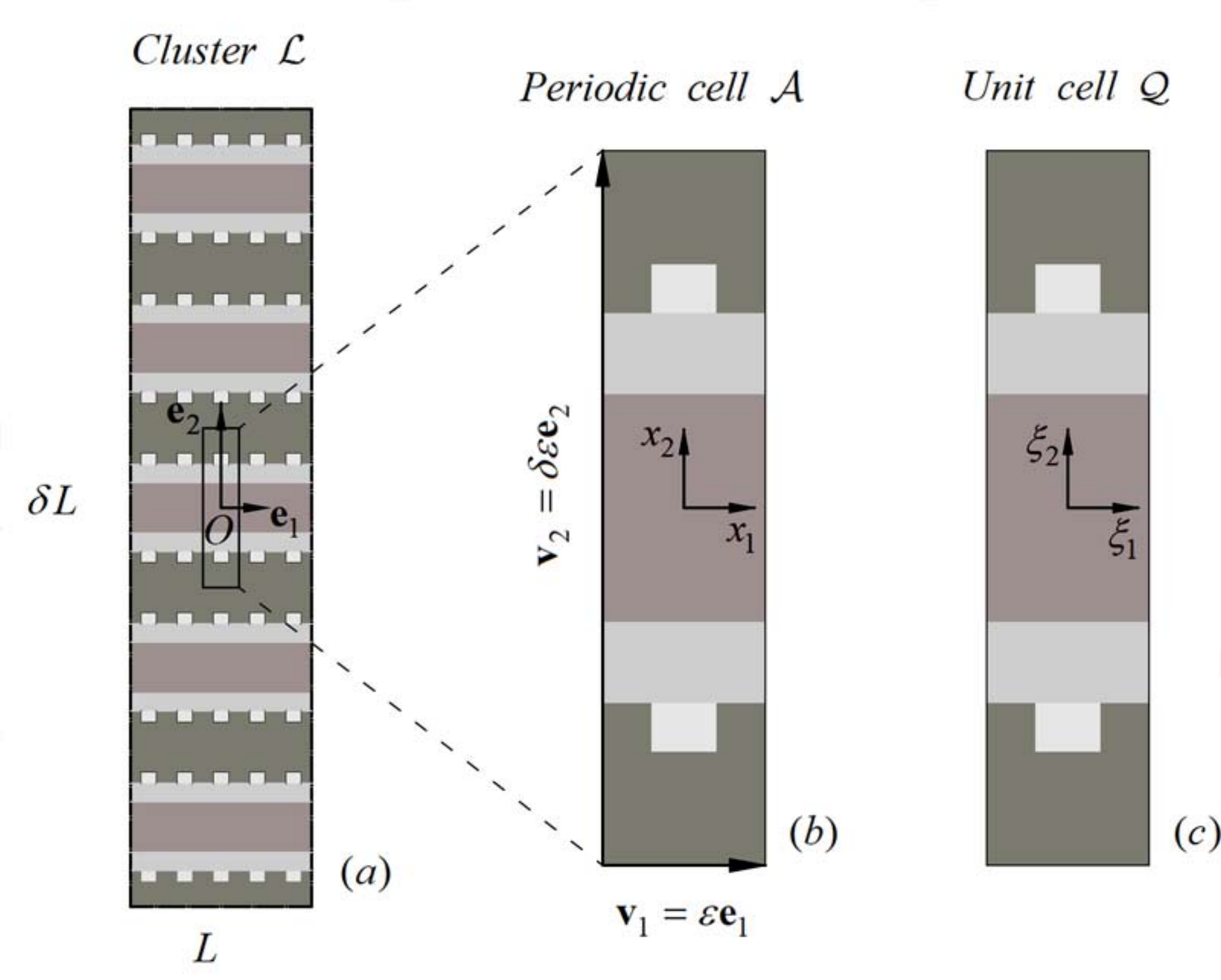}
  \caption{\it (a) Cluster $\mathcal{L}$ of heterogeneous SOFC having structural characteristic size $L$; (b) periodic cell $\mathcal{A}$ with microstructural characteristic size $\varepsilon$ and periodicity vectors $\mathbf{v}_1$ and $\mathbf{v}_2$; (c) unit cell $\mathcal{Q}$. }
  \label{Fig::periodic_SOFC}
\end{figure}

%
Under the assumption of small strains, the heterogeneous microstructured composite material depicted in figure \ref{Fig::periodic_SOFC} is described as a linear thermo-diffusive Cauchy medium \citep{Nowacki1974_1,Nowacki1974_2,Nowacki1974_3}.
In a two-dimensional perspective, as represented in figure \ref{Fig::periodic_SOFC}, vector $\mathbf{x}=x_1\,\mathbf{e}_1+x_2\,\mathbf{e}_2$ defines the position of each material point in the orthogonal reference system $\{O,\mathbf{e}_1,\mathbf{e}_2 \}$.
Micro fields characterizing the first-order continuum are the displacement field $\mathbf{u}(\mathbf{x},t)=u_i(\mathbf{x},t)\mathbf{e}_i$, relative temperature field $\theta(\mathbf{x},t)=T(\mathbf{x},t)-T_0$ with $T(\mathbf{x},t)$ the absolute temperature and $T_0$ a reference stress free temperature, and chemical potential field $\eta(\mathbf{x},t)$.
Being $\varepsilon$ the characteristic size of the microstructure, two periodicity vectors $\mathbf{v}_1=d_1\,\mathbf{e}_1=\varepsilon\,\mathbf{e}_1$ and $\mathbf{v}_2=d_2\,\mathbf{e}_2=\delta\varepsilon\,\mathbf{e}_2$ identify the periodic cell $\mathcal{A}=[0,\varepsilon]\times [0,\delta\varepsilon]$ (figure \ref{Fig::periodic_SOFC}-(b)).
Rescaling cell $\mathcal{A}$ by the length $\varepsilon$, the periodic microstructure is obtained by spanning the nondimensional unit cell $\mathcal{Q}=[0,1]\times[0,\delta]$, as depicted in figure \ref{Fig::periodic_SOFC}-(c).
The separation between the macro and the micro scales is mathematically described by two distinct variables, namely the macroscopic (or slow) one $\mathbf{x}\in\mathcal{A}$  and the microscopic (or fast) one $\tensor{\xi}=\mathbf{x}/\varepsilon\in \mathcal{Q}$ \citep{Bakhvalov1984,Smyshlyaev2000,PeerlingsFleck2004,Bacigalupo2014}.
Micro-stress tensor $\tensor{\sigma}(\mathbf{x},t)=\sigma_{ij}\,(\mathbf{x},t)\mathbf{e}_i\otimes\mathbf{e}_j$, micro heat flux vector $\mathbf{q}(\mathbf{x},t)=q_i\,(\mathbf{x},t)\mathbf{e}_i$, and mass flux vector $\mathbf{j}(\mathbf{x},t)=j_i\,(\mathbf{x},t)\mathbf{e}_i$ are determined by the coupled constitutive relations \citep{Nowacki1974_1,Nowacki1974_2,Nowacki1974_3}
%
%
\begin{subequations}
\begin{align}
\tensor{\sigma}(\mathbf{x},t)&=\mathfrak{C}^{(m,\varepsilon)} \tensor{\varepsilon}(\mathbf{x},t)
-
\tensor{\alpha}^{(m,\varepsilon)}
\theta(\mathbf{x},t)
-
\tensor{\beta}^{(m,\varepsilon)}
\eta(\mathbf{x},t),\label{eq:MicroStress}\\
\mathbf{q}(\mathbf{x},t)&=-
\tensor{K}^{(m,\varepsilon)}
\nabla\theta(\mathbf{x},t),\label{eq:MicroThermalFlux}\\
\mathbf{j}(\mathbf{x},t)&=-
\tensor{D}^{(m,\varepsilon)}
\nabla\eta(\mathbf{x},t),\label{eq:MicroMassFlux}
\end{align}
\end{subequations}
where symbol $\tensor{\varepsilon}(\mathbf{x},t)=sym \nabla \mathbf{u}(\mathbf{x},t)$ is the micro small strains tensor and superscript $m$ refers to the microscale.
In equations (\ref{eq:MicroStress})-(\ref{eq:MicroMassFlux})
$\mathfrak{C}^{(m,\varepsilon)}=C_{ijkl}^{m}\left(\tensor{\xi}=\frac{\mathbf{x}}{\varepsilon}\right)\mathbf{e}_i\otimes\mathbf{e}_j\otimes\mathbf{e}_k\otimes\mathbf{e}_l$ is the fourth order micro elasticity tensor having major and minor symmetries,
$\tensor{\alpha}^{(m,\varepsilon)}=\alpha_{ij}^{m}\left(\tensor{\xi}=\frac{\mathbf{x}}{\varepsilon}\right)\mathbf{e}_i\otimes\mathbf{e}_j$ is the symmetric second order micro thermal dilatation tensor,
$\tensor{\beta}^{(m,\varepsilon)}=\beta_{ij}^{m}\left(\tensor{\xi}=\frac{\mathbf{x}}{\varepsilon}\right)\mathbf{e}_i\otimes\mathbf{e}_j$ is the symmetric second order micro diffusive expansion tensor,
$\tensor{K}^{(m,\varepsilon)}=K_{ij}^{m}\left(\tensor{\xi}=\frac{\mathbf{x}}{\varepsilon}\right)\mathbf{e}_i\otimes\mathbf{e}_j$ is the symmetric second order micro heat conduction tensor, and
$\tensor{D}^{(m,\varepsilon)}=D_{ij}^{m}\left(\tensor{\xi}=\frac{\mathbf{x}}{\varepsilon}\right)\mathbf{e}_i\otimes\mathbf{e}_j$ is the symmetric second order micro mass diffusion tensor. Micro constitutive tensors are all $\mathcal{Q}$-periodic and dependent upon the fast variable $\tensor{\xi}$.
Local balance equations hold
%
%
\begin{subequations}
\begin{align}
&\nabla\cdot\tensor{\sigma}(\mathbf{x},t)+\mathbf{b}(\mathbf{x},t)=\rho^m \ddot{\mathbf{u}}(\mathbf{x},t),\label{eq:MicroEqMotion}\\
&\nabla\cdot\tensor{q}(\mathbf{x},t)-r(x)=
-\tensor{\alpha}^{m}\dot{\tensor{\varepsilon}}(\mathbf{x},t)
-\psi^{m}\dot{\eta}(\mathbf{x},t)
-p^{m}\dot{\theta}(\mathbf{x},t),\label{eq:MicroThermalConductivity}\\
&\nabla\cdot\mathbf{j}(\mathbf{x},t)-s(\mathbf{x},t)=
-\tensor{\beta}^{m}\dot{\tensor{\varepsilon}}(\mathbf{x},t)
-\psi^{m}\dot{\theta}(\mathbf{x},t)
-q^{m}\dot{\eta}(\mathbf{x},t),\label{eq:MicroDiffusionEquation}
\end{align}
\end{subequations}
where source terms depend exclusively upon the slow variable and  time and  are represented by body forces
$\mathbf{b}(\mathbf{x},t)$,  heat sources
$ r(\mathbf{x},t)$, and  mass sources
$ s(\mathbf{x},t)$.
Source terms are here assumed to be $\mathcal{L}$-periodic  and to have vanishing mean values on $\mathcal{L}$, where, indicating with $L$ the structural characteristic size, the portion $\mathcal{L}=[0,L]\times[0,\delta L]$ can be considered as  truly representative  of the whole medium.
In this regard, size $L$ has to be much greater than the microstructural one ($L\gg\varepsilon$) so that the scales separation condition is met.
In equations (\ref{eq:MicroEqMotion})-(\ref{eq:MicroDiffusionEquation}) inertial terms are $\mathcal{Q}$-periodic and represented by the mass density $\rho^{(m,\varepsilon)}=\rho^m(\tensor{\xi}=\frac{\mathbf{x}}{\varepsilon})$, material constant  $p^{(m,\varepsilon)}=p^m(\tensor{\xi}=\frac{\mathbf{x}}{\varepsilon})$  related to the specific heat at constant strain and to thermo-diffusive effects, and material constant $q^{(m,\varepsilon)}=q^m(\tensor{\xi}=\frac{\mathbf{x}}{\varepsilon})$ related to diffusive effect.
Finally, term
$\psi^{(m,\varepsilon)}=\psi^m(\tensor{\xi}=\frac{\mathbf{x}}{\varepsilon})$ is a $\mathcal{Q}$-periodic coupling constant measuring  the thermo-diffusive effect.
Substitution of constitutive equations (\ref{eq:MicroStress})-(\ref{eq:MicroMassFlux}) into local balance relations (\ref{eq:MicroEqMotion})-(\ref{eq:MicroDiffusionEquation}) leads to
%
%
\begin{subequations}
\begin{align}
&\nabla\cdot
\left(
\mathfrak{C}^m\left(\frac{\mathbf{x}}{\varepsilon}\right)\nabla\mathbf{u}(\mathbf{x},t)
\right)
-
\nabla\cdot
\left(\tensor{\alpha}^m\left(\frac{\mathbf{x}}{\varepsilon}\right)\theta(\mathbf{x},t)\
\right)
-
\nabla\cdot
\left(\tensor{\beta}^m\left(\frac{\mathbf{x}}{\varepsilon}\right)\eta(\mathbf{x},t)\
\right)
+
\mathbf{b}(\mathbf{x},t)=\rho^m\left(\frac{\mathbf{x}}{\varepsilon}\right)\ddot{\mathbf{u}}(\mathbf{x},t),\label{eq:FieldEquMicroStress}\\
&\nabla\cdot
\left(\tensor{K}^m\left(\frac{\mathbf{x}}{\varepsilon}\right)\nabla\theta(\mathbf{x},t)
\right)
-
\tensor{\alpha}^{m}\left(\frac{\mathbf{x}}{\varepsilon}\right)\nabla\dot{\tensor{u}}(\mathbf{x},t)
-
\psi^{m}\left(\frac{\mathbf{x}}{\varepsilon}\right)\dot{\eta}(\mathbf{x},t)
+
r(\mathbf{x},t)=
p^{m}\left(\frac{\mathbf{x}}{\varepsilon}\right)\dot{\theta}(\mathbf{x},t),
\label{eq:FieldEquMicroThermalFlux}\\
&\nabla\cdot
\left(
\mathbf{D}^m\left(\frac{\mathbf{x}}{\varepsilon}\right)\nabla\mathbf{\eta}(\mathbf{x},t)
\right)
-
\tensor{\beta}^{m}\left(\frac{\mathbf{x}}{\varepsilon}\right)\nabla\dot{\tensor{u}}(\mathbf{x},t)
-
\psi^{m}\left(\frac{\mathbf{x}}{\varepsilon}\right)\dot{\theta}(\mathbf{x},t)
+
s(\mathbf{x},t)=
q^{m}\left(\frac{\mathbf{x}}{\varepsilon}\right)\dot{\eta}(\mathbf{x},t).\label{eq:FieldEquMicroMassFlux}
\end{align}
\end{subequations}
For an ideally bonded interface $\Sigma$, the following continuity conditions hold
  %
  %
  %
\begin{subequations}
\begin{align}
&
\left.
\left[
\left[
u_i
\right]
\right]
\right|_{\mathbf{x}\in\Sigma}=0,
\hspace{0.38cm}
\left.
\left[
\left[
\left(
C_{ijkl}^m\frac{\partial u_k}{\partial x_l}
-
\alpha_{ij}^m\theta
-
\beta_{ij}^m\eta
\right)
n_j
\right]
\right]
\right|_{\mathbf{x}\in\Sigma}=0,\label{eq:InterfaceCondition1}\\
&
\left.
\left[
\left[
\theta
\right]
\right]
\right|_{\mathbf{x}\in\Sigma}=0,
\hspace{0.5cm}
\left.
\left[
\left[
K_{ij}^m\frac{\partial\theta}{\partial x_j}n_i
\right]
\right]
\right|_{\mathbf{x}\in\Sigma}=0,\label{eq:InterfaceCondition2}\\
&
\left.
\left[
\left[
\eta
\right]
\right]
\right|_{\mathbf{x}\in\Sigma}=0,
\hspace{0.5cm}
\left.
\left[
\left[
D_{ij}^m\frac{\partial\eta}{\partial x_j}n_i
\right]
\right]
\right|_{\mathbf{x}\in\Sigma}=0,
\label{eq:InterfaceCondition3}
\end{align}
\end{subequations}
where $[[f]]=f^i(\Sigma)-f^j(\Sigma)$ denotes the discontinuity of the values of a function $f$ at the interface $\Sigma$ between two different phases $i$ and $j$ of periodic cell $\mathcal{A}$ and $\mathbf{n}=n_j\,\mathbf{e}_j$ represents the outward normal to the interface $\Sigma$.
%
%
%
%
%
%
%
Taking into account  the $\mathcal{Q}$-periodicity of micro constitutive tensors and inertial terms, interface conditions (\ref{eq:InterfaceCondition1})-(\ref{eq:InterfaceCondition3}),  and the $\mathcal{L}$-periodicity of source terms, it results that the microscopic fields spatially  depend  on both the slow  and the fast variables $\mathbf{x}$ and $\tensor{\xi}$ and are expressed as  
\begin{equation}
\mathbf{u}=\mathbf{u}\left(\mathbf{x},\tensor{\xi}=\frac{\mathbf{x}}{\varepsilon},t\right),
\hspace{0.2 cm}
{\theta}={\theta}\left(\mathbf{x},\tensor{\xi}=\frac{\mathbf{x}}{\varepsilon},t\right),
\hspace{0.2 cm}
\eta={\eta}\left(\mathbf{x},\tensor{\xi}=\frac{\mathbf{x}}{\varepsilon},t\right).\end{equation}
Rapidly oscillating $\mathcal{Q}$-periodic coefficients of PDEs (\ref{eq:FieldEquMicroStress})-(\ref{eq:FieldEquMicroMassFlux}) make their analytical and/or numerical resolution  particularly labor intensive.
In this sense, homogenization techniques can reveal very useful in replacing the microstructured continuum with an equivalent homogeneous one. 
In what follows, field equations of a first-order thermo-diffusive equivalent  continuum will be characterized and the closed form of overall constitutive tensors will be obtained.
By means of a dynamic multi-field asymptotic homogenization technique, the global behavior of the composite material will be  concisely and accurately described, thus overcoming the computational burden of resolution of equations  (\ref{eq:FieldEquMicroStress})-(\ref{eq:FieldEquMicroMassFlux}) and facilitating their analytical resolution  on simple domains.
Macroscopic fields  of the equivalent homogenized medium, are denoted as $\mathbf{U}(\mathbf{x},t)=U_i(\mathbf{x},t)\mathbf{e}_i$ for the displacement, $\Theta(\mathbf{x},t)$ for the relative temperature and $\Upsilon(\mathbf{x},t)$ for chemical potential. They only depend in space upon the macroscopic slow variable $\mathbf{x}$ and they result to be $\mathcal{L}$-periodic if source terms are $\mathcal{L}$-periodic. 
%
%
%
\subsection{ Asymptotic expansion of field equations at the microscale for the thermo-diffusive medium}
\label{subsec::espansioniAsintotiche}
%
%
%
In accordance with the procedure described in \citep{Bensoussan1978,Bakhvalov1984}, an asymptotic expansion of the microfields $\mathbf{u}(\mathbf{x},\mathbf{x}/\varepsilon,t),\theta(\mathbf{x},\mathbf{x}/\varepsilon,t)$ and $\eta(\mathbf{x},\mathbf{x}/\varepsilon,t)$ is performed in powers of the micro structural size $\varepsilon$
\begin{subequations}
\begin{align}
&
u_h
\left(\mathbf{x},\frac{\mathbf{x}}{\varepsilon},t\right)
=
\sum_{l=0}^{+\infty}
\varepsilon^lu_h^{(l)}\left(\mathbf{x},\frac{\mathbf{x}}{\varepsilon},t\right)=
u_h^{(0)}\left(\mathbf{x},\frac{\mathbf{x}}{\varepsilon},t\right)
+
\varepsilon u_h^{(1)}\left(\mathbf{x},\frac{\mathbf{x}}{\varepsilon},t \right)
+
\varepsilon^2u_h^{(2)}\left(\mathbf{x},\frac{\mathbf{x}}{\varepsilon},t\right)
+
O(\varepsilon^3),\label{eq:AsymptoticExpansionMicroDisplacement}\\
&
\theta
\left(\mathbf{x},\frac{\mathbf{x}}{\varepsilon},t\right)
=
\sum_{l=0}^{+\infty}
\varepsilon^l \theta^{(l)}\left(\mathbf{x},\frac{\mathbf{x}}{\varepsilon},t\right)=
\theta^{(0)}\left(\mathbf{x},\frac{\mathbf{x}}{\varepsilon},t\right)
+
\varepsilon \theta^{(1)}\left(\mathbf{x},\frac{\mathbf{x}}{\varepsilon},t\right)
+
\varepsilon^2 \theta^{(2)}\left(\mathbf{x},\frac{\mathbf{x}}{\varepsilon},t\right)
+
O(\varepsilon^3),\label{eq:AsymptoticExpansionMicroTemperature}\\
&
\eta
\left(\mathbf{x},\frac{\mathbf{x}}{\varepsilon},t\right)
=
\sum_{l=0}^{+\infty}
\varepsilon^l \eta^{(l)}\left(\mathbf{x},\frac{\mathbf{x}}{\varepsilon},t\right)=
\eta^{(0)}\left(\mathbf{x},\frac{\mathbf{x}}{\varepsilon},t\right)
+
\varepsilon \eta^{(1)}\left(\mathbf{x},\frac{\mathbf{x}}{\varepsilon},t\right)
+
\varepsilon^2 \eta^{(2)}\left(\mathbf{x},\frac{\mathbf{x}}{\varepsilon},t\right)
+
O(\varepsilon^3).\label{eq:AsymptoticExpansionMicroChemicalPotential}
\end{align}
\end{subequations}
Taking into account the property
$$
\frac{D}{D x_j}f\left(\mathbf{x},\tensor{\xi}=\frac{\mathbf{x}}{\varepsilon},t\right)=
\left(\frac{\partial f}{\partial x_j}
+
\frac{1}{\varepsilon}
\left.
\frac{\partial f}{\partial \xi_j}
\right)
\right|_{\tensor{\xi=\frac{\mathbf{x}}{\varepsilon}}}=
\left.
\left(
\frac{\partial f}{\partial x_j}
+
\frac{1}{\varepsilon} f_{,j}
\right)
\right|_{\tensor{\xi}=\frac{\mathbf{x}}{\varepsilon}},
$$
asymptotic expansions (\ref{eq:AsymptoticExpansionMicroDisplacement})-(\ref{eq:AsymptoticExpansionMicroChemicalPotential}) are substituted into the local field equations (\ref{eq:FieldEquMicroStress})-(\ref{eq:FieldEquMicroMassFlux}).
From equation (\ref{eq:FieldEquMicroStress}) one has
\begin{eqnarray}
\label{eq:RecursiveDifferentialProblemMicroDisplacement}
&& \left\{
\varepsilon^{-2}
\left(
C_{ijkl}^m\, u_{k,l}^{(0)}
\right)_{,j}
+
\right.
 \nonumber\\
&&+ \varepsilon^{-1}
\left\{
\left[
C_{ijkl}^{m}
\left(
\frac{\partial u_{k}^{(0)}}{\partial x_l}+
u_{k,l}^{(1)}
\right)
\right]_{,j}+
\frac{\partial}{\partial x_j}
\left(
C_{ijkl}^{m}\,
u_{k,l}^{(0)}
\right)
-
\left(
\alpha_{ij}^m\,\theta^{(0)}\right)_{,j}
-
\left(
\beta_{ij}^m\,\eta^{(0)}\right)_{,j}
\right\}
\nonumber\\
&&+
\left\{
\left[
C_{ijkl}^{m}
\left(
\frac{\partial u_{k}^{(1)}}{\partial x_l}+
u_{k,l}^{(2)}
\right)
\right]_{,j}+
\frac{\partial}{\partial x_j}
\left[
C_{ijkl}^{m}\,
\left(
\frac{\partial u_k^{(0)}}{\partial x_l}
+
u_{k,l}^{(1)}
\right)
\right]
-
\left(
\alpha_{ij}^m\,\theta^{(1)}\right)_{,j}
-
\frac{\partial}{\partial x_j}
\left(
\alpha_{ij}^m\,\theta^{(0)}
\right)
+
\right.
\nonumber\\
&&-\left.
\left(
\beta_{ij}^m\,\eta^{(1)}\right)_{,j}
-
\frac{\partial}{\partial x_j}
\left(\beta_{ij}^m\,\eta^{(0)}\right)
\right\}+
\nonumber\\
&&+ \varepsilon \left\{
\left[
C_{ijkl}^{m}
\left(
\frac{\partial u_{k}^{(2)}}{\partial x_l}+
u_{k,l}^{(3)}
\right)
\right]_{,j}+
\frac{\partial}{\partial x_j}
\left[
C_{ijkl}^{m}\,
\left(
\frac{\partial u_k^{(1)}}{\partial x_l}
+
u_{k,l}^{(2)}
\right)
\right]
-
\left(
\alpha_{ij}^m\,\theta^{(2)}\right)_{,j}
-
\frac{\partial}{\partial x_j}
\left(
\alpha_{ij}^m\,\theta^{(1)}
\right)
+
\right.
\nonumber\\
&&-\left.
\left.
\left.
\left(
\beta_{ij}^m,\,\eta^{(2)}\right)_{,j}
-
\frac{\partial}{\partial x_j}
\left(\beta_{ij}^m\,\eta^{(1)}\right)
-\rho^m\frac{\partial^2 u_i^{(0)}}{\partial t^2}
-
\varepsilon \rho^m\frac{\partial^2 u_i^{(1)}}{\partial t^2}+O(\varepsilon^2)\right\}\right|_{\tensor{\xi=\frac{\mathbf{x}}{\varepsilon}}}
\right.
+b_i(\mathbf{x},t)= 0.
\end{eqnarray}
Analogously, field equation (\ref{eq:FieldEquMicroThermalFlux}) leads to
\begin{eqnarray}
\label{eq:RecursiveDifferentialProblemsMicroTemperature}
& &\left\{
\varepsilon^{-2}
\left( 
K_{ij}^m\, \theta_{,j}^{(0)}
\right)_{,i}
+ \varepsilon^{-1}
\left\{
\left[
K_{ij}^{m}
\left(
\frac{\partial \theta^{(0)}}{\partial x_j}+
\theta_{,j}^{(1)}
\right)
\right]_{,i}+
\frac{\partial}{\partial x_i}
\left(
K_{ij}^{m}\,
\theta_{,j}^{(0)}
\right)
-
\alpha_{ij}^m
\frac{\partial u_{i,j}^{(0)}}{\partial t}
\right.\right\}
+\nonumber\\
&&+
\left[
K_{ij}^{m}
\left(
\frac{\partial \theta^{(1)}}{\partial x_j}
+
\theta_{,j}^{(2)}
\right)
\right]_{,i}
+
\frac{\partial}{\partial x_i}
\left[
K_{ij}^{m}
\left(
\frac{\partial \theta^{(0)}}{\partial x_j}+
\theta_{,j}^{(1)}
\right)\right]
-
\alpha_{ij}^m
\left[
\frac{\partial^2 u_i^{(0)}}{\partial x_j\partial t}
+
\frac{\partial u_{i,j}^{(1)}}{\partial t}
\right]
-
\psi^m
\frac{\partial\eta^{(0)}}{\partial t}
 + \nonumber\\
&&+
\varepsilon
\left\{
\left[
K_{ij}^{m}
\left(
\frac{\partial \theta^{(2)}}{\partial x_j}
+
\theta_{,j}^{(3)}
\right)
\right]_{,i}
+
\frac{\partial}{\partial x_i}
\left[
K_{ij}^{m}
\left(\frac{\partial \theta^{(1)}}{\partial x_j}+\theta_{,j}^{(2)}
\right)
\right]
-
\alpha_{ij}^m
\left[
\frac{\partial^2 u_i^{(1)}}{\partial x_j \partial t}
+
\frac{\partial u_{i,j}^{(2)}}{\partial t}
\right]
-\psi^m
\frac{\partial \eta^{(1)}}{\partial t}
\right.+ \nonumber\\
&&-
\left.
\left.
p^m\frac{\partial \theta^{(0)}}{\partial t}-
\varepsilon p^m\frac{\partial \theta^{(1)}}{\partial t}
O(\varepsilon^2)\right\}\right|_{\tensor{\xi}=\frac{\mathbf{x}}{\varepsilon}}
+r(\mathbf{x},t)=0,
\end{eqnarray}
and equation (\ref{eq:FieldEquMicroMassFlux}) results
\begin{eqnarray}
\label{eq:RecursiveDifferentialProblemsMicroChemPot}
& &\left\{
\varepsilon^{-2}
\left( 
D_{ij}^m\, \eta_{,j}^{(0)}
\right)_{,i}
+ \varepsilon^{-1}
\left\{
\left[
D_{ij}^{m}
\left(
\frac{\partial \eta^{(0)}}{\partial x_j}+
\eta_{,j}^{(1)}
\right)
\right]_{,i}+
\frac{\partial}{\partial x_i}
\left(
D_{ij}^{m}\,
\eta_{,j}^{(0)}
\right)
-
\beta_{ij}^m
\frac{\partial u_{i,j}^{(0)}}{\partial t}
\right.\right\}
+\nonumber\\
&&+
\left[
D_{ij}^{m}
\left(
\frac{\partial \eta^{(1)}}{\partial x_j}
+
\eta_{,j}^{(2)}
\right)
\right]_{,i}
+
\frac{\partial}{\partial x_i}
\left[
D_{ij}^{m}
\left(
\frac{\partial \eta^{(0)}}{\partial x_j}+
\eta_{,j}^{(1)}
\right)\right]
-
\beta_{ij}^m
\left[
\frac{\partial^2 u_i^{(0)}}{\partial x_j\partial t}
+
\frac{\partial u_{i,j}^{(1)}}{\partial t}
\right]
-
\psi^m
\frac{\partial\theta^{(0)}}{\partial t}
 + \nonumber\\
&&+
\varepsilon
\left\{
\left[
D_{ij}^{m}
\left(
\frac{\partial \eta^{(2)}}{\partial x_j}
+
\eta_{,j}^{(3)}
\right)
\right]_{,i}
+
\frac{\partial}{\partial x_i}
\left[
D_{ij}^{m}
\left(\frac{\partial \eta^{(1)}}{\partial x_j}+\eta_{,j}^{(2)}
\right)
\right]
-
\beta_{ij}^m
\left[
\frac{\partial^2 u_i^{(1)}}{\partial x_j \partial t}
+
\frac{\partial u_{i,j}^{(2)}}{\partial t}
\right]
-\psi^m
\frac{\partial \theta^{(1)}}{\partial t}
\right.+ \nonumber\\
&&-
\left.
\left.
q^m\frac{\partial \eta^{(0)}}{\partial t}-
\varepsilon q^m\frac{\partial \eta^{(1)}}{\partial t}+
O(\varepsilon^2)
\right\}\right|_{\tensor{\xi}=\frac{\mathbf{x}}{\varepsilon}}
+s(\mathbf{x},t)=0.
\end{eqnarray}
Denoting with $\Sigma_1$  the interface between  two distinct phases in the unit cell $\mathcal{Q}$, asymptotic expansions (\ref{eq:AsymptoticExpansionMicroDisplacement})-(\ref{eq:AsymptoticExpansionMicroChemicalPotential}) allow rephrasing interface conditions (\ref{eq:InterfaceCondition1})-(\ref{eq:InterfaceCondition3}) over the unit cell $\mathcal{Q}$ in terms of the fast variable $\tensor{\xi}$ \citep{Bakhvalov1984}. In particular, equations (\ref{eq:InterfaceCondition1}) become 
%
%
%
\begin{eqnarray}
\label{eq:InterfaceConditionMicroDisplacementOnSigma1}
&&\left.
\left[
\left[
u_h^{(0)}
\right]
\right]
\right|_{\tensor{\xi}\in\Sigma_1}
+
\varepsilon
\left.
\left[
\left[
u_h^{(1)}
\right]
\right]
\right|_{\tensor{\xi}\in\Sigma_1}
+
O(\varepsilon^2)
=0,
\nonumber\\
&&
\frac{1}{\varepsilon}
\left.
\left[
\left[
C_{ijkl}^{m}\,u_{k,l}^{(0)}
n_j
\right]
\right]
\right|_{\tensor{\xi}\in\Sigma_1}
+
\left.
\left[
\left[
\left\{
C_{ijkl}^{m}
\left(
\frac{\partial u_k^{(0)}}{\partial x_l}+
u_{k,l}^{(1)}
\right)
-
\alpha_{ij}^m\,\theta^{(0)}
-
\beta_{ij}^m\,\eta^{(0)}
\right\}n_j
\right]
\right]
\right|_{\tensor{\xi}\in\Sigma_1}
+
\nonumber\\
&&+
\varepsilon
\left.
\left[
\left[
\left\{
C_{ijkl}^{m}
\left(
\frac{\partial u_k^{(1)}}{\partial x_l}+
u_{k,l}^{(2)}
\right)
-
\alpha_{ij}^m\,\theta^{(1)}
-
\beta_{ij}^m\,\eta^{(1)}
\right\}n_j
\right]
\right]
\right|_{\tensor{\xi}\in\Sigma_1}
+
O(\varepsilon^2) =0,
\end{eqnarray} 
equations (\ref{eq:InterfaceCondition2}) are written as
\begin{eqnarray}
\label{eq:InterfaceConditionMicroTemperatureOnSigma1}
&&\left.
\left[
\left[
\theta^{(0)}
\right]
\right]
\right|_{\tensor{\xi}\in\Sigma_1}
+
\varepsilon
\left.
\left[
\left[
\theta^{(1)}
\right]
\right]
\right|_{\tensor{\xi}\in\Sigma_1}
+
O(\varepsilon^2)
=0,
\nonumber\\
&&
\frac{1}{\varepsilon}
\left.
\left[
\left[
K_{ij}^{m}\,\theta_{,j}^{(0)}
\,n_i
\right]
\right]
\right|_{\tensor{\xi}\in\Sigma_1}
+
\left.
\left[
\left[
K_{ij}^{m}
\left(
\frac{\partial \theta^{(0)}}{\partial x_j}
+
\theta_{,j}^{(1)}
\right)
n_i
\right]
\right]
\right|_{\tensor{\xi}\in\Sigma_1}+
\nonumber\\
&&
+
\varepsilon
\left.
\left[
\left[
K_{ij}^{m}
\left(
\frac{\partial \theta^{(1)}}{\partial x_j}
+
\theta_{,j}^{(2)}
\right)
n_i
\right]
\right]
\right|_{\tensor{\xi}\in\Sigma_1}
+
O(\varepsilon^2)=0,
\end{eqnarray}
and interface conditions (\ref{eq:InterfaceCondition3}) involving chemical potential  turn into
\begin{eqnarray}
\label{eq:InterfaceConditionMicroChemPotOnSigma1}
&&\left.
\left[
\left[
\eta^{(0)}
\right]
\right]
\right|_{\tensor{\xi}\in\Sigma_1}
+
\varepsilon
\left.
\left[
\left[
\eta^{(1)}
\right]
\right]
\right|_{\tensor{\xi}\in\Sigma_1}
+
O(\varepsilon^2)
=0,
\nonumber\\
&&
\frac{1}{\varepsilon}
\left.
\left[
\left[
D_{ij}^{m}\,\eta_{,j}^{(0)}
\,n_i
\right]
\right]
\right|_{\tensor{\xi}\in\Sigma_1}
+
\left.
\left[
\left[
D_{ij}^{m}
\left(
\frac{\partial \eta^{(0)}}{\partial x_j}
+
\eta_{,j}^{(1)}
\right)
n_i
\right]
\right]
\right|_{\tensor{\xi}\in\Sigma_1}+
\nonumber\\
&&
+
\varepsilon
\left.
\left[
\left[
D_{ij}^{m}
\left(
\frac{\partial \eta^{(1)}}{\partial x_j}
+
\eta_{,j}^{(2)}
\right)
n_i
\right]
\right]
\right|_{\tensor{\xi}\in\Sigma_1}
+
O(\varepsilon^2)=0.
\end{eqnarray}
In the followings, recursive differential problems originating from equations (\ref{eq:RecursiveDifferentialProblemMicroDisplacement})-(\ref{eq:RecursiveDifferentialProblemsMicroChemPot}) are written explicitly at the different orders of length $\varepsilon$ till the order $\varepsilon^0$, leading to the definition of cell problems in Section \ref{Sec::CellProblems}.
%
%
%
\\
\\
{\itshape Recursive differential problems at the order $\varepsilon^{-2}$}
\\
\\
From equation (\ref{eq:RecursiveDifferentialProblemMicroDisplacement}), at the order $\varepsilon^{-2}$ one has the following differential problem
\begin{equation}
\label{eq:DiffProbMecc-2}
\left(
C_{ijkl}^m u_{k,l}^{(0)}
\right)_{,j}
=
f_i^{(0)}(\mathbf{x},t)
\end{equation}
with interface conditions
\\
\begin{equation}
\label{eq:InterfaceConditionMicroDisplacement-2}
\left.
\left[
\left[
u_k^{(0)}
\right]
\right]
\right|_{\tensor{\xi}\in\Sigma_1}=0,
\hspace{0.5cm}
\left.
\left[
\left[
C_{ijkl}^{m}\,u_{k,l}^{(0)}
n_j
\right]
\right]
\right|_{\tensor{\xi}\in\Sigma_1}=0.
\end{equation}
It results that $f_i^{(0)}=0$ in equation (\ref{eq:DiffProbMecc-2}) because of solvability condition of problem (\ref{eq:DiffProbMecc-2}) in the class of $\mathcal{Q}$-periodic functions and interface conditions (\ref{eq:InterfaceConditionMicroDisplacement-2}) \citep{Bakhvalov1984}, and the solution $u_k^{(0)}$ spatially depends  only upon the slow variable $\mathbf{x}$, being equal to the macroscopic field 
%
\begin{equation}
\label{eq:SolutionDiffProbMecc-2}
u_k^{(0)}(\mathbf{x},\tensor{\xi},t)={U}_k(\mathbf{x},t).
\end{equation}
At the order $\varepsilon^{-2}$, from  equation (\ref{eq:RecursiveDifferentialProblemsMicroTemperature}) one has
%
\begin{equation}
\label{eq:DiffProbTherm-2}
\left(
K_{ij}^m \theta_{,j}^{(0)}
\right)_{,i}
=
g^{(0)}(\mathbf{x},t),
\end{equation}
with relative interface conditions from (\ref{eq:InterfaceConditionMicroTemperatureOnSigma1}) that hold
\begin{equation}
\label{eq:InterfaceConditionMicroTemperature-2}
\left.
\left[
\left[
\theta^{(0)}
\right]
\right]
\right|_{\tensor{\xi}\in\Sigma_1}=0,
\hspace{0.5cm}
\left.
\left[
\left[
K_{ij}^{m}\,\theta_{,j}^{(0)}
\,n_i
\right]
\right]
\right|_{\tensor{\xi}\in\Sigma_1}=0.
\end{equation}
For the same reasons explicited above, the solution $\theta^{(0)}$ is equal to the macroscopic temperature field, namely
\begin{equation}
\label{eq:SolutionDiffProbTherm-2}
\theta^{(0)}\left(\mathbf{x},\tensor{\xi},t\right)=\Theta(\mathbf{x},t).
\end{equation}
Analogously, from equation (\ref{eq:RecursiveDifferentialProblemsMicroChemPot}) differential problem obtained at the order $\varepsilon^{-2}$ has the form
\begin{equation}
\label{eq:DiffProbDiffus-2}
\left(
D_{ij}^m \eta_{,j}^{(0)}
\right)_{,i}
=
h^{(0)}(\mathbf{x}),
\end{equation}
 with relative interface conditions from equation (\ref{eq:InterfaceConditionMicroChemPotOnSigma1}) that read 
\begin{equation}
\label{eq:InterfaceConditionMicroChemPot-2}
\left.
\left[
\left[
\eta^{(0)}
\right]
\right]
\right|_{\tensor{\xi}\in\Sigma_1}=0,
\hspace{0.5cm}
\left.
\left[
\left[
D_{ij}^{m}\,\eta_{,j}^{(0)}
\,n_i
\right]
\right]
\right|_{\tensor{\xi}\in\Sigma_1}=0.
\end{equation}
Once again, solution of (\ref{eq:DiffProbDiffus-2}) corresponds to the macroscopic chemical potential and it is expressed as
\begin{equation}
\label{eq:SolutionDiffProbDiffus-2}
\eta^{(0)}(\mathbf{x},\tensor{\xi},t)
=\Upsilon(\mathbf{x},t).
\end{equation}
%
%
%
\\
{\itshape  Recursive differential problems at the order $\varepsilon^{-1}$ }
\\
\\
Taking into account solutions (\ref{eq:SolutionDiffProbMecc-2}), (\ref{eq:SolutionDiffProbTherm-2}), and (\ref{eq:SolutionDiffProbDiffus-2}) of problems at the order $\varepsilon^{-2}$, at the order $\varepsilon^{-1}$ from equation (\ref{eq:RecursiveDifferentialProblemMicroDisplacement})  one has the following differential problem  
%
\begin{equation}
\label{eq:DiffProbMecc-1}
\left(
C_{ijkl}^m 
u_{k,l}^{(1)}
\right)_{,j}
+\left(
C_{ijkl}^m
\frac{\partial U_k}{\partial x_l}
\right)_{,j}
-
\alpha_{ij,j}^m \Theta
-
\beta_{ij,j}^m \Upsilon
=
f_i^{(1)}(\mathbf{x},t),
\end{equation}
with interface conditions expressed as
\begin{equation}
\label{eq:InterfaceConditionMicroDisplacement-1}
\left.
\left[
\left[
u_h^{(1)}
\right]
\right]
\right|_{\tensor{\xi}\in\Sigma_1}=0,
\hspace{0.5cm}
\left.
\left[
\left[
\left\{
C_{ijkl}^{m}
\left(
\frac{\partial U_k}{\partial x_l}+
u_{k,l}^{(1)}
\right)
-
\alpha_{ij}^m\,\Theta
-
\beta_{ij}^m\,\Upsilon
\right\}n_j
\right]
\right]
\right|_{\tensor{\xi}\in\Sigma_1}=0.
\end{equation}
Given the $Q$-periodicity of components $C^m_{ijkl}$, $\alpha^m_{ij}$, and $\beta^m_{ij}$, solvability condition of problem (\ref{eq:DiffProbMecc-1}) imposes that
\begin{equation}
f_i^{(1)}(\mathbf{x},t)
=
\left\langle
C_{ijkl,j}^{m}
\right\rangle
\frac{\partial U_k}{\partial x_l}
-
\left 
\langle
\alpha_{ij,j}^{m}
\right\rangle
\Theta
-
\left\langle
\beta_{ij,j}^{m}
\right\rangle
\Upsilon,
\end{equation}
%
%
where $\left\langle\left(\cdot\right)\right\rangle
=
\frac{1}{|\mathcal{Q}}|\int_{\mathcal{Q}}(\cdot)\,d\tensor{\xi}$ and $|\mathcal{Q}|=\delta$ denotes the area of the unit cell.
Solutions (\ref{eq:SolutionDiffProbMecc-2}), (\ref{eq:SolutionDiffProbTherm-2}), and (\ref{eq:SolutionDiffProbDiffus-2}), make the micro displacement solution at the order $\varepsilon^{-1}$ of the form
\begin{equation}
\label{eq:SolutionDiffProbMecc-1}
u_k^{(1)}
\left(
\mathbf{x},\tensor{\xi},t
\right)
=
N_{kpq_1}^{(1)}(\tensor{\xi})\frac{\partial U_p(\mathbf{x},t)}{\partial x_{q_1}}
+
\tilde{N}_{k}^{(1)}(\tensor{\xi}) \Theta(\mathbf{x},t)
+
\hat{N}_k^{(1)}(\tensor{\xi})\Upsilon(\mathbf{x},t),
\end{equation}
where $N_{kpq_1}^{(1)}$, $\tilde{N}_{k}^{(1)}$, and $\hat{N}_k^{(1)}$ are the first-order perturbation functions for the mechanical problem. These  are $\mathcal{Q}$-periodic functions and reflect the effects of the underlying microstructure being spatially dependent only upon $\tensor{\xi}$. 
At the order $\varepsilon^{-1}$, from equation (\ref{eq:RecursiveDifferentialProblemsMicroTemperature}) one obtains
\begin{equation}
\label{eq:DiffProbTherm-1}
\left(
K_{ij}^m 
\theta_{,j}^{(1)}
\right)_{,i}
+
\left(
K_{ij}^m
\frac{\partial \Theta}{\partial x_j}
\right)_{,i}
=
g^{(1)}(\mathbf{x},t),
\end{equation}
and relative interface conditions from (\ref{eq:InterfaceConditionMicroTemperatureOnSigma1}) read
 \begin{equation}
 \label{eq:InterfaceConditionMicroTemperature-1}
 \left.
\left[
\left[
\theta^{(1)}
\right]
\right]
\right|_{\tensor{\xi}\in\Sigma_1}=0,
\hspace{0.5cm}
\left.
\left[
\left[
K_{ij}^{m}
\left(
\frac{\partial \Theta}{\partial x_j}
+
\theta_{,j}^{(1)}
\right)
n_i
\right]
\right]
\right|_{\tensor{\xi}\in\Sigma_1}=0.
\end{equation}
Solvability of differential problem (\ref{eq:DiffProbTherm-1}), taking into account the $\mathcal{Q}$-periodicity of components $K_{ij}^m$  leads to 
\begin{equation}
g^{(1)}(\mathbf{x},t)=
\left\langle
K_{ij,j}^m
\right\rangle=0.
\end{equation}
Therefore, solution of (\ref{eq:DiffProbTherm-1}) has the form
\begin{equation}
\label{eq:SolutionDiffProbTherm-1}
\theta^{(1)}(\mathbf{x},\tensor{\xi},t)=M_{q_1}^{(1)}(\tensor{\xi})\frac{\partial\Theta}{\partial x_{q_1}},
\end{equation}
with perturbation function $M_{q_1}^{(1)}$.
Analogously to what done for thermal problem, from equation (\ref{eq:RecursiveDifferentialProblemsMicroChemPot}) diffusion problem at the order $\varepsilon^{-1}$ has the form 
\begin{equation}
\label{eq:DiffProbDiffus-1}
\left(
D_{ij}^m 
\eta^{(1)}_{,j}
\right)_{,i}
+
\left(
D_{ij}^m
\frac{\partial\Upsilon}{\partial x_j}
\right)_{,i}
=h^{(1)}(\mathbf{x},t),
\end{equation}
and its interface conditions read
 \begin{equation}
 \label{eq:InterfaceConditionMicroChemPot-1}
 \left.
\left[
\left[
\eta^{(1)}
\right]
\right]
\right|_{\tensor{\xi}\in\Sigma_1}=0,
\hspace{0.5cm}
\left.
\left[
\left[
D_{ij}^{m}
\left(
\frac{\partial \Upsilon}{\partial x_j}
+
\eta_{,j}^{(1)}
\right)
n_i
\right]
\right]
\right|_{\tensor{\xi}\in\Sigma_1}=0.
\end{equation}
Solvability condition for problem (\ref{eq:DiffProbDiffus-1}) imposes  that
\begin{equation}
h^{(1)}(\mathbf{x},t)=
\left\langle
D_{ij,i}^m
\right\rangle=0,
\end{equation}
and the solution $h^{(1)}(\mathbf{x},t)$ has the form
\begin{equation}
\label{eq:SolutionDiffProbDiffus-1}
\eta^{(1)}(\mathbf{x},\tensor{\xi},t)=W_{q_1}^{(1)}(\mathbf{\tensor{\xi}})\frac{\partial \Upsilon(\mathbf{x},t)}{\partial x_{q_1}},
\end{equation}
with first-order perturbation function $W_{q_1}^{(1)}$.
\\
\\
{\itshape Recursive differential problems at the order $\varepsilon^0$}
\\
\\
Bearing in mind  the two sets of solutions (\ref{eq:SolutionDiffProbMecc-2}), (\ref{eq:SolutionDiffProbTherm-2}), (\ref{eq:SolutionDiffProbDiffus-2}) and (\ref{eq:SolutionDiffProbMecc-1}), (\ref{eq:SolutionDiffProbTherm-1}), (\ref{eq:SolutionDiffProbDiffus-1}) of differential problems at the order $\varepsilon^{-2}$ and $\varepsilon^{-1}$, respectively, equation (\ref{eq:RecursiveDifferentialProblemMicroDisplacement}) at the order $\varepsilon^{0}$ yields 
\begin{eqnarray}
\label{eq:DiffProbMecc+0}
&&
\left(
C_{ijkl}^m 
+
u_{k,l}^{(2)}
\right)_{,j}
+
\left[
\left(
C_{ijkl}^m N_{kpq_1}^{(1)}
\right)_{,j}
+
C_{iq_1pl}^m
+
C^m_{ilkj}
N_{kpq_1,j}^{(1)}
\right]\frac{\partial^2 U_p}{\partial x_{q_1}\partial x_l}
+\nonumber\\
&&+
\left[
\left(
C_{ijkl}^{m}
\tilde{N}_k^{(1)}
\right)_{,j}
+
C_{ilkj}^{m}
\tilde{N}_{k,j}^{(1)}
-\left(\alpha_{ij}^{m}M_l^{(1)}
\right)_{,j}
-
\alpha_{il}^{m}
\right]
\frac{\partial\Theta}{\partial x_l}
+
\nonumber\\
&&+
\left[
\left(
C_{ijkl}^{m}
\hat{N}_k^{(1)}
\right)_{,j}
+
C_{ilkj}^{m}
\hat{N}_{k,j}^{(1)}
-
\left(\beta_{ij}^{m}W_l^{(1)}
\right)_{,j}
-
\beta_{il}^{m}
\right]
\frac{\partial\Upsilon}{\partial x_l}
-\rho^m\frac{\partial^2 U_i}{\partial t^2}=
f_i^{(2)}(\mathbf{x},t),
\end{eqnarray}
with interface conditions
\begin{eqnarray}
\label{eq:InterfaceConditionMicroDisplacement+0}
&&\left.
\left[
\left[
u_h^{(2)}
\right]
\right]
\right|_{\tensor{\xi}\in\Sigma_1}=0,
\nonumber\\
&&
\left.
\left[
\left[
\left\{
C_{ijkl}^{m}
\left(
u_{k,l}^{(2)}
+
N_{kpq_1}^{(1)}
\frac{\partial^2 U_p}{\partial x_{q_1}\partial x_l}
+
\tilde{N}_{k}^{(1)}
\frac{\partial \Theta}{\partial x_l}
+
\hat{N}_{k}^{(1)}
\frac{\partial \Upsilon}{\partial x_l}
\right)
\right.
\right.
\right.
\right.
+
\nonumber\\
&&
\left.
\left.
\left.
\left.
-\alpha_{ij}^{m}
\delta_{q_1l}
M_{q_1}^{(1)}
\frac{\partial \Theta}{\partial x_l}
-
\beta_{ij}^{m}
\delta_{q_1l}
W_{q_1}^{(1)}
\frac{\partial \Upsilon}{\partial x_l}
\right\}
n_j
\right]
\right]
\right|_{\tensor{\xi}\in\Sigma_1}=0.
\end{eqnarray}
Solvability condition for problem (\ref{eq:DiffProbMecc+0}) leads to the following condition for $f_i^{(2)}$
\begin{equation}
f_i^{(2)}(\mathbf{x},t)
=
\left\langle
C_{iq_1pl}^m 
+
C_{ilkj}^m N_{kpq_1,j}^{(1)}
\right\rangle
\frac{\partial^2 U_p}{\partial x_{q_1}\partial x_l}
+
\left\langle
C_{ilkj}^{m}
\tilde{N}_{k,j}^{(1)}
-\alpha_{il}^m
\right\rangle
\frac{\partial\Theta}{\partial x_l}
+
\left\langle
C_{ilkj}^{m}
\hat{N}_{k,j}^{(1)}
-
\beta_{il}^m
\right\rangle
\frac{\partial \Upsilon}{\partial x_l}
-
\left\langle
\rho^m
\right\rangle
\frac{\partial^2 U_i}{\partial t^2},
\end{equation}
and the solution has the form
\begin{equation}
\label{eq:SolutionDiffProbMecc0}
u_k^{(2)}(\mathbf{x},\tensor{\xi},t)=N_{kpq_1q_2}^{(2)}(\tensor{\xi})
\frac{\partial^2 U_p(\mathbf{x},t)}{\partial x_{q_1}\partial x_{q_2}}
+
\tilde{N}_{kq_1}^{(2)}(\tensor{\xi})
\frac{\partial\Theta(\mathbf{x},t)}{\partial x_{q_1}}
+
\hat{N}_{kq_1}^{(2)}(\tensor{\xi})
\frac{\partial\Upsilon(\mathbf{x},t)}{\partial x_{q_1}}
+
N^{(2,2)}_{kp}(\tensor{\xi})\frac{\partial^2 U_p(\mathbf{x},t)}{\partial t^2},
\end{equation}
where $N_{kpq_1q_2}^{(2)},\tilde{N}_{kq_1}^{(2)},\hat{N}_{kq_1}^{(2)}$, and $N^{(2,2)}_{kp}$ are the second order perturbation functions relative to the mechanical problem.
From equation ({\ref{eq:RecursiveDifferentialProblemsMicroTemperature}}), thermal problem at the order $\varepsilon^0$ reads  
\begin{eqnarray}
\label{eq:DiffProbTherm+0}
&&\left(
K_{ij}^m 
\theta_{,j}^{(2)}
\right)_{,i}
+
\left[
\left(
K_{ij}^m
M_{q_1}^{(1)}
\right)_{,i}
+
K_{q_1j}^{m}
+
K_{ji}^{m}
M_{q_1,i}^{(1)}
\right]
\frac{\partial^2\Theta}{\partial x_{q_1}\partial x_j}
-
\left(
\alpha_{ij}^{m}
N_{ipq_1,j}^{(1)}
+
\alpha_{pq_1}^{m}
\right)
\frac{\partial^2 U_p}{\partial x_{q_1}\partial t}
+\nonumber\\
&&-
\left(
\alpha_{ij}
\tilde{N}_{i,j}^{(1)}
+
p^m
\right)
\frac{\partial\Theta}{\partial t}
-
\left(
\alpha_{ij}^m
\hat{N}_{i,j}^{(1)}
+
\psi^m
\right)
\frac{\partial \Upsilon}{\partial t}
=
g^{(2)}(\mathbf{x},t),
\end{eqnarray}
and relative interface conditions have the following form
\begin{equation}
\label{eq:InterfaceConditionMicroTemperature+0}
\left.
\left[
\left[
\theta^{(2)}
\right]
\right]
\right|_{\tensor{\xi}\in\Sigma_1}=0,
\hspace{0.5cm}
\left.
\left[
\left[
K_{ij}^{m}
\left(
\theta_{,j}^{(2)}
+
M_{q_1}^{(1)}
\frac{\partial^2 \Theta}{\partial x_{q_1}\partial x_j}
\right)
n_i
\right]
\right]
\right|_{\tensor{\xi}\in\Sigma_1}=0.
\end{equation}
Solvability condition for (\ref{eq:DiffProbTherm+0}) entails that
\begin{eqnarray}
&&g^{(2)}(\mathbf{x},t)
=
\left\langle
\left(
K_{ij}^m M_{q_1}^{(1)}
\right)_{,i}
+ K_{q_1j}^m
+
K_{ji}^m
M_{q_1,i}^{(1)}
\right\rangle
\frac{\partial^2 \Theta}{\partial x_{q_1}\partial x_j}
-
\left\langle
\alpha_{ij}^{m}
N_{ipq_1,j}^{(1)}
+
\alpha_{pq_1}^{m}
\right\rangle
\frac{\partial^2 U_p}{\partial x_{q_1}\partial t}
+\nonumber\\
&&-
\left\langle
\alpha_{ij}
\tilde{N}_{i,j}^{(1)}
+
p^m
\right\rangle
\frac{\partial\Theta}{\partial t}
-
\left\langle
\alpha_{ij}^m
\hat{N}_{i,j}^{(1)}
+
\psi^m
\right\rangle
\frac{\partial \Upsilon}{\partial t},
\end{eqnarray}
and  solution reads
\begin{eqnarray}
\label{eq:SolutionDiffProbTherm+0}
\theta^{(2)}(\mathbf{x},\tensor{\xi},t)
=
M_{q_1q_2}^{(2)}(\tensor{\xi})
\frac{\partial^2\Theta(\mathbf{x},t)}{\partial x_{q_1}\partial x_{q_2}}
+
\tilde{M}_{pq_1}^{(2,1)}(\tensor{\xi})
\frac{\partial^2 U_p(\mathbf{x},t)}{\partial x_{q_1}\partial t}
+
M^{(2,1)}(\tensor{\xi})
\frac{\partial \Theta(\mathbf{x},t)}{\partial t}
+
\hat{M}^{(2,1)}(\tensor{\xi})
\frac{\partial \Upsilon(\mathbf{x},t)}{\partial t},
\end{eqnarray}
with  second order perturbation functions $M_{q_1q_2}^{(2)},\tilde{M}_{pq_1}^{(2,1)},M^{(2,1)}$ and $\hat{M}^{(2,1)}$.
Diffusion problem at the order $\varepsilon^{0}$ results from equation (\ref{eq:RecursiveDifferentialProblemsMicroChemPot}) and reads
\begin{eqnarray}
\label{eq:DiffProbDiffus+0}
&&\left(
D_{ij}^m 
\eta_{,j}^{(2)}
\right)_{,i}
+
\left[
\left(
D_{ij}^m
W_{q_1}^{(1)}
\right)_{,i}
+
D_{q_1j}^{m}
+
D_{ji}^{m}
W_{q_1,i}^{(1)}
\right]
\frac{\partial^2\Upsilon}{\partial x_{q_1}\partial x_j}
-
\left(
\beta_{ij}^{m}
N_{ipq_1,j}^{(1)}
+
\beta_{pq_1}^{m}
\right)
\frac{\partial^2 U_p}{\partial x_{q_1}\partial t}
+\nonumber\\
&&-
\left(
\beta_{ij}
\hat{N}_{i,j}^{(1)}
+
q^m
\right)
\frac{\partial\Upsilon}{\partial t}
-
\left(
\beta_{ij}^m
\tilde{N}_{i,j}^{(1)}
+
\psi^m
\right)
\frac{\partial \Theta}{\partial t}
=
h^{(2)}(\mathbf{x},t),
\end{eqnarray}
with relative interface conditions from (\ref{eq:InterfaceConditionMicroChemPotOnSigma1}) in the form
\begin{equation}
\label{eq:InterfaceConditionMicroChemPot+0}
\left.
\left[
\left[
\eta^{(2)}
\right]
\right]
\right|_{\tensor{\xi}\in\Sigma_1}=0,
\hspace{0.5cm}
\left.
\left[
\left[
D_{ij}^{m}
\left(
\eta_{,j}^{(2)}
+
W_{q_1}^{(1)}
\frac{\partial^2 \Upsilon}{\partial x_{q_1}\partial x_j}
\right)n_i
\right]
\right]
\right|_{\tensor{\xi}\in\Sigma_1}=0.
\end{equation}
Solvability condition for differential problem (\ref{eq:DiffProbDiffus+0}) imposes 
\begin{eqnarray}
&&h^{(2)}(\mathbf{x},t)
=
\left\langle
\left(
D_{ij}^m W_{q_1}^{(1)}
\right)_{,i}+
D_{q_1j}^m
+
D_{ji}^m
W_{q_1,i}^{(1)}
\right\rangle
\frac{\partial^2 \Upsilon}{\partial x_{q_1}\partial x_j}
-
\left\langle
\beta_{ij}^{m}
N_{ipq_1,j}^{(1)}
+
\beta_{pq_1}^{m}
\right\rangle
\frac{\partial^2 U_p}{\partial x_{q_1}\partial t}
+\nonumber\\
&&-
\left\langle
\beta_{ij}
\hat{N}_{i,j}^{(1)}
+
q^m
\right\rangle
\frac{\partial\Upsilon}{\partial t}
-
\left\langle
\beta_{ij}^m
\tilde{N}_{i,j}^{(1)}
+
\psi^m
\right\rangle
\frac{\partial \Theta}{\partial t},
\end{eqnarray}
with a solution of the form
\begin{eqnarray}
\label{eq:SolutionDiffProbDissus+0}
\eta^{(2)}(\mathbf{x},\tensor{\xi},t)
=
W_{q_1q_2}^{(2)}(\tensor{\xi})
\frac{\partial^2\Upsilon(\mathbf{x},t)}{\partial x_{q_1}\partial x_{q_2}}
+
\tilde{W}_{pq_1}^{(2,1)}(\tensor{\xi})
\frac{\partial^2 U_p(\mathbf{x},t)}{\partial x_{q_1}\partial t}
+
W^{(2,1)}(\tensor{\xi})
\frac{\partial \Upsilon(\mathbf{x},t)}{\partial t}
+
\hat{W}^{(2,1)}(\tensor{\xi})
\frac{\partial \Theta(\mathbf{x},t)}{\partial t},
\end{eqnarray}
where $W_{q_1q_2}^{(2)},\tilde{W}_{pq_1}^{(2,1)},W^{(2,1)}$ and $\hat{W}^{(2,1)}$ are the relative second order perturbation functions.
%
\section{Cell problems and perturbation functions }
\label{Sec::CellProblems}
Cell problems are non homogeneous recursive differential problems obtained inserting   into differential problems  (\ref{eq:RecursiveDifferentialProblemMicroDisplacement})-(\ref{eq:RecursiveDifferentialProblemsMicroChemPot}) the solutions obtained at the different orders of $\varepsilon$.
Cell problems are therefore expressed in terms of perturbation functions which depend on geometrical and physico-mechanical features of the microstructure and reflect the effects of material dishomogeneities on microfields.
Solutions of cell problems result to be regular, $\mathcal{Q}$-periodic functions because cell problems are elliptic differential problems in divergence form whose  terms have   vanishing mean values over $\mathcal{Q}$ \citep{Bakhvalov1984}. 
In order to guarantee the uniqueness of cell problems solution, the following normalization condition 
\begin{equation}
\left\langle
\left(
\cdot
\right)\right\rangle
=\frac{1}{|\mathcal{Q}|}
\int_{\mathcal{Q}}
(\cdot)\,d\tensor{\xi}=0
\end{equation}
is required to be fulfilled by all perturbation functions.
In what follows cell problems are described in detail for the mechanical, thermal and mass diffusion problems up to order $\varepsilon^0$. Higher order cell problems are obtained following the procedure described below, but their expression is not reported in the present note for brevity.
\\
\\
{\itshape Mechanical cell problems}
\\
\\
From equation (\ref{eq:DiffProbMecc-1}), in view of the form of solution (\ref{eq:SolutionDiffProbMecc-1}) one obtains the following three cell problems at the order $\varepsilon^{-1}$.
The first one and its relative interface conditions are expressed in terms of perturbation function $N_{kpq_1}^{(1)}$ and read
\begin{eqnarray}
\label{eq:FirstCellProblemDisplacement-1}
&&
\left(C_{ijkl}^{m}N_{kpq_1,l}^{(1)}
\right)_{,j}
+
C_{ijpq_1,j}^{m}
=0,
\nonumber\\
&&
\left.
\left[
\left[
N_{kpq_1}^{(1)}
\right]
\right]
\right|_{\tensor{\xi}\in\Sigma_1}=0,
\nonumber\\
&&
\left.
\left[
\left[
C_{ijkl}^{m}
\left(
N_{kpq_1,l}^{(1)}
+
\delta_{lq_1}
\delta_{kp}
\right)
n_j
\right]
\right]
\right|_{\tensor{\xi}\in\Sigma_1},
=0,
\end{eqnarray}
where symbol $\delta_{lq_1}$ denotes the Kronecker delta function.
The second cell problem and its interface conditions are expressed in terms of $\tilde{N}_{k}^{(1)}$ and have the form
\begin{eqnarray}
\label{eq:SecondCellProblemDisplacement-1}
&&
\left(
C_{ijkl}^{m}
\tilde{N}_{k,l}^{(1)}
\right)_{,j}
-
\alpha_{ij,j}^{m}
=0,
\nonumber\\
&&
\left.
\left[
\left[
\tilde{N}_{k}^{(1)}
\right]
\right]
\right|_{\tensor{\xi}\in\Sigma_1}=0,
\nonumber\\
&&
\left.
\left[
\left[
\left(
C_{ijkl}^{m}
\tilde{N}_{k,l}^{(1)}
-
\alpha_{ij}^m
\right)
n_j
\right]
\right]
\right|_{\tensor{\xi}\in\Sigma_1}
=0.
\end{eqnarray}
Finally, the third cell problem is in terms of $\hat{N}_{k}^{(1)}$ and it is expressed in the following way, together with relative interface conditions
\begin{eqnarray}
\label{eq:ThirdCellProblemDisplacement-1}
&&
\left(
C_{ijkl}^{m}
\hat{N}_{k,l}^{(1)}
\right)_{,j}
-
\beta_{ij,j}^m=0,
\nonumber\\
&&
\left.
\left[
\left[
\hat{N}_{k}^{(1)}
\right]
\right]
\right|_{\tensor{\xi}\in\Sigma_1}=0,
\nonumber\\
&&
\left.
\left[
\left[
\left(
C_{ijkl}^{m}
\hat{N}_{k,l}^{(1)}
-
\beta_{ij}^m
\right)
n_j
\right]
\right]
\right|_{\tensor{\xi}\in\Sigma_1}
=0.
\end{eqnarray}
%
%
When perturbation functions $N_{kpq_1}^{(1)},\tilde{N}_{k}^{(1)},$ and $\hat{N}_{k}^{(1)}$ are determined as solutions of relative cell problems at the order $\varepsilon^{-1}$, from equation (\ref{eq:DiffProbMecc+0}) and in consideration of the form of the solution (\ref{eq:SolutionDiffProbMecc0}) one obtains the following four cell problems at the order $\varepsilon^0$.
The first one is written in a symmetrized from with respect to indices $q_1$ and $q_2$ and, together with its interface conditions, is here formulated in terms of second order perturbation function $N_{kpq_1q_2}^{(2)}$ and reads
\begin{eqnarray}
\label{eq:FirstCellProblemDisplacement+0}
&&
\left(
C_{ijkl}^{m}
N_{kpq_1q_2,l}^{(2)}
\right)_{,j}
+
\frac{1}{2}
\left[
\left(
C_{ijkq_2}^{m}
N_{kpq_1}^{(1)}+
C_{ijkq_1}^{m}
N_{kpq_2}^{(1)}
\right)_{,j}
+
C_{iq_1pq_2}^{m}+
C_{iq_2pq_1}^{m}
+
\right.
\nonumber\\
&&+
\left.
C_{iq_2kj}^{m}
N_{kpq_1,j}^{(1)}+
C_{iq_1kj}^{m}
N_{kpq_2,j}^{(1)}
\right]
=
\frac{1}{2}
\left\langle
C_{iq_1pq_2}^{m}
+
C_{iq_2kj}^{m}
N_{kpq_1,j}^{(1)}
+
C_{iq_2pq_1}^{m}
+
C_{iq_1kj}^{m}
N_{kpq_2,j}^{(1)}
\right\rangle,
\nonumber\\
&&
\left.
\left[
\left[
N_{kpq_1q_2}^{(2)}
\right]
\right]
\right|_{\tensor{\xi}\in\Sigma_1}=0,
\nonumber\\
&&
\left.
\left[
\left[
C_{ijkl}^{m}
\left\{
N_{kpq_1q_2,l}^{(2)}+
\frac{1}{2}
\left(
\delta_{q_2l}
N_{kpq_1}^{(1)}+
\delta_{q_1l}
N_{kpq_2}^{(1)}
\right)
\right\}
n_j
\right]
\right]
\right|_{\tensor{\xi}\in\Sigma_1}=0.
\end{eqnarray}
The second cell problem  deriving from (\ref{eq:DiffProbMecc+0}) and its interface condition involve perturbation function $\tilde{N}_{kq_1}^{(2)}$ and are expressed as
\begin{eqnarray}
\label{eq:SecondCellProblemDisplacement+0}
&&\left(
C_{ijkl}^{m}
\tilde{N}_{kq_1,l}^{(2)}
\right)_{,j}
+
\left[
\left(
C_{ijkq_1}^{m}
\tilde{N}_{k}^{(1)}
\right)_{,j}
+
C_{iq_1kj}^{m}
\tilde{N}_{k,j}^{(1)}
-
\left(
\alpha_{ij}^{m}
M_{q_1}^{(1)}
\right)_{,j}
-
\alpha_{iq_1}^{m}
\right]
=
\left\langle
C_{iq_1kj}^{m}
\tilde{N}_{k,j}^{(1)}
-
\alpha_{iq_1}^{m}
\right\rangle,
\nonumber\\
&&
\left.
\left[
\left[
\tilde{N}_{kq_1}^{(2)}
\right]
\right]
\right|_{\tensor{\xi}\in\Sigma_1}=0,
\nonumber\\
&&\left.
\left[
\left[
\left\{
C_{ijkl}^{m}
\left(
\tilde{N}_{kq_1}^{(2)}
+
\delta_{q_1l}
\tilde{N}_{k}^{(1)}
\right)
-
\alpha_{ij}^{m}
M_{q_1}^{(1)}
\right\}
n_j
\right]
\right]
\right|_{\tensor{\xi}\in\Sigma_1}=0.
\end{eqnarray}
The form of the third cell problem from (\ref{eq:DiffProbMecc+0}) and its interface conditions in terms of $\hat{N}_{kq_1}^{(2)}$ is
\begin{eqnarray}
\label{eq:ThirdCellProblemDisplacement+0}
&&\left(
C_{ijkl}^{m}
\hat{N}_{kq_1,l}^{(2)}
\right)_{,j}
+
\left[
\left(
C_{ijkq_1}^{m}
\hat{N}_{k}^{(1)}
\right)_{,j}
+
C_{iq_1kj}^{m}
\hat{N}_{k,j}^{(1)}
-
\left(
\beta_{ij}^{m}
W_{q_1}^{(1)}
\right)_{,j}
-
\beta_{iq_1}^{m}
\right]
=
\left\langle
C_{iq_1kj}^{m}
\hat{N}_{k,j}^{(1)}
-
\beta_{iq_1}^{m}
\right\rangle,
\nonumber\\
&&
\left.
\left[
\left[
\hat{N}_{kq_1}^{(2)}
\right]
\right]
\right|_{\tensor{\xi}\in\Sigma_1}=0,
\nonumber\\
&&
\left.
\left[
\left[
\left\{
C_{ijkl}^{m}
\left(
\hat{N}_{kq_1}^{(2)}
+
\delta_{q_1l}
\hat{N}_{k}^{(1)}
\right)
-
\beta_{ij}^{m}
W_{q_1}^{(1)}
\right\}
n_j
\right]
\right]
\right|_{\tensor{\xi}\in\Sigma_1}=0.
\end{eqnarray}
The last mechanical cell problem at the order $\varepsilon^0$ and its interface conditions have the following expression in terms of perturbation function $N_{kp}^{(2,2)}$
\begin{eqnarray}
\label{eq:FourthCellProblemDisplacement+0}
&&\left(
C_{ijkl}^{m}
N_{kp,l}^{(2,2)}
\right)_{,j}
-
\rho^m
\delta_{ip}=
-
\left\langle
\rho^m
\right\rangle
\delta_{ip},
\nonumber\\
&&
\left.
\left[
\left[
{N}_{kp}^{(2,2)}
\right]
\right]
\right|_{\tensor{\xi}\in\Sigma_1}=0,
\nonumber\\
&&
\left.
\left[
\left[
C_{ijkl}^{m}
\hat{N}_{kp}^{(2,2)}
n_j
\right]
\right]
\right|_{\tensor{\xi}\in\Sigma_1}=0.
\end{eqnarray}
%
%
%
%
%
\\
\\
{\itshape Thermal cell problems}
\\
\\
From equation (\ref{eq:DiffProbTherm-1}) and taking into account solution (\ref{eq:SolutionDiffProbTherm-1}), one derives the following   cell problem at the order $\varepsilon^{-1}$ which, together with relative interface conditions, provides perturbation function $M_{q_1}^{(1)}$
\begin{eqnarray}
\label{eq:CellProblTherm-1}
&&\left(
K_{ij}^m
M_{q_1,j}^{(1)}
\right)_{,i}
+
K_{iq_1,i}^m
=0,
\nonumber\\
&&
\left.
\left[
\left[
M_{q_1}^{(1)}
\right]
\right]
\right|_{\tensor{\xi}\in\Sigma_1}=0,
\nonumber\\
&&
\left.
\left[
\left[
K_{ij}^{m}
\left(
M_{q_1,j}^{(1)}
+
\delta_{q_1j}
\right)
n_i
\right]
\right]
\right|_{\tensor{\xi}\in\Sigma_1}=0.
\end{eqnarray}
Once first-order perturbation function  $M_{q_1}^{(1)}$ is known, four cell problems are derived at the order $\varepsilon^0$ from equation (\ref{eq:DiffProbTherm+0}), bearing in mind solution (\ref{eq:SolutionDiffProbTherm+0}). The first one provides second order perturbation function $M_{q_1q_2}^{(2)}$ and it is here written in a symmetrized form with respect to indices $q_1$ and $q_2$, together with relative interface conditions
\begin{eqnarray}
\label{eq:FirstCellProblTherm+0}
&&
\left(
K_{ij}^m
M_{q_1q_2,j}^{(2)}
\right)_{,i}
+
\frac{1}{2}
\left[
\left(
K_{iq_2}^m M_{q_1}^{(1)}
\right)_{,i}
+
K_{q_1q_2}^m
+
K_{q_2i}^m
M_{q_1,i}^{(1)}+
\right.
\nonumber\\
&&
\left.
\left(
K_{iq_1}^m M_{q_2}^{(1)}
\right)_{,i}
+
K_{q_2q_1}^m
+
K_{q_1i}^m
M_{q_2,i}^{(1)}
\right]
=
\frac{1}{2}
\left\langle
K_{q_1q_2}^{m}
+
K_{q_2i}^{m}
M_{q_1,i}^{(1)}
+
K_{q_2q_1}^{m}
+
K_{q_1i}^{m}
M_{q_2,i}^{(1)}
\right\rangle,
\nonumber\\
&&
\left.
\left[
\left[
M_{q_1q_2}^{(2)}
\right]
\right]
\right|_{\tensor{\xi}\in\Sigma_1}=0,
\nonumber\\
&&
\left.
\left[
\left[
K_{ij}^{m}
\left\{
M_{q_1q_2,j}^{(2)}
+
\frac{1}{2}
\left(
\delta_{jq_2}
M_{q_1}^{(1)}
+
\delta_{jq_1}
M_{q_2}^{(1)}
\right)
\right\}
n_i
\right]
\right]
\right|_{\tensor{\xi}\in\Sigma_1}=0.
\end{eqnarray}
Perturbation function $\tilde{M}_{pq_1}^{(2,1)}$ is provided by the following cell problem and relative interface conditions
\begin{eqnarray}
\label{eq:SecondCellProblTherm+0}
&&\left(
K_{ij}^m
\tilde{M}_{pq_1,j}^{(2,1)}
\right)_{,i}
-
\left(
\alpha_{ij}^{m}
N_{ipq_1,j}^{(1)}
+
\alpha_{pq_1}^{m}
\right)
=
-
\left\langle
\alpha_{ij}^{m}
N_{ipq_1,j}^{(1)}
+
\alpha_{pq_1}^{m}
\right\rangle,
\nonumber\\
&&
\left.
\left[
\left[
\tilde{M}_{pq_1}^{(2,1)}
\right]
\right]
\right|_{\tensor{\xi}\in\Sigma_1}=0,
\nonumber\\
&&
\left.
\left[
\left[
K_{ij}^{m} \tilde{M}_{pq_1,j}^{(2,1)} n_i
\right]
\right]
\right|_{\tensor{\xi}\in\Sigma_1}=0.
\end{eqnarray}
The third cell problem and its interface conditions have the following expressions in terms of perturbation function $M^{(2,1)}$
\begin{eqnarray}
\label{eq:ThirdCellProblTherm+0}
&&\left(
K_{ij}^{m}
M_{,j}^{(2,1)}
\right)_{,i}
-
\left(
\alpha_{ij}^{m}
\tilde{N}_{i,j}^{(1)}
+
p^m
\right)
=
-
\left\langle
\alpha_{ij}^{m}
\tilde{N}_{i,j}^{(1)}
+
p^m
\right\rangle,
\nonumber\\
&&
\left.
\left[
\left[
{M}^{(2,1)}
\right]
\right]
\right|_{\tensor{\xi}\in\Sigma_1},
\nonumber\\
&&
\left.
\left[
\left[
K_{ij}^{m} M_{,j}^{2,1} n_i
\right]
\right]
\right|_{\tensor{\xi}\in\Sigma_1}=0.
\end{eqnarray}
Finally, the fourth cell problem and its interface conditions at the order $\varepsilon^0$ read
\begin{eqnarray}
&&
\left(
K_{ij}^{m}
\hat{M}_{,j}^{(2,1)}
\right)_{,i}
-
\left(
\alpha_{ij}^{m}
\hat{N}_{i,j}^{(1)}
+
\psi^m
\right)
=
-
\left\langle
\alpha_{ij}^{m}
\hat{N}_{i,j}^{(1)}
+\psi^m
\right\rangle,
\nonumber\\
&&
\left.
\left[
\left[
\hat{M}^{(2,1)}
\right]
\right]
\right|_{\tensor{\xi}\in\Sigma_1}=0,
\nonumber\\
&&
\left.
\left[
\left[
K_{ij}^{m} \hat{M}_{,j}^{(2,1)} n_i
\right]
\right]
\right|_{\tensor{\xi}\in\Sigma_1}=0,
\label{eq:FourthCellProblemTemperature+0}
\end{eqnarray}
in terms of $\hat{M}^{(2,1)}$.
\\
\\
{\itshape Mass diffusion cell problems}
\\
\\
Analogously to what done for the thermal problem, at the order $\varepsilon^{-1}$, from equation (\ref{eq:DiffProbDiffus-1}) and taking into account solution (\ref{eq:SolutionDiffProbDiffus-1}), one obtains the following cell problem and its interface conditions in terms of perturbation  function $W_{q_1}^{(1)}$
\begin{eqnarray}
\label{eq:CellProblDiffus-1}
&&\left(
D_{ij}^m
W_{q_1,j}^{(1)}
\right)_{,i}
+
D_{iq_1,i}^m
=0,
\nonumber\\
&&
\left.
\left[
\left[
W_{q_1}^{(1)}
\right]
\right]
\right|_{\tensor{\xi}\in\Sigma_1}=0,
\nonumber\\
&&
\left.
\left[
\left[
D_{ij}^{m}
\left(
W_{q_1,j}^{(1)}
+
\delta_{q_1j}
\right)
n_i
\right]
\right]
\right|_{\tensor{\xi}\in\Sigma_1}=0.
\end{eqnarray}
At the order $\varepsilon^0 $, the following four cell problems arise, once first-order perturbation function $W_{q_1}^{(1)}$ is computed as solutioon of (\ref{eq:CellProblDiffus-1}).
The first cell problem provides second order perturbation function $W_{q_1q_2}^{(2)}$ and it is expressed in the  following way, symmetrized with respect to indices $q_1$ and $q_2$, together with its interface conditions 
\begin{eqnarray}
\label{eq:FirstCellProblDiffus+0}
&&
\left(
D_{ij}^m
W_{q_1q_2,j}^{(2)}
\right)_{,i}
+
\frac{1}{2}
\left[
\left(
D_{iq_2}^m W_{q_1}^{(1)}
\right)_{,i}
+
D_{q_1q_2}^m
+
D_{q_2i}^m
W_{q_1,i}^{(1)}+
\right.
\nonumber\\
&&
\left.
\left(
D_{iq_1}^m W_{q_2}^{(1)}
\right)_{,i}
+
D_{q_2q_1}^m
+
D_{q_1i}^m
W_{q_2,i}^{(1)}
\right]
=
\frac{1}{2}
\left\langle
D_{q_1q_2}^{m}
+
D_{q_2i}^{m}
W_{q_1,i}^{(1)}
+
D_{q_2q_1}^{m}
+
D_{q_1i}^{m}
W_{q_2,i}^{(1)}
\right\rangle,
\nonumber\\
&&
\left.
\left[
\left[
W_{q_1q_2}^{(2)}
\right]
\right]
\right|_{\tensor{\xi}\in\Sigma_1}=0,
\nonumber\\
&&
\left.
\left[
\left[
D_{ij}^{m}
\left\{
W_{q_1q_2,j}^{(2)}
+
\frac{1}{2}
\left(
\delta_{jq_2}
W_{q_1}^{(1)}
+
\delta_{jq_1}
W_{q_2}^{(1)}
\right)
\right\}
n_i
\right]
\right]
\right|_{\tensor{\xi}\in\Sigma_1}=0.
\end{eqnarray}
The second cell problem and its interface conditions have the following form in terms of $\tilde{W}_{pq_1}^{(2,1)}$ 
\begin{eqnarray}
\label{eq:SecondCellProblDiffus+0}
&&\left(
D_{ij}^m
\tilde{W}_{pq_1,j}^{(2,1)}
\right)_{,i}
-
\left(
\beta_{ij}^{m}
N_{ipq_1,j}^{(1)}
+
\beta_{pq_1}^{m}
\right)
=
-
\left\langle
\beta_{ij}^{m}
N_{ipq_1,j}^{(1)}
+
\beta_{pq_1}^{m}
\right\rangle,
\nonumber\\
&&
\left.
\left[
\left[
\tilde{W}_{pq_1}^{(2,1)}
\right]
\right]
\right|_{\tensor{\xi}\in\Sigma_1}=0,
\nonumber\\
&&
\left.
\left[
\left[
D_{ij}^{m} \tilde{W}_{pq_1,j}^{(2,1)} n_i
\right]
\right]
\right|_{\tensor{\xi}\in\Sigma_1}=0.
\end{eqnarray}
Second order perturbation function $W^{(2,1)}$ is provided by the resolution of the following cell problem with relative interface conditions
\begin{eqnarray}
\label{eq:ThirdCellProblDiffus+0}
&&\left(
D_{ij}^{m}
W_{,j}^{(2,1)}
\right)_{,i}
-
\left(
\beta_{ij}^{m}
\hat{N}_{i,j}^{(1)}
+
q^m
\right)
=
-
\left\langle
\beta_{ij}^{m}
\hat{N}_{i,j}^{(1)}
+
q^m
\right\rangle,
\nonumber\\
&&\left.
\left[
\left[
{W}^{(2,1)}
\right]
\right]
\right|_{\tensor{\xi}\in\Sigma_1}=0,
\nonumber\\
&&
\left.
\left[
\left[
D_{ij}^{m} W_{,j}^{(2,1)} n_i
\right]
\right]
\right|_{\tensor{\xi}\in\Sigma_1}=0.
\end{eqnarray}
Finally, the last cell problem at the order $\varepsilon^0$ is expressed in the following way, together with interface conditions, in terms of perturbation function $\hat{W}^{(2,1)}$
\begin{eqnarray}
\label{eq:FourthCellProblDiffus+0}
&&
\left(
D_{ij}^{m}
\hat{W}_{,j}^{(2,1)}
\right)_{,i}
-
\left(
\beta_{ij}^{m}
\tilde{N}_{i,j}^{(1)}
+
\psi^m
\right)
=
-
\left\langle
\beta_{ij}^{m}
\tilde{N}_{i,j}^{(1)}
+\psi^m
\right\rangle,
\nonumber\\
&&
\left.
\left[
\left[
\hat{W}^{(2,1)}
\right]
\right]
\right|_{\tensor{\xi}\in\Sigma_1}=0,
\nonumber\\
&&
\left.
\left[
\left[
D_{ij}^{m} \hat{W}_{,j}^{(2,1)} n_i
\right]
\right]
\right|_{\tensor{\xi}\in\Sigma_1}=0.
\end{eqnarray}
%
%
%
%
\section{Down-scaling and up-scaling relations}
\label{Sec::DownScalingUpScaling}
%
When perturbation functions are known from the resolution of relative cell problems at the different orders of $\varepsilon$ as detailed in Section \ref{Sec::CellProblems}, from equations (\ref{eq:AsymptoticExpansionMicroDisplacement})-(\ref{eq:AsymptoticExpansionMicroChemicalPotential})
microscopic fields $\mathbf{u}(\mathbf{x},\mathbf{x}/	\varepsilon,t)
,\theta(\mathbf{x},\mathbf{x}/	\varepsilon,t)$ and $\eta(\mathbf{x},\mathbf{x}/	\varepsilon,t)$ are expressed as asymptotic expansions in powers of micro characteristic size $\varepsilon$ in terms of such $\mathcal{Q}$-periodic perturbation functions and in terms of macrofields $\mathbf{U}(\mathbf{x},t), \Theta(\mathbf{x},t)$ and $\Upsilon(\mathbf{x},t)$ and their gradients. Considering the form of solutions (\ref{eq:SolutionDiffProbMecc-1}),(\ref{eq:SolutionDiffProbTherm-1}), (\ref{eq:SolutionDiffProbDiffus-1}) at the order $\varepsilon^{-1}$ and (\ref{eq:SolutionDiffProbMecc0}), (\ref{eq:SolutionDiffProbTherm+0}), (\ref{eq:SolutionDiffProbDissus+0}) at the order $\varepsilon^0$, the following down-scaling relations are obtained for the three microfields
\begin{subequations}
\label{eq:DownScalingRelations}
\begin{align}
&
u_k\left(\mathbf{x},\frac{\mathbf{x}}{	\varepsilon},t\right)=
\left[
U_k(\mathbf{x},t)
+
\varepsilon
\left(
N_{kpq_1}^{(1)}(\tensor{\xi})
\frac{\partial U_p(\mathbf{x},t)}{\partial x_{q_1}}
+
\tilde{N}_{k}^{(1)}(\tensor{\xi})
\Theta(\mathbf{x},t)
+
\hat{N}_k^{(1)}(\tensor{\xi})
\Upsilon(\mathbf{x},t)
\right)
+
\right.
\nonumber\\
&
\left.
\left.
+
\varepsilon^2
\left(
N_{kpq_1q_2}^{(2)}(\tensor{\xi})
\frac{\partial^2 U_p(\mathbf{x},t)}{\partial x_{q_1}\partial x_{q_2}}
+
\tilde{N}_{kq_1}^{(2)}(\tensor{\xi})
\frac{\partial\Theta(\mathbf{x},t)}{\partial x_{q_1}}
+
\hat{N}_{kq_1}^{(2)}(\tensor{\xi})
\frac{\partial\Upsilon(\mathbf{x},t)}{\partial x_{q_1}}
+
N_{kp}^{(2,2)}(\tensor{\xi})
\frac{\partial^2 U_p(\mathbf{x},t)}{\partial t^2}
\right)
+
O(\varepsilon^3)
\right]
\right|_{\tensor{\xi}=\frac{\mathbf{x}}{\varepsilon}},
\\
&
\theta\left(\mathbf{x},\frac{\mathbf{x}}{	\varepsilon},t\right)
=
\left[
\Theta(\mathbf{x},t)
+
\varepsilon
M_{q_1}^{(1)}(\tensor{\xi})
\frac{\partial\Theta(\mathbf{x},t)}{\partial x_{q_1}}
+
\varepsilon^2
\left(
M_{q_1q_2}^{(2)}(\tensor{\xi})
\frac{\partial^2\Theta(\mathbf{x},t)}{\partial x_{q_1}\partial x_{q_2}}
+
\tilde{M}_{pq_1}^{(2,1)}(\tensor{\xi})
\frac{\partial^2 U_p(\mathbf{x},t)}{\partial x_{q_1}\partial t}
+
\right.
\right.
\nonumber\\
&
\left.
\left.
\left.
+
M^{(2,1)}(\tensor{\xi})
\frac{\partial \Theta(\mathbf{x},t)}{\partial t}
+
\hat{M}^{(2,1)}(\tensor{\xi})
\frac{\partial\Upsilon(\mathbf{x},t)}{\partial t}
\right)+O(\varepsilon^3)
\right]
\right|_{\tensor{\xi}=\frac{\mathbf{x}}{\varepsilon}},
\\
&
\eta\left(\mathbf{x},\frac{\mathbf{x}}{	\varepsilon},t\right)
=
\left[
\Upsilon(\mathbf{x},t)
+
\varepsilon\,
W_{q_1}^{(1)}(\tensor{\xi})
\frac{\partial\Upsilon(\mathbf{x},t)}{\partial x_{q_1}}
+
\varepsilon^2
\left(
W_{q_1q_2}^{(2)}(\tensor{\xi})
\frac{\partial^2\Upsilon(\mathbf{x},t)}{\partial x_{q_1}\partial x_{q_2}}
+
\tilde{W}_{pq_1}^{(2,1)}(\tensor{\xi})
\frac{\partial^2 U_p(\mathbf{x},t)}{\partial x_{q_1}\partial t}
+
\right.
\right.
\nonumber\\
&
\left.
\left.
\left.
+
W^{(2,1)}(\tensor{\xi})
\frac{\partial \Upsilon(\mathbf{x},t)}{\partial t}
+
\hat{W}^{(2,1)}(\tensor{\xi})
\frac{\partial\Theta(\mathbf{x},t)}{\partial t}
\right)+O(\varepsilon^3)
\right]
\right|_{\tensor{\xi}=\frac{\mathbf{x}}{\varepsilon}}.
\end{align}
\end{subequations}
In equations (\ref{eq:DownScalingRelations}) microstructural heterogeneities are taken into account by the  $\mathcal{Q}$-periodic perturbation functions, which depend exclusively upon the fast variable $\tensor{\xi}$, while the $\mathcal{L}$-periodic  macrofields depend solely upon the slow variable $\mathbf{x}$.
Up-scaling relations are the ones that provide macroscopic fields $\mathbf{U}(\mathbf{x},t),\Theta(\mathbf{x},t)$ and $\Upsilon(\mathbf{x},t)$ in terms of the corresponding microscopic quantities. In particular, macro fields are expressed as mean values of micro fields over the unit cell $\mathcal{Q}$
\begin{eqnarray}
\label{eq:Up-scaling}
&&U_k(\mathbf{x},t)\doteq
\left\langle
u_k\left(\mathbf{x,\frac{\mathbf{x}}{\varepsilon}}+\tensor{\zeta},t\right)
\right\rangle_{\tensor{\zeta}},
\nonumber\\
&&\Theta(\mathbf{x},t)\doteq
\left\langle
\theta\left(\mathbf{x,\frac{\mathbf{x}}{\varepsilon}}+\tensor{\zeta},t\right)
\right\rangle_{\tensor{\zeta}},
\nonumber\\
&&\Upsilon(\mathbf{x},t)\doteq
\left\langle
\eta\left(\mathbf{x,\frac{\mathbf{x}}{\varepsilon}}+\tensor{\zeta},t\right)
\right\rangle_{\tensor{\zeta}},
\end{eqnarray}
where variable $\tensor{\zeta}\in\mathcal{Q}$ is a translation variable  such that $\varepsilon\tensor{\zeta}\in\mathcal{A}$ describes the translation of the body with respect to $\mathcal{L}$-periodic source terms, thus removing rapid fluctuations of coefficients \citep{Smyshlyaev2000,Bacigalupo2014}. Invariance property
\begin{equation}
\left\langle
g\left(
\tensor{\xi}+
\tensor{\zeta}
\right)
\right\rangle_{\tensor{\zeta}}
=
\frac{1}{\delta}
\int_{\mathcal{Q}}
g\left(
\tensor{\xi}+
\tensor{\zeta}
\right)
d\tensor{\zeta}
=
\frac{1}{\delta}
\int_{\mathcal{Q}}
g\left(
\tensor{\xi}+
\tensor{\zeta}
\right)
d\tensor{\xi}
\label{eq:InvarianceProperty}
\end{equation}
is proved to hold for all functions with $\mathcal{Q}$-periodicity. 
%
\section{Overall constitutive tensors and field equations of the first order homogenized thermo-diffusive medium}
\label{Sec::FieldEquatuationsCauchyHomMedium}
%
%
Average field equations of infinite order are determined from the substitution of down-scaling relations (\ref{eq:DownScalingRelations}) into local balance equations (\ref{eq:FieldEquMicroStress})-(\ref{eq:FieldEquMicroMassFlux}) and ordering at the different orders of $\varepsilon$.  They are expressed in the following form
\begin{subequations}
\begin{align}
&n_{ipq_1q_2}^{(2)}
\frac{\partial^2 U_p}{\partial x_{q_1}\partial x_{q_2}}
-
\tilde{n}_{iq_1}^{(2)}
\frac{\partial\Theta}{\partial x_{q_1}}
-
\hat{n}_{iq_1}^{(2)}
\frac{\partial \Upsilon}{\partial x_{q_1}} 
-
n^{(2,2)}
\frac{\partial^2 U_i}{\partial t^2}+ O(\varepsilon) + b_i(\mathbf{x},t)=0,
\label{eq:AverageFieldEquationsInfiniteOrder1}
\\
&m_{q_1q_2}^{(2)}
\frac{\partial^2 \Theta}{\partial x_{q_1}\partial x_{q_2}}
-
\tilde{m}_{pq_1}^{(2,1)}
\frac{\partial^2U_p}{\partial x_{q_1}\partial  t}
-
m^{(2,1)}
\frac{\partial \Theta}{\partial t}
-
\hat{m}^{(2,1)}
\frac{\partial \Upsilon}{\partial t} + O(\varepsilon) + r(\mathbf{x},t)=0,
\label{eq:AverageFieldEquationsInfiniteOrder2}
\\
&
w_{q_1q_2}^{(2)}
\frac{\partial^2 \Upsilon}{\partial x_{q_1}\partial x_{q_2}}
-
\tilde{w}_{pq_1}^{(2,1)}
\frac{\partial^2 U_p}{\partial x_{q_1}\partial  t}
-
w^{(2,1)}
\frac{\partial \Upsilon}{\partial t}
-
\hat{w}^{(2,1)}
\frac{\partial \Theta}{\partial t} + O(\varepsilon) + s(\mathbf{x},t)=0.
\label{eq:AverageFieldEquationsInfiniteOrder3}
\end{align}
\label{eq:AverageFieldEquationsInfiniteOrder}
\end{subequations}
Coefficients of macro fields gradients in expressions (\ref{eq:AverageFieldEquationsInfiniteOrder}) are defined as mean values over $\mathcal{Q}$ of linear combinations of perturbation functions and microscopic constitutive tensors components. They are the known terms of the corresponding cell problems and, at the order $\varepsilon^0$, they read
\begin{subequations}
\label{eq:OverallConstitutiveTensorsFirstOrder}
\begin{align}
&
{n}_{ipq_1q_2}^{(2)}=
\frac{1}{2}
\left\langle
C_{iq_2pq_1}^m
+
C_{iq_2kl}^m
N_{kpq_1,l}^{(1)}
+
C_{iq_1pq_2}^m
+
C_{iq_1kl}^m
N_{kpq_2,l}^{(1)}
\right\rangle,
\\
&
\tilde{n}_{iq_1}^{(2)}=
\left\langle
\alpha_{iq_1}^{m}
-
C_{iq_1kj}^{m}
\tilde{N}_{k,j}^{(1)}
\right\rangle,
\\
&
\hat{n}_{iq_1}^{(2)}=
\left\langle
\beta_{iq_1}^{m}
-
C_{iq_1kj}^{m}
\hat{N}_{k,j}^{(1)}
\right\rangle,
\\
&
{n}^{(2,2)}
=
\left\langle
\rho^m
\right\rangle,
\\
&
m_{q_1q_2}^{(2)}=
\frac{1}{2}
\left\langle
K_{q_1q_2}^m + K_{q_2j}\,M_{q_1,j}^{(1)}+
K_{q_2q_1}^m + K_{q_1j}\,M_{q_2,j}^{(1)}
\right\rangle,
\\
&
\tilde{m}_{pq_1}^{(2,1)}
=
\left\langle
\alpha_{pq_1}^{m}
+
\alpha_{iq_2}^{m}
N_{ipq_1,q_2}^{(1)}
\right\rangle,
\\
&
{m}^{(2,1)}
=
\left\langle
p^m+\alpha_{q_1q_2}^m\tilde{N}_{q_1,q_2}^{(1)}
\right\rangle,
\\
&
\hat{m}^{(2,1)}
=
\left\langle
\psi^{m}
+
\alpha_{q_1q_2}^{m}
\hat{N}_{q_1,q_2}^{(1)}
\right\rangle,
\\
&
w_{q_1q_2}^{(2)}=
\frac{1}{2}
\left\langle
D_{q_1q_2}^m + D_{q_2j}\,W_{q_1,j}^{(1)}+
D_{q_2q_1}^m + D_{q_1j}\,W_{q_2,j}^{(1)}
\right\rangle,
\\
&
\tilde{w}_{pq_1}^{(2,1)}
=
\left\langle
\beta_{pq_1}^{m}
+
\beta_{iq_2}^{m}
N_{ipq_1,q_2}^{(1)}
\right\rangle,
\\
&
{w}^{(2,1)}
=
\left\langle
q^m+\beta_{q_1q_2}^m\hat{N}_{q_1,q_2}^{(1)}
\right\rangle,
\\
&
\hat{w}^{(2,1)}
=
\left\langle
\psi^{m}
+
\beta_{q_1q_2}^{m}
\tilde{N}_{q_1,q_2}^{(1)}
\right\rangle.
\end{align}
\end{subequations}
If one performs the following asymptotic expansions of the macro fields $\mathbf{U}(\mathbf{x},t),\Theta(\mathbf{x},t)$ and $\Upsilon(\mathbf{x},t)$ in powers of characteristic length $\varepsilon$
\begin{subequations}
\label{eq:AsympExpansMacroFields}
\begin{align}
&
U_k(\mathbf{x},t)=\sum_{j=0}^{+\infty}\varepsilon^j U_k^{(j)}(\mathbf{x},t),
\\
&
\Theta(\mathbf{x},t)=\sum_{j=0}^{+\infty}\varepsilon^j \Theta^{(j)}(\mathbf{x},t),
\\
&
\Upsilon(\mathbf{x},t)=\sum_{j=0}^{+\infty}\varepsilon^j \Upsilon^{(j)}(\mathbf{x},t),
\end{align}
\end{subequations}%
a formal solution of the average field equations of infinite order (\ref{eq:AverageFieldEquationsInfiniteOrder}) can be obtained. In particular, substituting expansions (\ref{eq:AsympExpansMacroFields}) into (\ref{eq:AverageFieldEquationsInfiniteOrder}), and reordering at the different orders of $\varepsilon$, one obtains the following three sets of recursive differential problems in terms of the macroscopic fields.
Equation (\ref{eq:AverageFieldEquationsInfiniteOrder1}) becomes
\begin{eqnarray}
\label{eq:RecursiveDifferentialProblemMacroDisplacement}
&&n_{ipq_1q_2}^{(2)}
\left(
\frac{\partial^2 U_p^{(0)}}{\partial x_{q_1}\partial x_{q_2}}
+
\varepsilon\,
\frac{\partial^2 U_p^{(1)}}{\partial x_{q_1}\partial x_{q_2}}
+
\varepsilon^2\,
\frac{\partial^2 U_p^{(2)}}{\partial x_{q_1}\partial x_{q_2}}
+...
\right)+
\varepsilon\,n_{ipq_1...q_3}^{(3)}
\left(
\frac{\partial^3 U_p^{(0)}}{\partial x_{q_1}...\partial x_{q_3}}
+
\varepsilon\,
\frac{\partial^3 U_p^{(1)}}{\partial x_{q_1}...\partial x_{q_3}}
+
\right.
\nonumber\\
&&
\left.
+\varepsilon^2\,
\frac{\partial^3 U_p^{(2)}}{\partial x_{q_1}...\partial x_{q_3}}
+...
\right)+
\varepsilon^2\,n_{ipq_1...q_4}^{(4)}
\left(
\frac{\partial^4 U_p^{(0)}}{\partial x_{q_1}...\partial x_{q_4}}
+
\varepsilon\,
\frac{\partial^4 U_p^{(1)}}{\partial x_{q_1}...\partial x_{q_4}}
+
\varepsilon^2\,
\frac{\partial^4 U_p^{(2)}}{\partial x_{q_1}...\partial x_{q_4}}
+...
\right)+
\nonumber\\
&&
-\tilde{n}_{iq_1}^{(2)}
\left(
\frac{\partial \Theta^{(0)}}{\partial x_{q_1}}
+
\varepsilon\,
\frac{\partial \Theta^{(1)}}{\partial x_{q_1}}
+
\varepsilon^2\,
\frac{\partial \Theta^{(2)}}{\partial x_{q_1}}
+...
\right)
-
\varepsilon\,\tilde{n}_{iq_1q_2}^{(3)}
\left(
\frac{\partial^2 \Theta^{(0)}}{\partial x_{q_1}\partial x_{q_2}}
+
\varepsilon\,
\frac{\partial^2 \Theta^{(1)}}{\partial x_{q_1}\partial x_{q_2}}
+
\right.
\nonumber\\
&&
\left.
+\varepsilon^2\,
\frac{\partial^2 \Theta^{(2)}}{\partial x_{q_1}\partial x_{q_2}}
+...
\right)
-
\varepsilon^2\,\tilde{n}_{iq_1...q_3}^{(4)}
\left(
\frac{\partial^3 \Theta^{(0)}}{\partial x_{q_1}...\partial x_{q_3}}
+
\varepsilon\,
\frac{\partial^3 \Theta^{(1)}}{\partial x_{q_1}...\partial x_{q_3}}
+
\varepsilon^2\,
\frac{\partial^3 \Theta^{(2)}}{\partial x_{q_1}...\partial x_{q_3}}
+...
\right)+
\nonumber\\
&&
-\hat{n}_{iq_1}^{(2)}
\left(
\frac{\partial \Upsilon^{(0)}}{\partial x_{q_1}}
+
\varepsilon\,
\frac{\partial \Upsilon^{(1)}}{\partial x_{q_1}}
+
\varepsilon^2\,
\frac{\partial \Upsilon^{(2)}}{\partial x_{q_1}}
+...
\right)
-
\varepsilon\,\hat{n}_{iq_1q_2}^{(3)}
\left(
\frac{\partial^2 \Upsilon^{(0)}}{\partial x_{q_1}\partial x_{q_2}}
+
\varepsilon\,
\frac{\partial^2 \Upsilon^{(1)}}{\partial x_{q_1}\partial x_{q_2}}
+
\right.
\nonumber\\
&&
\left.
+\varepsilon^2\,
\frac{\partial^2 \Upsilon^{(2)}}{\partial x_{q_1}\partial x_{q_2}}
+...
\right)
-
\varepsilon^2\,\hat{n}_{iq_1...q_3}^{(4)}
\left(
\frac{\partial^3 \Upsilon^{(0)}}{\partial x_{q_1}...\partial x_{q_3}}
+
\varepsilon\,
\frac{\partial^3 \Upsilon^{(1)}}{\partial x_{q_1}...\partial x_{q_3}}
+
\varepsilon^2\,
\frac{\partial^3 \Upsilon^{(2)}}{\partial x_{q_1}...\partial x_{q_3}}
+...
\right)+
\nonumber\\
&&
-{n}_i^{(2,2)}
\left(
\frac{\partial^2 U_i^{(0)}}{\partial t^2}
+
\varepsilon\,
\frac{\partial^2 U_i^{(1)}}{\partial t^2}
+
\varepsilon^2\,
\frac{\partial^2 U_i^{(2)}}{\partial t^2}
+...
\right)
-
\varepsilon\,{n}_{iq_1}^{(3,2)}
\left(
\frac{\partial^3 U_i^{(0)}}{\partial x_{q_1}\partial t^2}
+
\varepsilon\,
\frac{\partial^3 U_i^{(1)}}{\partial x_{q_1}\partial t^2}
+
\right.
\nonumber\\
&&
\left.
+\varepsilon^2\,
\frac{\partial^3 U_i^{(2)}}{\partial x_{q_1}\partial t^2}
+...
\right)
-
\varepsilon^2\,{n}_{iq_1q_2}^{(4,2)}
\left(
\frac{\partial^4 U_i^{(0)}}{\partial x_{q_1}\partial x_{q_2}\partial t^2}
+
\varepsilon\,
\frac{\partial^4 U_i^{(1)}}{\partial x_{q_1}\partial x_{q_2}\partial t^2}
+
\varepsilon^2\,
\frac{\partial^4 U_i^{(2)}}{\partial x_{q_1}\partial x_{q_2}\partial t^2}
+...
\right)+
...+b_i(\mathbf{x},t)=0.
\nonumber\\
\end{eqnarray}
From equation (\ref{eq:AverageFieldEquationsInfiniteOrder2}) one obtains
\begin{eqnarray}
\label{eq:RecursiveDifferentialProblemMacroTemperature}
&&m_{q_1q_2}^{(2)}
\left(
\frac{\partial^2\Theta^{(0)}}{\partial x_{q_1}\partial x_{q_2}}
+
\varepsilon\,
\frac{\partial^2 \Theta^{(1)}}{\partial x_{q_1}\partial x_{q_2}}
+
\varepsilon^2\,
\frac{\partial^2 \Theta^{(2)}}{\partial x_{q_1}\partial x_{q_2}}
+...
\right)+
\varepsilon\,m_{q_1...q_3}^{(3)}
\left(
\frac{\partial^3 \Theta^{(0)}}{\partial x_{q_1}...\partial x_{q_3}}
+
\varepsilon\,
\frac{\partial^3 \Theta^{(1)}}{\partial x_{q_1}...\partial x_{q_3}}
+
\right.
\nonumber\\
&&
\left.
+\varepsilon^2\,
\frac{\partial^3 \Theta^{(2)}}{\partial x_{q_1}...\partial x_{q_3}}
+...
\right)+
\varepsilon^2\,m_{q_1...q_4}^{(4)}
\left(
\frac{\partial^4 \Theta^{(0)}}{\partial x_{q_1}...\partial x_{q_4}}
+
\varepsilon\,
\frac{\partial^4 \Theta^{(1)}}{\partial x_{q_1}...\partial x_{q_4}}
+
\varepsilon^2\,
\frac{\partial^4 \Theta^{(2)}}{\partial x_{q_1}...\partial x_{q_4}}
+...
\right)+
\nonumber\\
&&
-\tilde{m}_{pq_1}^{(2,1)}
\left(
\frac{\partial^2 U_p^{(0)}}{\partial x_{q_1}\partial t}
+
\varepsilon\,
\frac{\partial^2 U_p^{(1)}}{\partial x_{q_1}\partial t}
+
\varepsilon^2\,
\frac{\partial^2 U_p^{(2)}}{\partial x_{q_1}\partial t}
+...
\right)
-
\varepsilon\,\tilde{m}_{pq_1q_2}^{(3,1)}
\left(
\frac{\partial^3 U_p^{(0)}}{\partial x_{q_1}\partial x_{q_2}\partial t}
+
\varepsilon\,
\frac{\partial^3 U_p^{(1)}}{\partial x_{q_1}\partial x_{q_2}\partial t}
+
\right.
\nonumber\\
&&
\left.
+\varepsilon^2\,
\frac{\partial^3 U_p^{(2)}}{\partial x_{q_1}\partial x_{q_2}\partial t}
+...
\right)
-
\varepsilon^2\,\tilde{m}_{pq_1...q_3}^{(4,1)}
\left(
\frac{\partial^4 U_p^{(0)}}{\partial x_{q_1}...\partial x_{q_3}\partial t}
+
\varepsilon\,
\frac{\partial^4 U_p^{(1)}}{\partial x_{q_1}...\partial x_{q_3}\partial t}
+
\varepsilon^2\,
\frac{\partial^4 U_p^{(2)}}{\partial x_{q_1}...\partial x_{q_3}\partial t}
+...
\right)+
\nonumber\\
&&
-{m}^{(2,1)}
\left(
\frac{\partial \Theta^{(0)}}{\partial t}
+
\varepsilon\,
\frac{\partial \Theta^{(1)}}{\partial t}
+
\varepsilon^2\,
\frac{\partial \Theta^{(2)}}{\partial t}
+...
\right)
-
\varepsilon\,{m}_{q1}^{(3,1)}
\left(
\frac{\partial^2 \Theta^{(0)}}{\partial x_{q_1}\partial t}
+
\varepsilon\,
\frac{\partial^2 \Theta^{(1)}}{\partial x_{q_1}\partial t}
+
\right.
\nonumber\\
&&
\left.
+\varepsilon^2\,
\frac{\partial^2 \Theta^{(2)}}{\partial x_{q_1}\partial t}
+...
\right)
-
\varepsilon^2\,{m}_{q_1q_2}^{(4,1)}
\left(
\frac{\partial^3 \Theta^{(0)}}{\partial x_{q_1}\partial x_{q_2}\partial t}
+
\varepsilon\,
\frac{\partial^3 \Theta^{(1)}}{\partial x_{q_1}\partial x_{q_2}\partial t}
+
\varepsilon^2\,
\frac{\partial^3 \Theta^{(2)}}{\partial x_{q_1}\partial x_{q_2}\partial t}
+...
\right)+
\nonumber\\
&&
-\hat{m}^{(2,1)}
\left(
\frac{\partial \Upsilon^{(0)}}{\partial t}
+
\varepsilon\,
\frac{\partial \Upsilon^{(1)}}{\partial t}
+
\varepsilon^2\,
\frac{\partial \Upsilon^{(2)}}{\partial t}
+...
\right)
-
\varepsilon\,\hat{m}_{q_1}^{(3,1)}
\left(
\frac{\partial^2 \Upsilon^{(0)}}{\partial x_{q_1}\partial t}
+
\varepsilon\,
\frac{\partial^2 \Upsilon^{(1)}}{\partial x_{q_1}\partial t}
+
\right.
\nonumber\\
&&
\left.
+\varepsilon^2\,
\frac{\partial^2 \Upsilon^{(2)}}{\partial x_{q_1}\partial t}
+...
\right)
-
\varepsilon^2\,\hat{m}_{q_1q_2}^{(4,1)}
\left(
\frac{\partial^3 \Upsilon^{(0)}}{\partial x_{q_1}\partial x_{q_2}\partial t}
+
\varepsilon\,
\frac{\partial^3 \Upsilon^{(1)}}{\partial x_{q_1}\partial x_{q_2}\partial t}
+
\varepsilon^2\,
\frac{\partial^3 \Upsilon^{(2)}}{\partial x_{q_1}\partial x_{q_2}\partial t}
+...
\right)+
...+r(\mathbf{x},t)=0.
\nonumber\\
\end{eqnarray}
Finally, equation (\ref{eq:AverageFieldEquationsInfiniteOrder3}) reads
\begin{eqnarray}
\label{eq:RecursiveDifferentialProblemMacroChemPot}
&&w_{q_1q_2}^{(2)}
\left(
\frac{\partial^2\Upsilon^{(0)}}{\partial x_{q_1}\partial x_{q_2}}
+
\varepsilon\,
\frac{\partial^2 \Upsilon^{(1)}}{\partial x_{q_1}\partial x_{q_2}}
+
\varepsilon^2\,
\frac{\partial^2 \Upsilon^{(2)}}{\partial x_{q_1}\partial x_{q_2}}
+...
\right)+
\varepsilon\,m_{q_1...q_3}^{(3)}
\left(
\frac{\partial^3 \Upsilon^{(0)}}{\partial x_{q_1}...\partial x_{q_3}}
+
\varepsilon\,
\frac{\partial^3 \Upsilon^{(1)}}{\partial x_{q_1}...\partial x_{q_3}}
+
\right.
\nonumber\\
&&
\left.
+\varepsilon^2\,
\frac{\partial^3 \Upsilon^{(2)}}{\partial x_{q_1}...\partial x_{q_3}}
+...
\right)+
\varepsilon^2\,m_{q_1...q_4}^{(4)}
\left(
\frac{\partial^4 \Upsilon^{(0)}}{\partial x_{q_1}...\partial x_{q_4}}
+
\varepsilon\,
\frac{\partial^4 \Upsilon^{(1)}}{\partial x_{q_1}...\partial x_{q_4}}
+
\varepsilon^2\,
\frac{\partial^4 \Upsilon^{(2)}}{\partial x_{q_1}...\partial x_{q_4}}
+...
\right)+
\nonumber\\
&&
-\tilde{w}_{pq_1}^{(2,1)}
\left(
\frac{\partial^2 U_p^{(0)}}{\partial x_{q_1}\partial t}
+
\varepsilon\,
\frac{\partial^2 U_p^{(1)}}{\partial x_{q_1}\partial t}
+
\varepsilon^2\,
\frac{\partial^2 U_p^{(2)}}{\partial x_{q_1}\partial t}
+...
\right)
-
\varepsilon\,\tilde{w}_{pq_1q_2}^{(3,1)}
\left(
\frac{\partial^3 U_p^{(0)}}{\partial x_{q_1}\partial x_{q_2}\partial t}
+
\varepsilon\,
\frac{\partial^3 U_p^{(1)}}{\partial x_{q_1}\partial x_{q_2}\partial t}
+
\right.
\nonumber\\
&&
\left.
+\varepsilon^2\,
\frac{\partial^3 U_p^{(2)}}{\partial x_{q_1}\partial x_{q_2}\partial t}
+...
\right)
-
\varepsilon^2\,\tilde{w}_{pq_1...q_3}^{(4,1)}
\left(
\frac{\partial^4 U_p^{(0)}}{\partial x_{q_1}...\partial x_{q_3}\partial t}
+
\varepsilon\,
\frac{\partial^4 U_p^{(1)}}{\partial x_{q_1}...\partial x_{q_3}\partial t}
+
\varepsilon^2\,
\frac{\partial^4 U_p^{(2)}}{\partial x_{q_1}...\partial x_{q_3}\partial t}
+...
\right)+
\nonumber\\
&&
-{w}^{(2,1)}
\left(
\frac{\partial \Upsilon^{(0)}}{\partial t}
+
\varepsilon\,
\frac{\partial \Upsilon^{(1)}}{\partial t}
+
\varepsilon^2\,
\frac{\partial \Upsilon^{(2)}}{\partial t}
+...
\right)
-
\varepsilon\,{w}_{q_1}^{(3,1)}
\left(
\frac{\partial^2 \Upsilon^{(0)}}{\partial x_{q_1}\partial t}
+
\varepsilon\,
\frac{\partial^2 \Upsilon^{(1)}}{\partial x_{q_1}\partial t}
+
\right.
\nonumber\\
&&
\left.
+\varepsilon^2\,
\frac{\partial^2 \Upsilon^{(2)}}{\partial x_{q_1}\partial t}
+...
\right)
-
\varepsilon^2\,{w}_{q_1q_2}^{(4,1)}
\left(
\frac{\partial^3 \Upsilon^{(0)}}{\partial x_{q_1}\partial x_{q_2}\partial t}
+
\varepsilon\,
\frac{\partial^3 \Upsilon^{(1)}}{\partial x_{q_1}\partial x_{q_2}\partial t}
+
\varepsilon^2\,
\frac{\partial^3 \Upsilon^{(2)}}{\partial x_{q_1}\partial x_{q_2}\partial t}
+...
\right)+
\nonumber\\
&&
-\hat{w}^{(2,1)}
\left(
\frac{\partial \Theta^{(0)}}{\partial t}
+
\varepsilon\,
\frac{\partial \Theta^{(1)}}{\partial t}
+
\varepsilon^2\,
\frac{\partial \Theta^{(2)}}{\partial t}
+...
\right)
-
\varepsilon\,\hat{w}_{q_1}^{(3,1)}
\left(
\frac{\partial^2 \Theta^{(0)}}{\partial x_{q_1}\partial t}
+
\varepsilon\,
\frac{\partial^2 \Theta^{(1)}}{\partial x_{q_1}\partial t}
+
\right.
\nonumber\\
&&
\left.
+\varepsilon^2\,
\frac{\partial^2 \Theta^{(2)}}{\partial x_{q_1}\partial t}
+...
\right)
-
\varepsilon^2\,\hat{w}_{q_1q_2}^{(4,1)}
\left(
\frac{\partial^3 \Theta^{(0)}}{\partial x_{q_1}\partial x_{q_2}\partial t}
+
\varepsilon\,
\frac{\partial^3 \Theta^{(1)}}{\partial x_{q_1}\partial x_{q_2}\partial t}
+
\varepsilon^2\,
\frac{\partial^3 \Theta^{(2)}}{\partial x_{q_1}\partial x_{q_2}\partial t}
+...
\right)+
...+s(\mathbf{x},t)=0.
\nonumber\\
\end{eqnarray}
Truncating at the order $\varepsilon^0$, from equation (\ref{eq:RecursiveDifferentialProblemMacroDisplacement})  the following macro differential problem is derived
\begin{eqnarray}
\label{eq:ZerothOrderProblemMacroDisplacement}
n_{ipq_1q_2}^{(2)}
\frac{\partial^2 U_p^{(0)}}{\partial x_{q_1}\partial x_{q_2}}
-
\tilde{n}_{iq_1}^{(2)}
\frac{\partial\Theta^{(0)}}{\partial x_{q_1}}
-
\hat{n}_{iq_1}^{(2)}
\frac{\partial\Upsilon^{(0)}}{\partial x_{q_1}}
-
n^{(2,2)}
\frac{\partial^2 U_i^{(0)}}{\partial t^2}
+
b_i(\mathbf{x},t)=0.
\end{eqnarray}
Analogously, macro differential problem obtained truncating equation (\ref{eq:RecursiveDifferentialProblemMacroTemperature}) at the order $\varepsilon^0$ has the form
\begin{eqnarray}
\label{eq:ZerothOrderProblemMacroTemperature}
m_{q_1q_2}^{(2)}
\frac{\partial^2 \Theta^{(0)}}{\partial x_{q_1}\partial x_{q_2}}
-
\tilde{m}_{pq_1}^{(2,1)}
\frac{\partial^2 U_p^{(0)}}{\partial x_{q_1}\partial t}
-
{m}^{(2,1)}
\frac{\partial\Theta^{(0)}}{\partial t}
-
\hat{m}^{(2,1)}
\frac{\partial \Upsilon^{(0)}}{\partial t}
+
r(\mathbf{x},t)=0.
\end{eqnarray}
Third macro problem from equation (\ref{eq:RecursiveDifferentialProblemMacroChemPot}) reads
\begin{eqnarray}
\label{eq:ZerothOrderProblemMacroChemPot}
w_{q_1q_2}^{(2)}
\frac{\partial^2 \Upsilon^{(0)}}{\partial x_{q_1}\partial x_{q_2}}
-
\tilde{w}_{pq_1}^{(2,1)}
\frac{\partial^2 U_p^{(0)}}{\partial x_{q_1}\partial t}
-
{w}^{(2,1)}
\frac{\partial\Upsilon^{(0)}}{\partial t}
-
\hat{w}^{(2,1)}
\frac{\partial \Theta^{(0)}}{\partial t}
+
s(\mathbf{x},t)=0.
\end{eqnarray}
The following normalization conditions
\begin{equation}
\label{eq:NormalizationConditionsMacroFields}
\frac{1}{\delta L^2}
\,
\int_{\mathcal{L}}U_p^{(m)}(\mathbf{x},t)\,d\mathbf{x}=0,
\hspace{0,2cm}
\frac{1}{\delta L^2}
\,
\int_{\mathcal{L}}\Theta^{(m)}(\mathbf{x},t)\,d\mathbf{x}=0,
\hspace{0,2cm}
\frac{1}{\delta L^2}
\,
\int_{\mathcal{L}}\Upsilon^{(m)}(\mathbf{x},t)\,d\mathbf{x}=0,
\end{equation}
are demanded to be satisfied by
macro fields $\mathbf{U}(\mathbf{x},t)^{(m)}, \Theta^{(m)}(\mathbf{x},t)$ and $\Upsilon^{(m)}(\mathbf{x},t)$, in the case of $\mathcal{L}$-periodic source terms, for each $m\in\mathbb{Z}$. In this case, macro fields result to be $\mathcal{L}$-periodic, too.
If source terms are not $\mathcal{L}$-periodic, normalization conditions (\ref{eq:NormalizationConditionsMacroFields}) need to be substituted by appropriate boundary conditions  to compute the macro fields.
In fact, $\mathcal{L}$-periodicity is not a mandatory requirement for source terms. These lasts are only required to show a variability much greater than the characteristic microstructural length $\varepsilon$ in order to preserve the separation of scales. 
In order to derive governing field equations for the first-order homogenized  continuum,
zeroth order differential problems
(\ref{eq:ZerothOrderProblemMacroDisplacement})-(\ref{eq:ZerothOrderProblemMacroChemPot}) need to be expressed in terms of components $C_{iq_1pq_2},\alpha_{iq_1},\beta_{iq_1},K_{q_1q_2},D_{q_1q_2}$ of overall constitutive tensors, in terms of overall thermo-diffusive coupling constant $\psi$ and overall inertial terms $\rho, p$ and $q$.
Relations between  components of the relative overall constitutive tensors $\mathfrak{C}, \mathbf{K}$, and $\mathbf{D}$ and the ones  of tensors $\mathbf{n}^{(2)},\mathbf{m}^{(2)}$, and $\mathbf{w}^{(2)}$  are detailed in \citep{fantoni2017multi} and read
\begin{equation}
n^{(2)}_{ipq_1q_2}=\frac{1}{2}\left(C_{pq_1iq_2}+C_{pq_2iq_1}\right),
\hspace{0.2 cm}
m^{(2)}_{q_1q_2}=K_{q_1q_2},
\hspace{0.2 cm}
w^{(2)}_{q_1q_2}=D_{q_1q_2}.
\label{eq:correspondenceWithOverallTensors1}
\end{equation}
Symmetries and positive definition of tensors $\mathbf{n}^{(2)}=n_{ipq_1q_1}^{(2)} \mathbf{e}_i\otimes\mathbf{e}_p\otimes\mathbf{e}_{q_1}\otimes\mathbf{e}_{q2}$, $\mathbf{m}^{(2)}=m^{(2)}_{q_1q_2}\mathbf{e}_{q_1}\otimes\mathbf{e}_{q_2}$, and $\mathbf{w}^{(2)}=w^{(2)}_{q_1q_2}\mathbf{e}_{q_1}\otimes\mathbf{e}_{q_2}$ are accurately provided in   the above mentioned references, where is proved that such tensors can be expressed as
\begin{eqnarray}
n_{ipq_1q_2}^{(2)}&=& 
\frac{1}{2}
\left\langle
\left.
C_{rjkl}^{m}
\left(
N_{riq_2,j}^{(1)}+
\delta_{ir}\delta_{jq_2}
\right)
\left(
N_{kpq_1,l}^{(1)}+
\delta_{pk}\delta_{lq_1}\right)
+
\right.
\right.
\nonumber\\
&+&
\left.
\left.
C_{rjkl}^{m}
\left(
N_{riq_1,j}^{(1)}+
\delta_{ir}\delta_{jq_1}
\right)
\left(
N_{kpq_2,l}^{(1)}+
\delta_{pk}\delta_{lq_2}\right)
\right.
\right\rangle,
\nonumber\\
{m}_{q_1q_2}^{(2)}
&=& K_{q_1q_2}=
\left\langle
K_{ij}^m\left(
M_{q_2,i}^{(1)}+\delta_{iq_2}
\right)
\left(
M_{q_1,j}^{(1)}+\delta_{jq_1}
\right)
\right\rangle,
\nonumber\\
{w}_{q_1q_2}^{(2)}
&=&D_{q_1q_2}=
\left\langle
D_{ij}^m
\left(
W_{q_2,i}^{(1)}+\delta_{iq_2}
\right)
\left(
W_{q_1,j}^{(1)}+\delta_{jq_1}
\right)
\right\rangle.
\label{eq:correspondenceWithOverallTensors2}
\end{eqnarray} 
A comparison between the first of equations (\ref{eq:correspondenceWithOverallTensors1}) and the first of (\ref{eq:correspondenceWithOverallTensors2}) leads to the expression of components of overall elastic tensor $\mathfrak{C}$, namely
\begin{equation}
C_{pq_1iq_2}=
\left\langle
C_{rjkl}^m
\left(
N_{riq_2,j}^{(1)}
+
\delta_{ir}
\delta_{jq_2}
\right)
\left(
N_{kpq_1,l}^{(1)}
+
\delta_{pk}
\delta_{lq_1}
\right)
\right\rangle.
\end{equation}
In Appendix A equalities $\alpha_{pq_1}=\tilde{n}^{(2)}_{pq_1}=\tilde{m}^{(2,1)}_{pq_1}$, $\beta_{pq_1}=\hat{n}^{(2)}_{pq_1}=\tilde{w}^{(2,1)}_{pq_1}=$ and $\psi=\hat{m}^{(2,1)}=\hat{w}^{(2,1)}$ are proved in detail.
Equalities between scalars $n^{(2,2)}=\rho$, $m^{(2,1)}=p$, and $w^{(2,1)}=q$, involving overall inertial terms,  trivially follow.
Field equations for the equivalent first-order (Cauchy) thermo-diffusive medium are therefore expressed in the form
\begin{subequations}
\begin{align}
&
C_{iq_1pq_2}\,\frac{\partial^2 U_p(\mathbf{x},t)}{\partial x_{q_1}\partial x_{q_2}}
-
\alpha_{iq_1}\,\frac{\partial \Theta(\mathbf{x},t)}{\partial x_{q_1}}
-
\beta_{iq_1}\,\frac{\partial \Upsilon(\mathbf{x},t)}{\partial x_{q_1}}
-
\rho
\frac{\partial^2 U_i(\mathbf{x},t)}{\partial t^2}
+
b_i(\mathbf{x},t)=0,
\label{eq:CauchyFieldEquationDisplacement}\\
&
K_{q_1q_2}\,\frac{\partial^2 \Theta(\mathbf{x},t)}{\partial x_{q_1}\partial x_{q_2}}
-
\alpha_{pq_1}
\frac{\partial^2 U_p(\mathbf{x},t)}{\partial x_{q_1}\partial t}
-\psi
\frac{\partial\Upsilon(\mathbf{x},t)}{\partial t}
-
p
\frac{\partial\Theta(\mathbf{x},t)}{\partial t}
+
r(\mathbf{x},t)=0,
\label{eq:CauchyFieldEquationTemperature}\\
&
D_{q_1q_2}\,\frac{\partial^2 \Upsilon(\mathbf{x},t)}{\partial x_{q_1}\partial x_{q_2}}
-
\beta_{pq_1}
\frac{\partial^2 U_p(\mathbf{x},t)}{\partial x_{q_1}\partial t}
-\psi
\frac{\partial\Theta(\mathbf{x},t)}{\partial t}
-
q
\frac{\partial\Upsilon(\mathbf{x},t)}{\partial t}
+
s(\mathbf{x},t)=0,
\label{eq:CauchyFieldEquationChemPot}
\end{align}
\label{eq:ConstitutiveMacroscopicFieldEquations}
\end{subequations}
where macro fields correspond to the zeroth order ones, namely
\begin{equation}
U_p(\mathbf{x},t)\approx U_p^{(0)}(\mathbf{x},t),
\hspace{0.2cm}
\Theta(\mathbf{x},t)\approx \Theta^{(0)}(\mathbf{x},t),
\hspace{0.2cm}
\Upsilon(\mathbf{x},t)\approx \Upsilon^{(0)}(\mathbf{x},t).
\end{equation}
%
%
%

\section{Complex frequency band structure of the equivalent thermo-diffusive medium}
\label{Sec::PropagazioneContinuoalPrimoOrdine}
A two-sided Laplace transform of a real valued time dependent function $f:\mathbb{R}\rightarrow\mathbb{R}$ is defined in the following way \citep{Paley1934}
\begin{equation}
\label{eq:LaplaceTrasformata}
\mathcal{L}(f(t))=\hat{f}(\omega)=\int_{-\infty}^{+\infty}f(t)\,e^{-\omega\,t}\,dt
\end{equation}
with the Laplace argument $\omega\in\mathbb{C}$ and the Laplace transform a complex valued function $\hat{f}:\mathbb{C}\rightarrow\mathbb{C}$.
Taking into account the following  derivation rule 
\begin{equation}
\label{eq:DerivataLaplaceTrasformata}
\mathcal{L}
\left(\frac{ \partial^n f(t)}{\partial t^n}\right)=\omega^n\hat{f}(\omega)
\end{equation}
and performing Laplace transform of field equations (\ref{eq:ConstitutiveMacroscopicFieldEquations}), one obtains the following generalized Christoffel equations  for the first-order equivalent medium
%
%
%
%
\begin{subequations}
\label{eq:ChristoffelEquations}
\begin{align}
&C_{iq_1pq_2}\frac{\partial^2 \hat{U}_p(\mathbf{x},\omega)}{\partial x_{q_1}\partial x_{q_2}}
-
\alpha_{iq_1}\frac{\partial \hat{\Theta}(\mathbf{x},\omega)}{\partial x_{q_1}}
-
\beta_{iq_1}\frac{\partial \hat{\Upsilon}(\mathbf{x},\omega)}{\partial x_{q_1}}
-
\rho\omega^2\hat{U}_i (\mathbf{x},\omega)
+
\hat{b}_i(\mathbf{x},\omega)=0,\label{eq:ChristoffelSpostamento}
\\
&K_{q_1q_2}\,\frac{\partial^2 \hat{\Theta}(\mathbf{x},\omega)}{\partial x_{q_1}\partial x_{q_2}}
-
\alpha_{pq_1}\omega
\frac{\partial \hat{U}_p(\mathbf{x},\omega)}{\partial x_{q_1}}
-\psi\omega\,
\hat{\Upsilon}(\mathbf{x},\omega)
-
p\omega\,
\hat{\Theta}(\mathbf{x},\omega)
+
\hat{r}(\mathbf{x},\omega)=0,\label{eq:ChristoffelTemperatura}\\
&D_{q_1q_2}\,\frac{\partial^2 \hat{\Upsilon}(\mathbf{x},\omega)}{\partial x_{q_1}\partial x_{q_2}}
-
\beta_{pq_1}\omega
\frac{\partial \hat{U}_p(\mathbf{x},\omega)}{\partial x_{q_1}}
-\psi\omega\,
\hat{\Theta}(\mathbf{x},\omega)
-
q\omega\,
\hat{\Upsilon}(\mathbf{x},\omega)
+
\hat{s}(\mathbf{x},\omega)=0,\label{eq:ChristoffelPotenziale}.
\end{align}
\end{subequations}
Fourier transform of a real valued, space varying function $f$ has the following definition \citep{Paley1934}
\begin{equation}
\label{eq:FourierTrasformata}
\mathcal{F}(f(\mathbf{x}))=\check{f}(\mathbf{k})=\int_{-\infty}^{+\infty}\int_{-\infty}^{+\infty}f(\mathbf{x})
\,e^{-i\mathbf{k}\mathbf{x}}\,d\mathbf{x},
\end{equation}
where  Fourier argument $\mathbf{k}\in\mathbb{R}^2$ and $i$ is the imaginary unit such that $i^2=-1$.
Fourier transform of equations (\ref{eq:ChristoffelEquations}), bearing in mind derivation rule 
\begin{equation}
\label{eq:DerivataFourierTrasformata}
\mathcal{F}
\left(\frac{ \partial^n f(\mathbf{x})}{\partial x_j^n}\right)=(i\,k_j)^n\check{f}(\mathbf{k})
\end{equation}
leads to the following  equations
\begin{subequations}
\label{eq:ChristoffelEqsTrasformate}
\begin{align}
&-k_{q_1}k_{q_2}C_{iq_1pq_2}\check{\hat{U}}_p(\mathbf{k},\omega)
-
\alpha_{iq_1}ik_{q_1}\check{\hat{\Theta}}(\mathbf{k},\omega)
-
\beta_{iq_1}ik_{q_1}\check{\hat{\Upsilon}}(\mathbf{k},\omega)
-
\rho\omega^2\check{\hat{U}}_i(\mathbf{k},\omega)
+
\check{\hat{b}}_i(\mathbf{k},\omega)=0,\label{eq:ChristoffelTrasformataSpostamento}\\
&-k_{q_1}k_{q_2}K_{q_1q_2}\check{\hat{\Theta}}(\mathbf{k},\omega)
-
\alpha_{pq_1}ik_{q_1}\omega\check{\hat{U}}_p(\mathbf{k},\omega)
-
\psi\omega\check{\hat{\Upsilon}}(\mathbf{k},\omega)
-
p\omega\check{\hat{\Theta}}(\mathbf{k},\omega)
+
\check{\hat{r}}(\mathbf{k},\omega)=0,\label{eq:ChristoffelTrasformataTemperatura}\\
&-k_{q_1}k_{q_2}D_{q_1q_2}\check{\hat{\Upsilon}}(\mathbf{k},\omega)
-
\beta_{pq_1}ik_{q_1}\omega\check{\hat{U}}_p(\mathbf{k},\omega)
-
\psi\omega\check{\hat{\Theta}}(\mathbf{k},\omega)
-
q\omega\check{\hat{\Upsilon}}(\mathbf{k},\omega)
+
\check{\hat{s}}(\mathbf{k},\omega)=0.\label{eq:ChristoffelTrasformataPotenziale}
\end{align}
\end{subequations}
With the aim of studying the propagation of free waves inside the equivalent thermo-diffusive material, source terms are put equal to zero ($\check{\hat{\mathbf{b}}}=\mathbf{0}$, $\check{\hat{r}}=0$, $\check{\hat{s}}=0$)
in equations (\ref{eq:ChristoffelEqsTrasformate}).
Waves propagating inside the medium will be damped in time and dispersive, because of the structure of governing field equations (\ref{eq:ConstitutiveMacroscopicFieldEquations}).
Governing equations in the transformed space and frequency domain (\ref{eq:ChristoffelEqsTrasformate}) can be written in absolute notation as
\begin{subequations}
\label{eq:ChristoffelEqsTrasformateNotazioneAssoluta}
\begin{align}
&
\left(
\tilde{\mathfrak{C}}(\mathbf{k}\otimes\mathbf{k})
+
\rho\omega^2\mathbf{I}
\right)
\check{\hat{\mathbf{U}}}(\mathbf{k},\omega)
+
i\tensor{\alpha}\mathbf{k}
\check{\hat{{\Theta}}}(\mathbf{k},\omega)
+
i\tensor{\beta}\mathbf{k}
\check{\hat{{\Upsilon}}}(\mathbf{k},\omega)=\mathbf{0},
\label{eq:ChristoffelTrasNotAssSpostamento}\\
&
\left(
{\mathbf{k}}:(\mathbf{k}\otimes\mathbf{k})
+
p\omega
\right)
\check{\hat{{\Theta}}}(\mathbf{k},\omega)
+
i\omega(\tensor{\alpha}\mathbf{k}).\,
\check{\hat{\tensor{U}}}(\mathbf{k},\omega)
+
\psi\omega
\check{\hat{{\Upsilon}}}(\mathbf{k},\omega)={0},
\label{eq:ChristoffelTrasNotAssTemperatura}\\
&
\left(
{\tensor{D}}:(\mathbf{k}\otimes\mathbf{k})
+
q\omega
\right)
\check{\hat{{\Upsilon}}}(\mathbf{k},\omega)
+
i\omega(\tensor{\beta}\mathbf{k}).\,
\check{\hat{\tensor{U}}}(\mathbf{k},\omega)
+
\psi\omega
\check{\hat{{\Theta}}}(\mathbf{k},\omega)={0},
\label{eq:ChristoffelTrasNotAssPotensiale}
\end{align}
\end{subequations}
and in matrix notation as
%
%
%
\begin{equation}
\label{eq:ChristoffelEqsTrasformateNotazioneMatriciale}
\left(
\begin{array}{ccc}
\tilde{\mathfrak{C}}(\mathbf{k}\otimes\mathbf{k})+\rho\omega^2\mathbf{I}&
i\tensor{\alpha}\mathbf{k}&
i\tensor{\beta}\mathbf{k}\\
i\omega(\tensor{\alpha}\mathbf{k})^T&
\mathbf{K}:(\mathbf{k}\otimes\mathbf{k})+p\,\omega&
\psi\omega\\
i\omega(\tensor{\beta\mathbf{k}})^T&
\psi\omega&
\tensor{D}:(\mathbf{k}\otimes\mathbf{k})+q\,\omega
\end{array}
\right)
\left(
\begin{array}{c}
\check{\hat{\mathbf{U}}}\\
\check{\hat{{\Theta}}}\\
\check{\hat{{\Upsilon}}}
\end{array}
\right)=
\left(
\begin{array}{c}
\mathbf{0}\\
0
\\
0
\end{array}
\right),
\end{equation}
 where $\tilde{\mathfrak{C}}=\tilde{C}_{ipq_1q_2}\mathbf{e}_i\otimes\mathbf{e}_p\otimes\mathbf{e}_{q_1}\otimes\mathbf{e}_{q_2}$
and
$\tilde{C}_{ipq_1q_2}=C_{iq_1pq_2}$ and $\mathbf{I}$ is the identity operator.
 Equation (\ref{eq:ChristoffelEqsTrasformateNotazioneMatriciale}) represents a quadratic generalized eigenvalue problem that can be written in a concise form as 
 \begin{equation}
 \left(
 \mathbf{H}_2\omega^2+\mathbf{H}_1\omega+\mathbf{H}_0
 \right)
 \mathbf{Z}
 =
 \mathbf{0},
 \label{eq:GeneralizedEigenValueProblemCompactForm}
 \end{equation} 
 where $\omega$ corresponds to the generalized eigenvalue and $\mathbf{Z}=(\check{\hat{\mathbf{U}}}\hspace{0.1cm} \check{\hat{\mathbf{\Theta}}} \hspace{0.1cm}\check{\hat{\mathbf{\Upsilon}}})^T$ is  the generalized eigenvector.  
 Generalized eigenvalue $\omega$ is the complex angular frequency of the damped wave  and its real and imaginary parts describe the damping and the propagation modes of dispersive Bloch waves propagating inside the medium, respectively.
 Vector  $\mathbf{Z}$, which collect the macrofields in the transformed space and frequency domain, is the polarization vector of the damped wave, while
 $\mathbf{k}=k_1\,\mathbf
 e_1+k_2\,\mathbf{e}_2 \in \mathcal{B}$ represents the wave vector, with $k_1$ and $k_2$ the wave numbers and $\mathcal{B}=[-\pi/d_1,\pi/d_1]\times[-\pi/d_2,\pi/d_2]$ the first Brillouin zone associated to periodic cell $\mathcal{A}$.
 Complex frequencies $\omega$ related to problem (\ref{eq:GeneralizedEigenValueProblemCompactForm}) are computed as the roots of the characteristic equation
 \begin{equation}
 det(\mathbf{H})=0,
 \label{eq:characteristic_equation}
\end{equation}  
with matrix $\mathbf{H}=\mathbf{H}_2\,\omega^2+\mathbf{H}_1\,\omega+\mathbf{H}_0$, thus defining the complex frequency band structure of the periodic thermo-diffusive homogenized medium.
 Complex algebraic operators $\mathbf{H}_2$, $\mathbf{H}_1$ and $\mathbf{H}_0$ are such that $\mathbf{H}_2$ is constant with respect to $\mathbf{k}$, while $\mathbf{H}_1$ and $\mathbf{H}_0$ quadratically and linearly depend upon $\mathbf{k}$. Consequently, complex angular frequency $\omega$ depends upon $\mathbf{k}$, thus defining the complex dispersion curves characterizing the equivalent medium.
 %
 \subsection{Asymptotic approximation of the complex spectrum}
 \label{SubSec::AsympApproxOfComplexSpectrum}
%
After representing the wave vector components in a polar coordinate system as $k_1=r \,cos(\phi)$ and $k_2=r\, sin(\phi)$, with $r=||\mathbf{k}||_2=\sqrt{k_1^2+k_2^2}$ the radial coordinate and $\phi$ the angular coordinate, for a given value of $\phi$, characteristic equation (\ref{eq:characteristic_equation}) can be written in the form $F(\omega(r),r)=0$.
Since the characteristic function $F(\omega(r),r)$ substantially depends upon the $r$ variable,
in order to find an explicit solution of the  characteristic equation $F(\omega(r),r)=0$, an asymptotic expansion of function $\omega(r)$ is performed in powers of $r$, which essentially acts as a single perturbation parameter. Asymptotic expansion reads
%
 \begin{equation}
 \omega(r)=\omega^{[0]}+\sum_{n\in\mathbb{N}^*}\omega^{[n]}r^n=\omega^{[0]}+\omega^{[1]}\,r
 +\omega^{[2]}\,r^2 + ... + \omega^{[n]}\,r^n + ...
 \label{eq:asymptotic_expansion_eigenvalue}
 \end{equation}
Assuming sufficient regularity for dispersion function $\omega(r)$, expansion (\ref{eq:asymptotic_expansion_eigenvalue}) locally approximates the exact eigenvalue $\omega$ in the vicinity of the reference point $r=0$.
 Once multiplied by factorial $n!$,  coefficient $\omega^{[n]}$ of (\ref{eq:asymptotic_expansion_eigenvalue}) represents the unknown $r$-derivative of order $n$ of the exact, but implicit equation $F(\omega(r),r)=0$. 
 In this regard, approximation (\ref{eq:asymptotic_expansion_eigenvalue}) of dispersion function $\omega(r)$ is tangent to the exact  dispersion curve in $r=0$, while, for increasing values of parameter $r$, the accuracy of approximation (\ref{eq:asymptotic_expansion_eigenvalue}) is expected to  diminish.
 Established the series (\ref{eq:asymptotic_expansion_eigenvalue}), characteristic function $F(\omega(r),r)$ can be regarded as  a composite single variable function $G(r)$ and its Taylor expansion reads
 \begin{equation}
 G(r)=G^{[0]}
 +
 \sum_{n\in\mathbb{N}^*}
 \frac{G^{[n]}}
 {n!}\,r^n
 =
 G^{[0]}
 +
 G^{[1]}\,r 
 +
 \frac{G^{[2]}}{2}\,r^2
 +
 ...
 +
 \frac{G^{[n]}}{n!}\,r^n
 +
 ...
 \label{eq:TaylorExpansionOfG}
 \end{equation}
 in powers of radial coordinate $r$.
 %
 %
 %
 %
 %
 %
%
%
%
%
Beginning with the {\itshape generating solution}  at the order $r^0$, which defines the six 
known 
eigenvalues $\omega^{[0]}=0$ as solutions of $G^{[0]}(0)=0$, equating to zero each coefficient $G^{[n]}$ at the order $r^n$, the approximate characteristic equation $G(r)=0$ results asymptotically satisfied. The procedure gives rise to  a chain of $n$-ordered equations called {\itshape perturbation equations}, each one characterized by a single unknown, namely one of the higher order sensitivities $\omega^{[n]}$.
Higher order coefficients $G^{[n]}$ of (\ref{eq:TaylorExpansionOfG}) represents the $r$-derivative of  order $n$ of function $G(r)$ evaluated at $r=0$, thus requiring the recursive implementation of the chain rule in order to obtain  the differentiation of a composite function.
Lowest order characteristic polynomials $G^{[1]}$ and $G^{[2]}$ have the form 
 \begin{eqnarray}
 &&
 r^1: \hspace{0.2cm} 
 G^{[1]}
 =
 \omega^{[1]}
 \frac{\partial F(\omega,r)}{\partial \omega} 
 +
 \frac{\partial F(\omega,r)}{\partial r},
 \nonumber\\
 &&
 r^2: \hspace{0.2cm}
 G^{[2]}
 =
 2
 \,\omega^{[2]}
 \frac{\partial F(\omega,r)}{\partial \omega}
 +
 {\omega^{[1]}}^2
\frac{\partial^2 F(\omega, r)}{\partial \omega^2}
+
2\,
\omega^{[1]}\,
 \frac{\partial^2 F(\omega,r)}{\partial \omega \partial r}
 +
\frac{\partial^2 F(\omega, r)}{\partial r^2} ,
\end{eqnarray}
 where the partial derivatives of function $F$ are  evaluated at $\omega=\omega^{[0]}$ and $r=0$. The generalization of the chain rule to higher order derivatives
 can be found in \cite{BacigalupoLepidi2016}, formula (23),  where it is expressed in a recursive form of the generic $n_{th}$ sensitivity. %
The solution scheme needed to accomplish a fourth order approximation
\begin{equation}
\omega_i(r)=\omega_i^{[0]} + \omega_i^{[1]}\,r+\omega_i^{[2]}\,r^2+\omega_i^{[3]}\,r^3+\omega_i^{[4]}\,r^4+O(r^5)
\label{eq:fourthorderapproximationOfTheEigenvalue}
\end{equation}
for all the six eigenvalues ($i=1,...,6$) is described in table \ref{Table::SolutionSchemePerturbativeApproach} and it is valid for any angular coordinate $\phi$.
As evident from table \ref{Table::SolutionSchemePerturbativeApproach}, when sensitivity $\omega_i^{[n]}$ has a multiplicity $m>1$, the successive $m-1$ perturbation equations result to be indeterminate and sensitivity $\omega_i^{[n+1]}$ is computed as the solution of the next $m_{th}$ perturbation problem.
Perturbative technique described in the present Section  allows obtaining a parametric approximation of the  complex eigenspectrum of polinomial operator $\mathbf{H}(\omega(r),r)$ at $r=0$, from which the explicit dependence of complex dispersion functions $\omega(r)$ upon the overall constitutive parameters of the homogenized medium is obtained in a compact form.
Such explicit expression of sensitivities $\omega_i^{[n]}$, with $n=0,...,4$ and $i=1,...,6$ is reported in Appendices B and C for angular coordinate $\phi=0$.
\begin{table}[h!]
\caption{ \it Solution scheme to compute sensitivities $\omega_i^{[n]}$, with $n=1,...,4$ and $i=1,...,6$ from perturbation equation $G^{[n]}=0$ at the order $r^n$. Symbol ``-" has the meaning of ``indeterminate", while symbol ``..." means ``higher order unknowns". }
\renewcommand{\arraystretch}{1.5}
\begin{center}
\begin{tabular}{c;{0.01mm/3pt}c;{0.01mm/3pt}c;{0.01mm/3pt}c;{0.01mm/3pt}c;{0.01mm/3pt}c|c|c|c|c|c}
\bottomrule[1.5pt]
$r^0$ & $r ^1$ & $r ^2$ & $r^3$ & $r^4$ & $r^5$ & $r^6$ & $r^7$ & $r^8$ & $r^9$ & $r^{10}$\\
\bottomrule[1.5pt]
 &  &  &  &  &  & \multirow{2}{*}{ $\omega^{[1]}_{1,2}$} & \multirow{2}{*}{-} & $\omega^{[2]}_1$ & $\omega^{[3]}_1$ & $\omega^{[4]}_1$\\
 & & & & & & & & $\omega^{[2]}_2$ & $\omega^{[3]}_2$ & $\omega^{[4]}_2$\\
 \cline{7-11}
$\omega^{[0]}_{1...6}$& - & -& -& -& -& $\omega^{[1]}_3$ & $\omega^{[2]}_3$ & $\omega^{[3]}_3$ & $\omega^{[4]}_3$ & ...\\
\cline{7-11}
&  & & & & & $\omega^{[1]}_4$ & $\omega^{[2]}_4$ & $\omega^{[3]}_4$ & $\omega^{[4]}_4$ & ...\\
\cline{7-11}
&  & & & & & $\omega^{[1]}_5$ & $\omega^{[2]}_5$ & $\omega^{[3]}_5$ & $\omega^{[4]}_5$ & ...\\
\cline{7-11}
&  & & & & & $\omega^{[1]}_6$ & $\omega^{[2]}_6$ & $\omega^{[3]}_6$ & $\omega^{[4]}_6$ & ...\\
\bottomrule[1.5pt]
\end{tabular}
\end{center}
\label{Table::SolutionSchemePerturbativeApproach}
\end{table}%

\newpage
\section{Benchmark test: dispersion properties of SOFC-like devices}
\label{Sec:Benchmark}
\begin{figure}[h!]
  \centering
  \begin{tabular}{c c }
 \hspace{-0.75cm}
  \includegraphics[width=8cm]{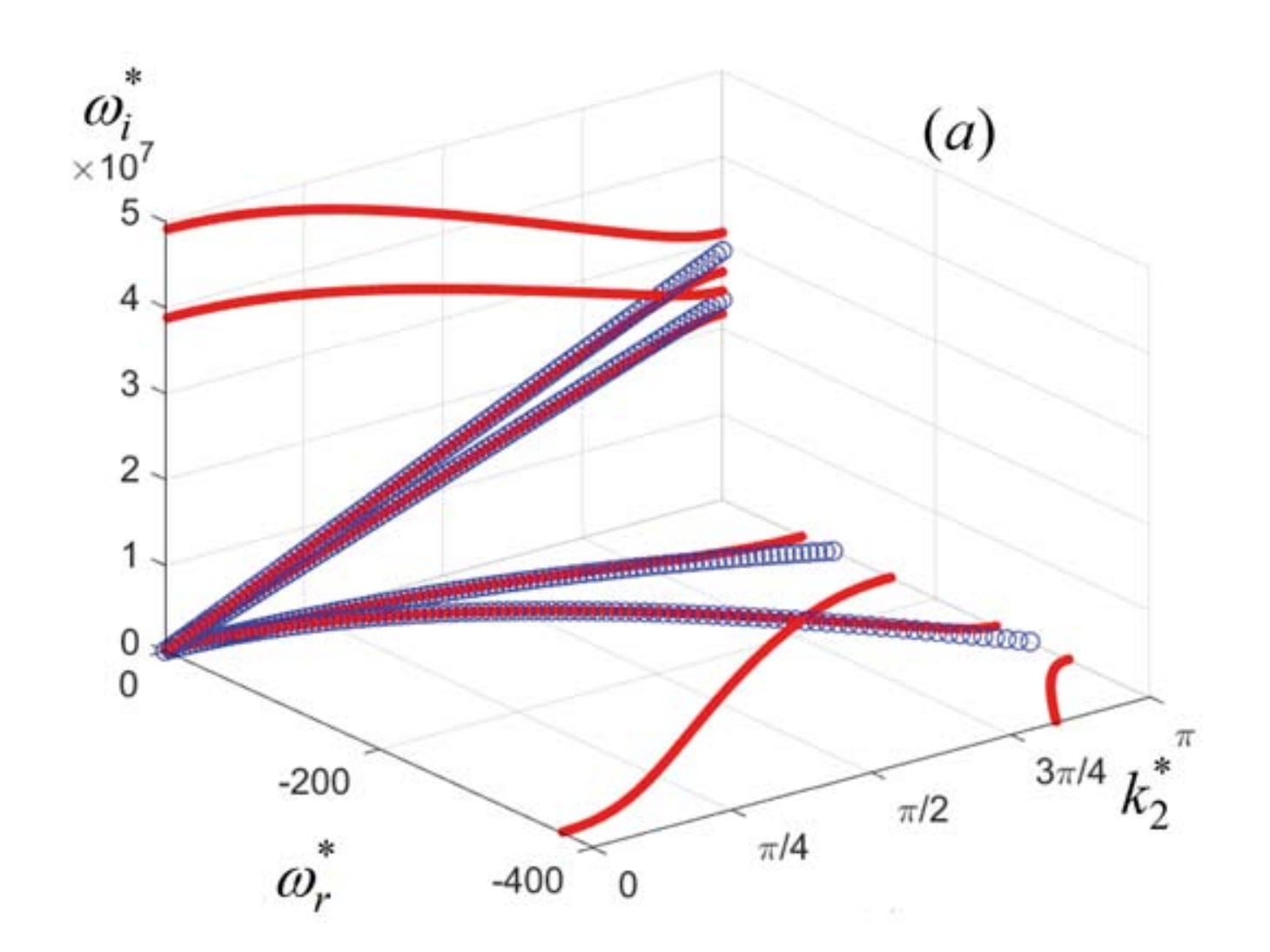}
  &
 \hspace{-1cm}
  \includegraphics[width=8cm]{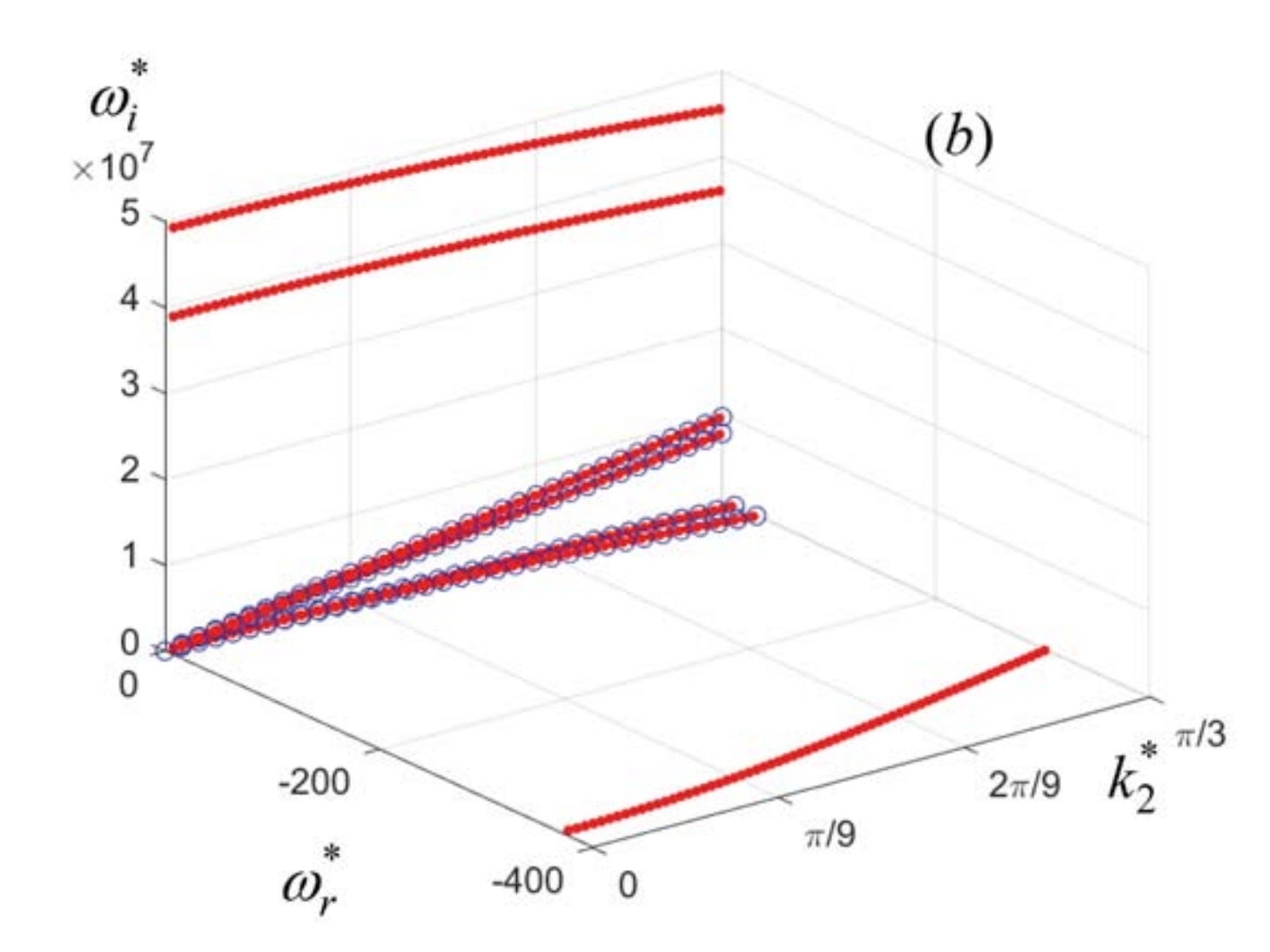}
   \\
 \hspace{-0.75cm}
  \includegraphics[width=8cm]{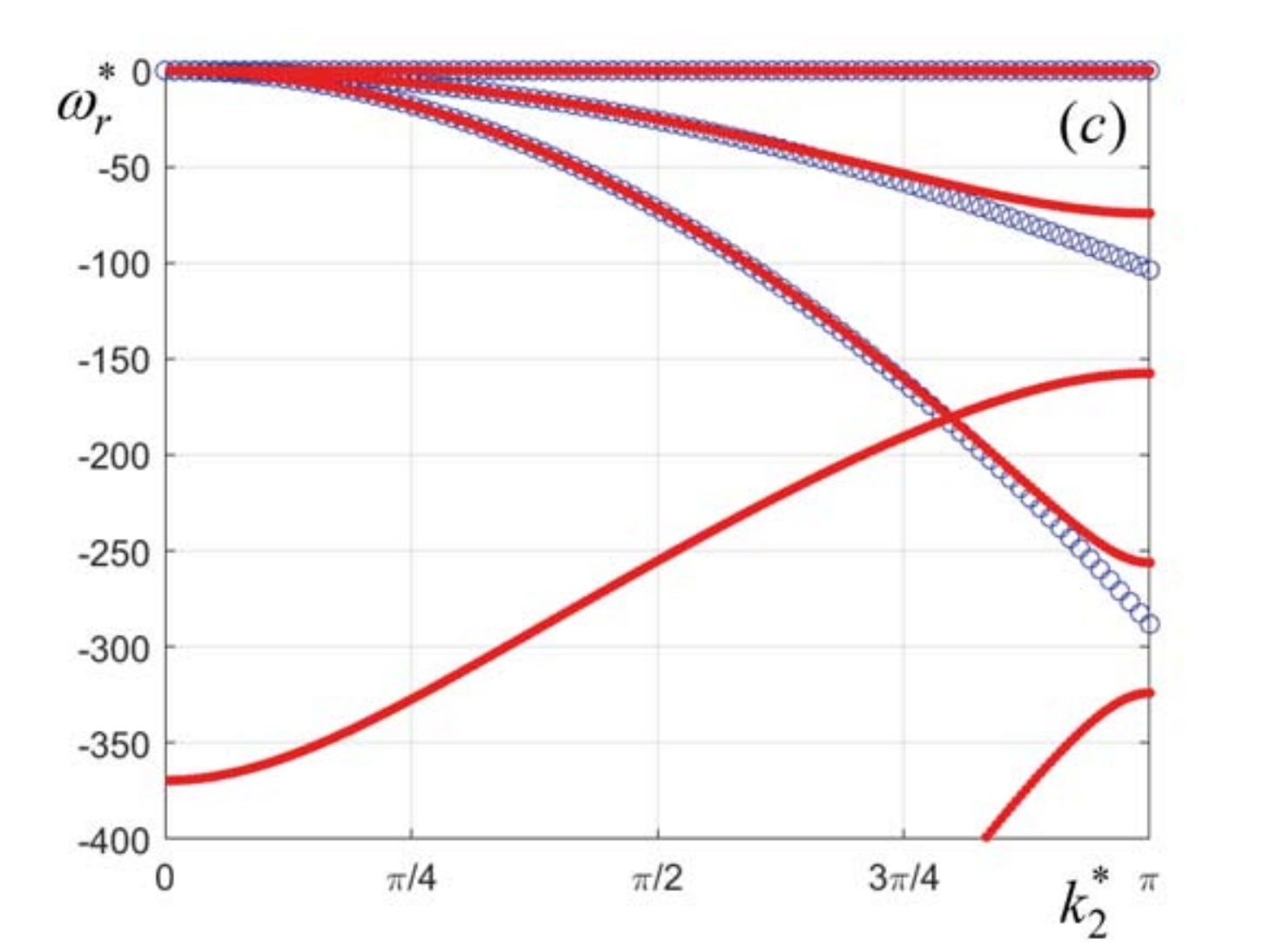}
  &
 \hspace{-1cm}
  \includegraphics[width=8cm]{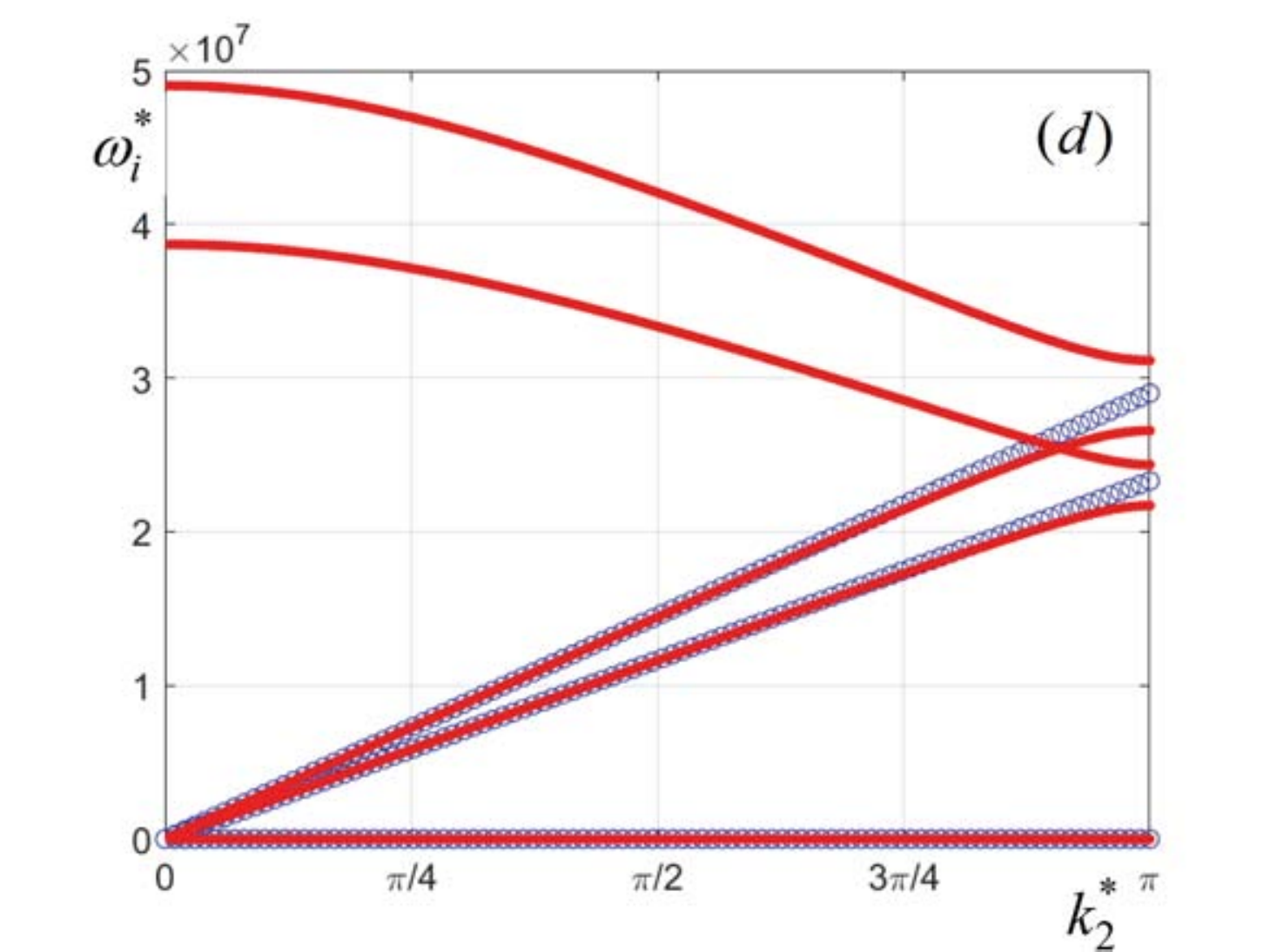}
  \end{tabular}
  \caption{\it Complex frequency spectrum of the heterogeneous SOFC-like material (red curves) and of the first-order equivalent medium (blue curves) along direction $\mathbf{m}_2$ for vanishing coupling coefficients. (a) $\omega^*_r$ and $\omega^*_i$ vs $k^*_2$; (b) zoomed view of the 3D spectrum considering $0\leq k_2^* \leq \pi/3$; (c)damping modes in the plane $\omega^*_r-k^*_2$; (d) propagation modes in the plane $\omega^*_i-k^*_2$. }
  \label{Fig::spettroInKappa2}
\end{figure}

\begin{figure}[h!]
  \centering
  \begin{tabular}{c c }
 \hspace{-0.75cm}
  \includegraphics[width=8cm]{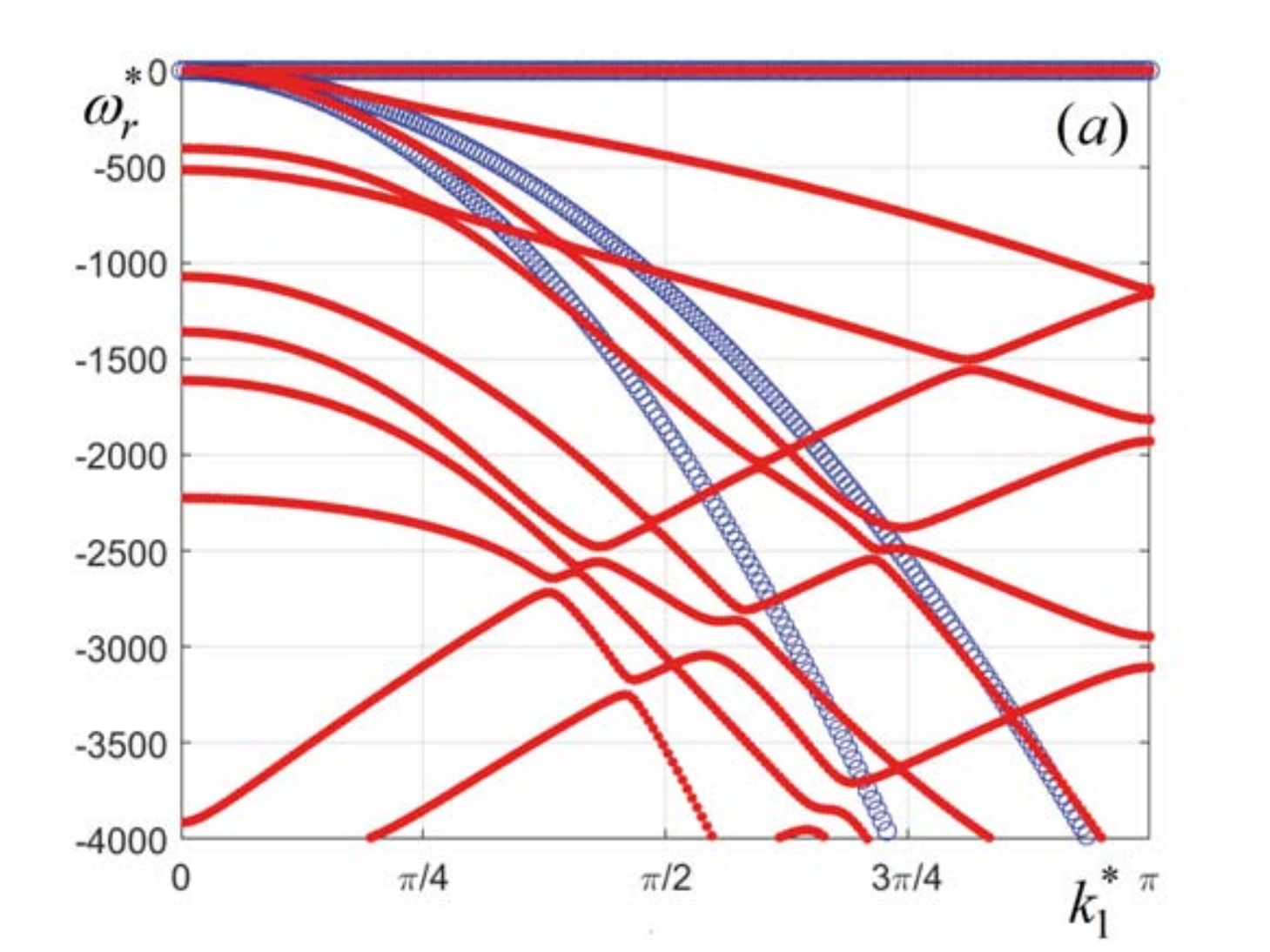}
  &
 \hspace{-1cm}
  \includegraphics[width=8cm]{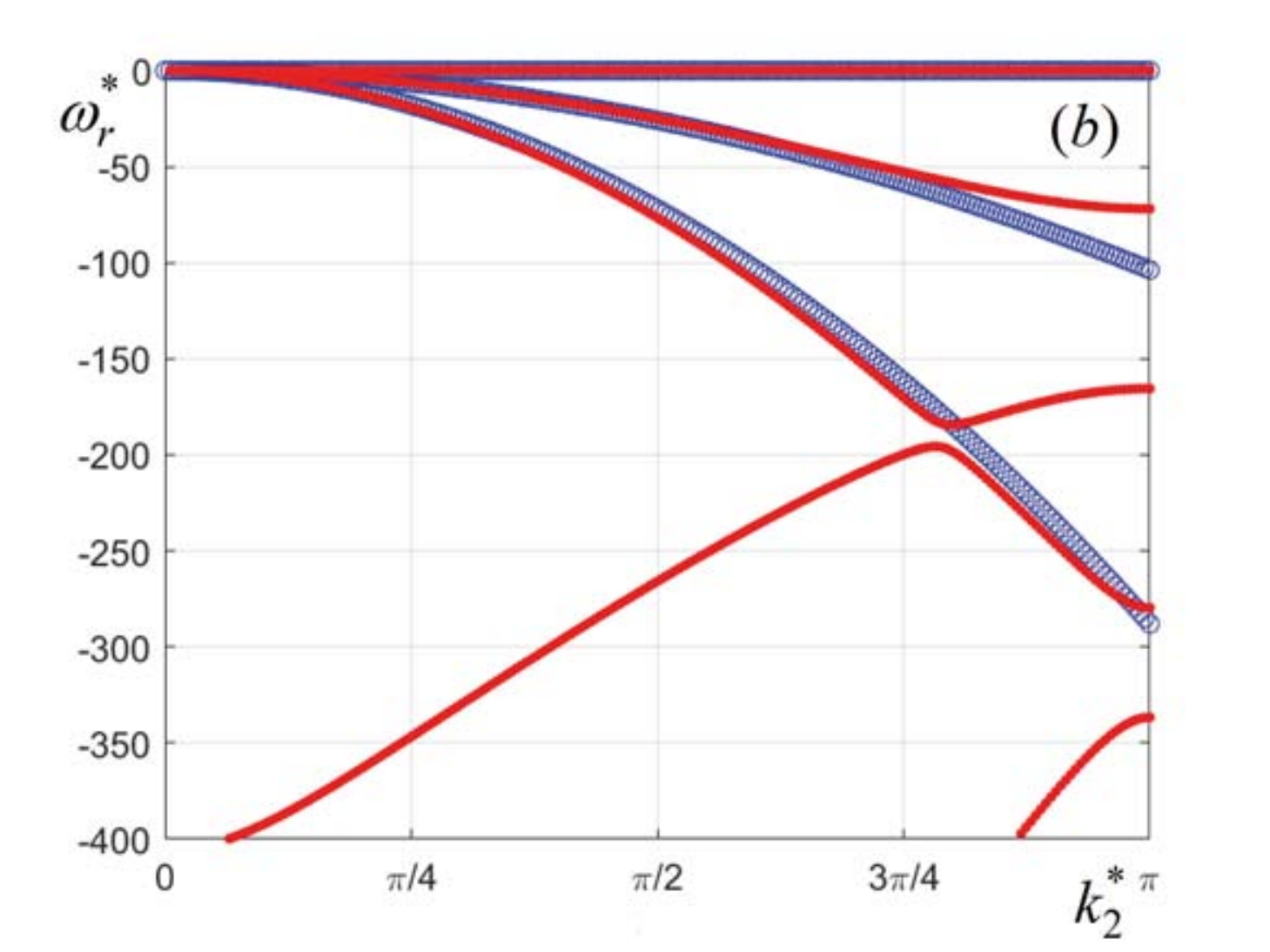}
  \end{tabular}
  \caption{\it Real part of  complex frequency for the heterogeneous SOFC-like material (red curves) and for the first-order equivalent medium (blue curves) in the case of non vanishing  thermo-diffusive coupling coefficient $\psi$. (a) Propagation along direction $m_1$; (b) propagation along direction $\mathbf{m}_2$.}
  \label{Fig::semiaccoppiato}
\end{figure}

One considers a multi-phase laminate, generated by the spatial repetition SOFC-like cell, whose periodic cell $\mathcal{A}$ is represented in figure \ref{Fig::periodic_SOFC}-(b) and has dimensions  $d_1=100\mu m$ and $d_2=440 \mu m$. All phases are assumed to be linear isotropic and a plane problem characterized by conditions $\tensor{\sigma}\mathbf{e}_3=\mathbf{0}$, $\mathbf{q}\cdot\mathbf{e}_3=0$, and $\mathbf{j}\cdot\mathbf{e}_3=0$ is considered, where $\mathbf{e_3}$ is a unit vector perpendicular to $\mathbf{e}_1$ and $\mathbf{e}_2$ to form a right handed base. Under these conditions the non vanishing components of micro constitutive tensors are
\begin{eqnarray}
&&C^m_{1111}=C^m_{2222}=\frac{E}{1-\nu^2},
\hspace{0.2 cm}
C^m_{1122}= \frac{\nu E}{1-\nu^2},
\hspace{0.2 cm}
C^m_{1212}=\frac{E}{2(1+\nu)},
\nonumber\\
&&K^m_{11}=K^m_{22}=K,
\hspace{0.2cm}
D_{11}^m=D_{22}^m=D,
\nonumber\\
&&\alpha_{11}^m=\alpha_{22}^m=\alpha\frac{1-2\nu}{1-\nu},
\hspace{0.2 cm}
\beta_{11}^m=\beta_{22}^m=\beta\frac{1-2\nu}{1-\nu}
\end{eqnarray}
where $E$ is the Young modulus, $\nu$  is the Poisson ratio, $K$ is the thermal conductivity constant, $D$ is the mass diffusivity constant, $\alpha$ is the thermal dilatation constant, and $\beta$ is the diffusive expansion constant.
%
%
%
%
%
\begin{figure}[h!]
  \centering
  \begin{tabular}{c c }
 \hspace{-2cm}
  \includegraphics[width=7cm]{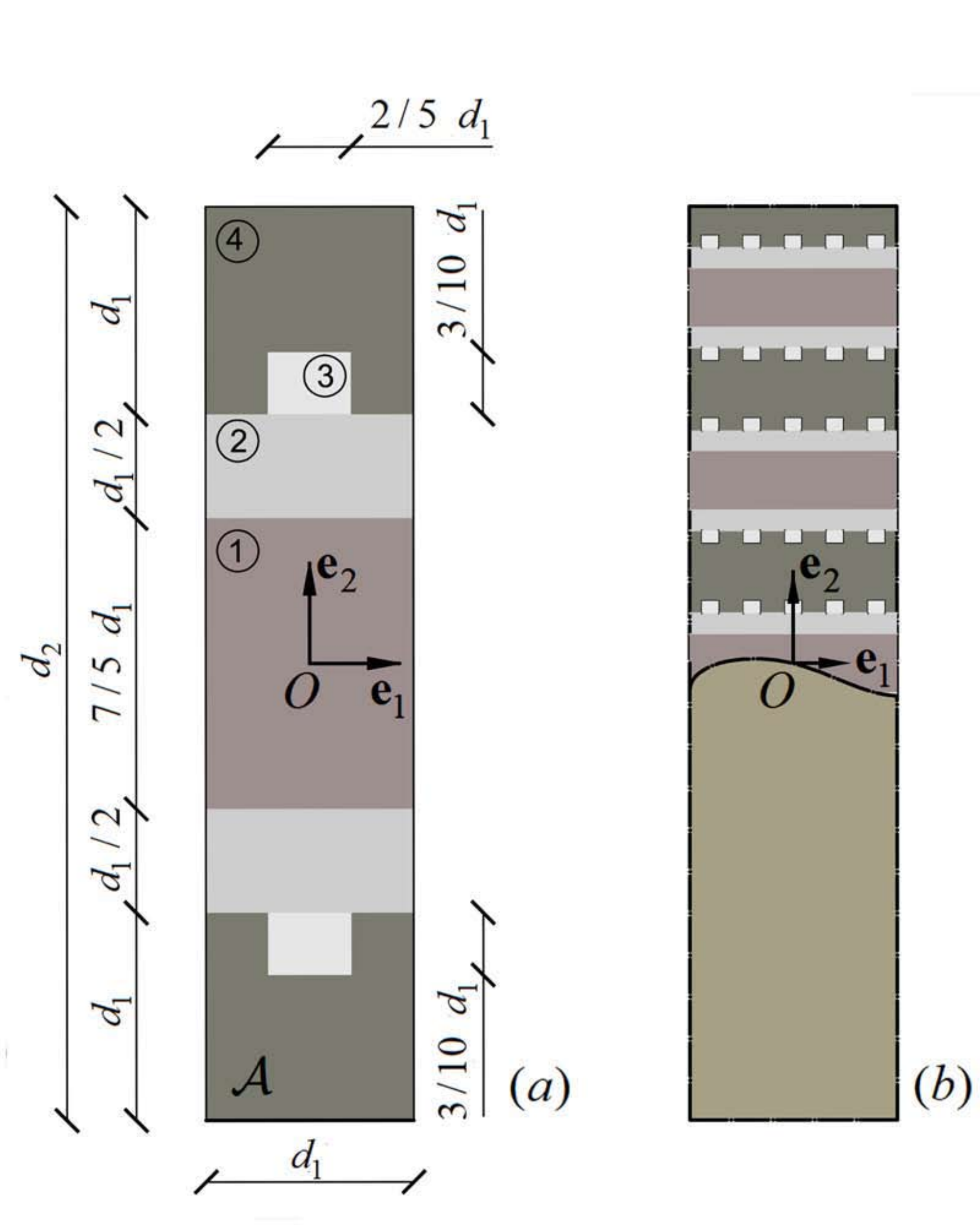}
  &
 \hspace{-0.5cm}
  \includegraphics[width=6cm]{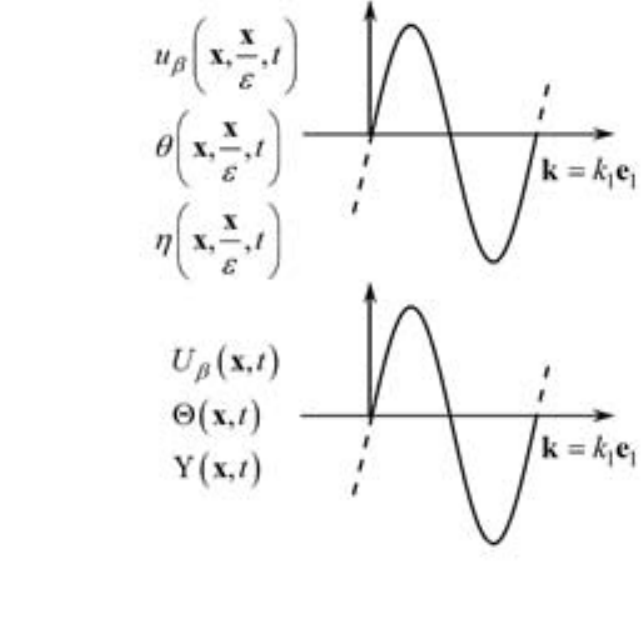}
  \end{tabular}
  \caption{\it (a) draft of the SOFC-like periodic cell $\mathcal{A}$ having dimensions $d_1$ and $d_2$; (b) Waves propagation along direction $\mathbf{e}_1$ in the heterogeneous and in the equivalent first-order medium.  }
  \label{Fig::spettroInKappa1}
\end{figure}
Ceramic electrolyte (phase 1) is considered made by yttria-stabilized zirconia (YSZ) having $E=155\, GPa$, $\nu=0.3$, $K=0.0856\, W/(m K^2)$, $D = 0.614134\, kg\,s/m^3$, $\alpha=4.22375 \cdot  10^3 N/(m^2 K)$, $\beta=4.22375 \cdot  10^2\,kg/m^3$  and inertial terms $\rho^m=5900\, kg/m^3$,  $p^m=8412.8 \,N/(m^2 K^2)$ and $q^m=84128\,kg^2/(m^3\,J) $. Thermo-diffusive coupling constant $\psi^m$ is assumed to be equal to $280.427\,kg/(m^3\,K)$.
Electrodes (phase 2) are considered made by nickel oxide (NiO) with $E=50\,GPa$, $\nu=0.25$, $K=0.1570 \, W/(m K^2)$, $D=1.26072\,kg\,s/m^3$, $\alpha=1090\cdot 10^3 N/(m^2 K)$, $\beta=1090\cdot 10^2\,kg/m^3 $, $\rho^m= 6810\,kg/m^3$,  $p^m=14008 \,N/(m^2 K^2)$, $q^m=140080\,kg^2/(m^3\,J)$  and $\psi^m=1400.8\,kg/(m^3\,K)$.
Finally, steel is supposed to constitute the conductive interconnections (phase 4) with $E=2.01 \cdot 10^{2}\,GPa$, $\nu=0.3$, $K=0.05 \, W/(m K^2)$, $D=6.8495\cdot10^{-6}\,kg\,s/m^3$, $\alpha=5477\cdot 10^3 N/(m^2 K)$, $\beta=5477\cdot 10^2\,kg/m^3$, $\rho^m=8000\,kg/m^3$, $p^m=13699 \,N/(m^2 K^2)$, $q^m=1369.9 \,kg^2/(m^3\,J)$ and vanishing constant $\psi^m$.
%
%
%
All constitutive properties of phase 3 representing  the flow channels, are assumed to be equal to $1/10$ of the corresponding  constitutive properties of electrodes.
Perturbation functions $N_{kpq_1}^{(1)}$, $\tilde{N}_k^{(1)}$, $\hat{N}_k^{(1)}$, $M_{q_1}^{(1)}$, and $W_{q_1}^{(1)}$ have been obtained by numerically solving  cell problems (\ref{eq:FirstCellProblemDisplacement-1}), (\ref{eq:SecondCellProblemDisplacement-1}), (\ref{eq:ThirdCellProblemDisplacement-1}), (\ref{eq:CellProblTherm-1}), and (\ref{eq:CellProblDiffus-1}) at the order $\varepsilon^{-1}$.
Numerical resolution has been obtained by means of  a finite element procedure over the unit cell $\mathcal{Q}$, as detailed in Appendix E.
Once perturbation functions are known, the overall constitutive tensors (\ref{eq:OverallConstitutiveTensorsFirstOrder}) are computed for the first-order thermo-diffusive homogenized medium and, exploiting the formalism described in Appendix D, they result
\begin{eqnarray}
&&\mathfrak{C}=
\left[\begin{array}{c c c}
1.3927 & 0.3003 & 0 \\
  0.3003 & 1.0595 & 0 \\
  0 & 0 & 2\cdot 0.6775
\end{array}
\right]10^{11}\, \frac{N}{m^2},
\hspace{0.2 cm}
\tensor{K}=
\left[
\begin{array}{c c}
8.32 &0\\
0 & 6.45 
\end{array}
\right]10^{-2}\, \frac{W}{m K^2},
\nonumber\\
&&\tensor{D}=
\left[
\begin{array}{c c}
5.392  &0\\
0 & 2.324 
\end{array}
\right]10^{-1}\, \frac{kg\,s}{m^3},
\hspace{0.2 cm}
\tensor{\alpha}=
\left[
\begin{array}{c}
1.8454 \\
1.4822 \\
\sqrt{2}\cdot 0.0007 
\end{array}
\right] 10^6\,\frac{N}{m^2 \,K},
\hspace{0.2 cm}
\tensor{\beta}=
\left[
\begin{array}{c}
1.8454 \\
1.4822 \\
\sqrt{2}\cdot  0.0007
\end{array}
\right] 10^5\,\frac{kg}{m^3},
\nonumber\\
&&
\rho= 6662.1 \, \frac{kg}{m^3},
\hspace{0.2cm}
p=11428 \, \frac{N}{m^2\,K^2},
\hspace{0.2 cm}
q= 114280\, \frac{kg}{m^3\,J},
\hspace{0.2 cm}
\psi= 4.1523\cdot 10^5 \,\frac{kg}{m^3\,K}.
\label{eq:OverallConstitutiveTensorsBenchmark}
\end{eqnarray}
 Generalized quadratic eigenvalue problem (\ref{eq:GeneralizedEigenValueProblemCompactForm}) has been solved in order to investigate the complex frequency spectrum of the periodic thermo-diffusive material varying the wave propagation direction $\mathbf{k}$. Defining the unit vector of propagation $\mathbf{m}=\mathbf{k}/|| \mathbf{k} ||_2$, two unit vectors of propagation are taken into account in the present example, namely $\mathbf{m}_1=\mathbf{e_1}$ parallel to the SOFC layering, and $\mathbf{m}_2=\mathbf{e}_2$ perpendicular to the first one.
 Dimensionless wave vector $\mathbf{k}^*=k_1^*\,\mathbf{e}_1+k_2^*\mathbf{e}_2$ is conveniently introduced, where dimensionless wave numbers $k_1^*=k_1\,d_1$ and $k_2^*=k_2\,d_2$ belong to the dimensionless first  Brillouin zone $\mathcal{B}^*=[-\pi,\pi]\times[-\pi,\pi]$.
MATLAB$^{\circledR}$ has been used as a tool to solve the quadratic eigenvalue problem. 
It has been enhanced with the  Advanpix Multiprecision Computing Toolbox 
 which enables computing using an arbitrary precision.
Matrices $\mathbf{H}_2$, $\mathbf{H}_1$ and $\mathbf{H}_0$ of problem (\ref{eq:GeneralizedEigenValueProblemCompactForm}), in fact, result to be neither symmetric nor Hermitian and their entries are  characterized by having absolute values differing by several orders of magnitudes. In this case  the use of higher precision with respect to the standard double one, together with  sparse representation of matrices,  revealed to be crucial to get to the right final result.
 Figures \ref{Fig::spettroInKappa1} and \ref{Fig::spettroInKappa2} represent the complex spectrum obtained along directions $\mathbf{m}_1$ and $\mathbf{m}_2$, respectively.
 In particular, defining a reference frequency $\omega_{ref}=1\,rad/s$ the dimensionless real part $\omega^*_r=\omega_r/\omega_{ref}$ and the dimensionless positive imaginary part $\omega^*_i=\omega_i/\omega_{ref}$ of the complex angular frequency, related to the attenuation and propagation mode, respectively, are represented in the two perpendicular directions as functions of the correspondent dimensionless wave number.
Assuming $\tensor{\alpha}=\mathbf{0}$, $\tensor{\beta}=\mathbf{0}$ and $\psi=0$ in equations (\ref{eq:ChristoffelEqsTrasformateNotazioneMatriciale}), blue curves of figures \ref{Fig::spettroInKappa1} and \ref{Fig::spettroInKappa2} are the dispersion curves of the homogenized first-order thermo-diffusive medium, computed as solutions of the quadratic generalized eigenvalue problem (\ref{eq:GeneralizedEigenValueProblemCompactForm}). 
This last  gives rise to two pure damping modes, represented by the two parabolas  in the plane $\omega^*_i=0$, and four pure propagation curves, complex conjugate in twos, plotted in the plane $\omega^*_r=0$. 
 They are all acoustic branches departing from the origin of the reference system.
 Red curves in figures \ref{Fig::spettroInKappa1} and \ref{Fig::spettroInKappa2} describe the low frequencies branches of the complex frequency Floquet-Bloch spectrum relative to the heterogeneous thermo-diffusive SOFC-like material where all the four phases are characterized by vanishing coupling tensors $\tensor{\alpha}^m$ and $\tensor{\beta}^m$ and vanishing coupling constant $\psi^m$.
Thanks to the periodicity of the medium, dispersion curves for the heterogeneous material have been obtained by solving  the generalized quadratic eigenvalue problem  (\ref{eq:GeneralizedchristoffelEquationsEterogeneo}) over the periodic cell $\mathcal{A}$, where this last is subjected to Floquet-Bloch, or quasi-periodicity, boundary conditions \citep{Floquet1883,Bloch1929,Brillouin1953,Mead1973,Langley1993}  .
The procedure adopted to obtain the complex frequency band structure for the heterogeneous material is outlined in detail in Appendix E.
Figures \ref{Fig::spettroInKappa1}-(b) and \ref{Fig::spettroInKappa2}-(b) are a zoom of the correspondent three dimensional spectra \ref{Fig::spettroInKappa1}-(a) and \ref{Fig::spettroInKappa2}-(a) considering $0\leq k^*_i \leq \pi/3\,\, (i=1,2)$. 
Plane $\omega^*_i=0$ is represented in figure   \ref{Fig::spettroInKappa1}-(c) along direction $\mathbf{m}_1$ and in figure \ref{Fig::spettroInKappa2}-(c) along  $\mathbf{m}_2$.
Analogously, planes $\omega^*_r=0$ are plotted in figures \ref{Fig::spettroInKappa1}-(d) and \ref{Fig::spettroInKappa2}-(d). 
  \begin{figure}[h!]
  \centering
  \begin{tabular}{c c }
 \hspace{-0.75cm}
  \includegraphics[width=8cm]{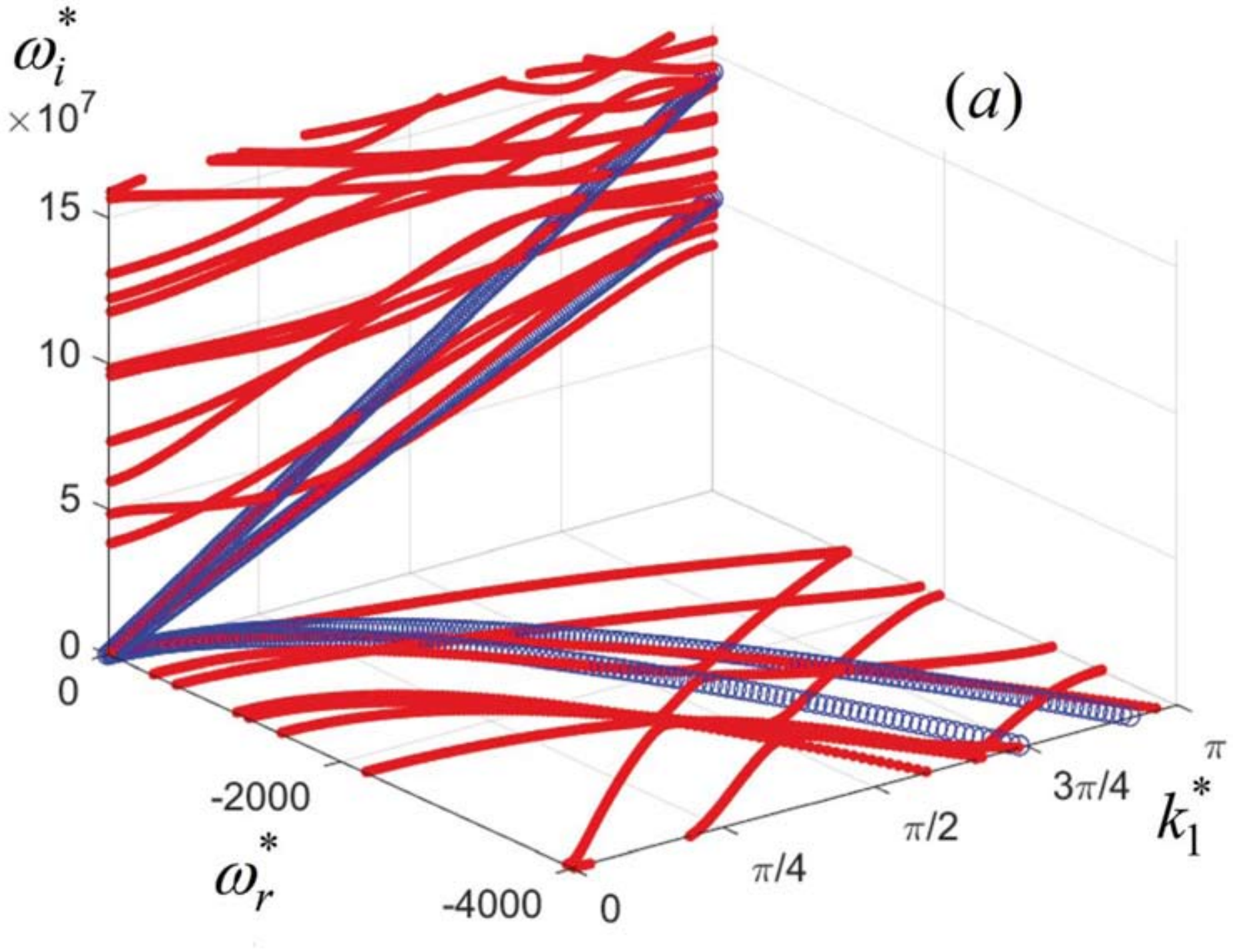}
  &
 \hspace{-1cm}
  \includegraphics[width=8cm]{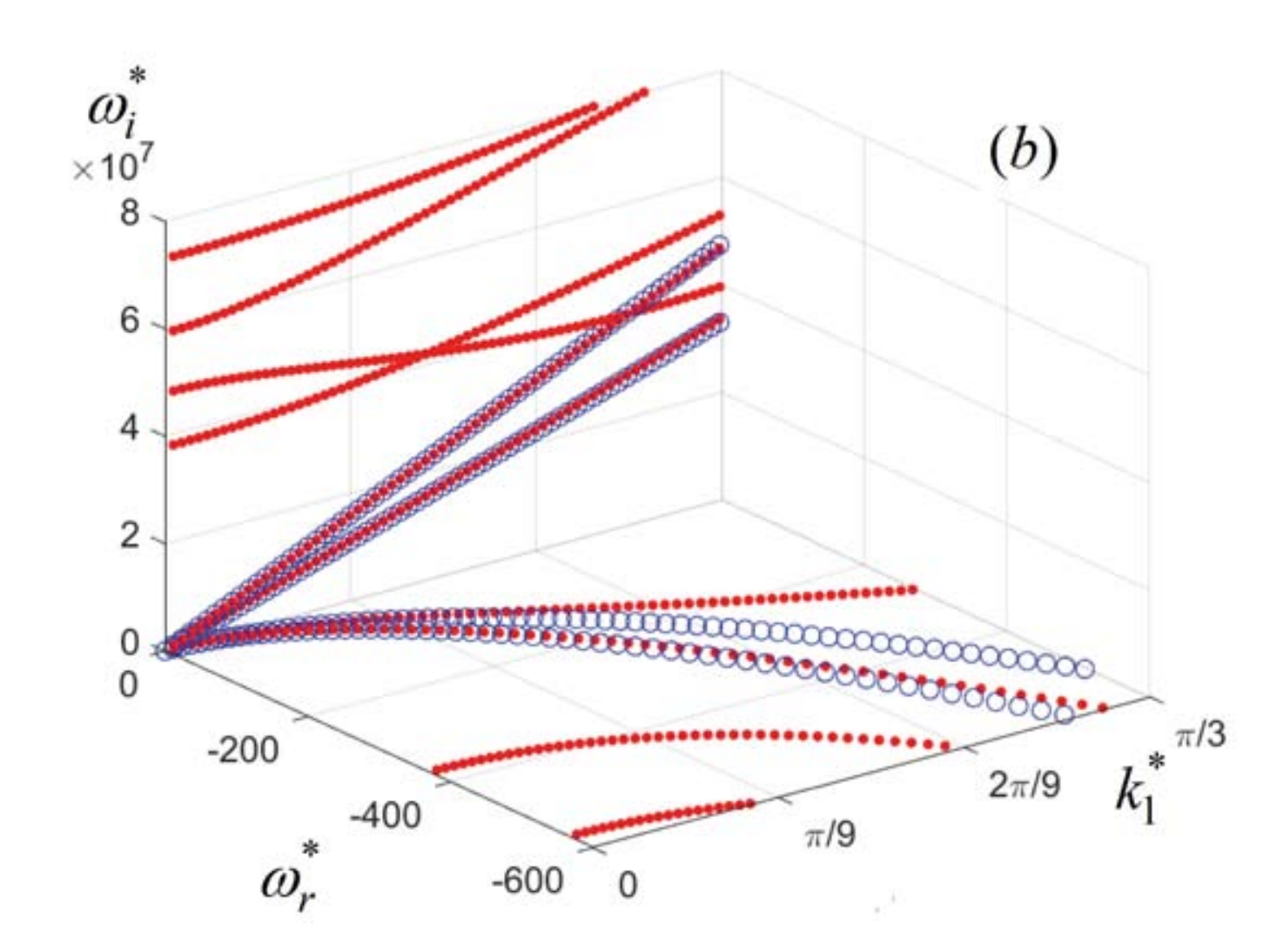}
  \\
   %
 \hspace{-0.75cm}
  \includegraphics[width=8cm]{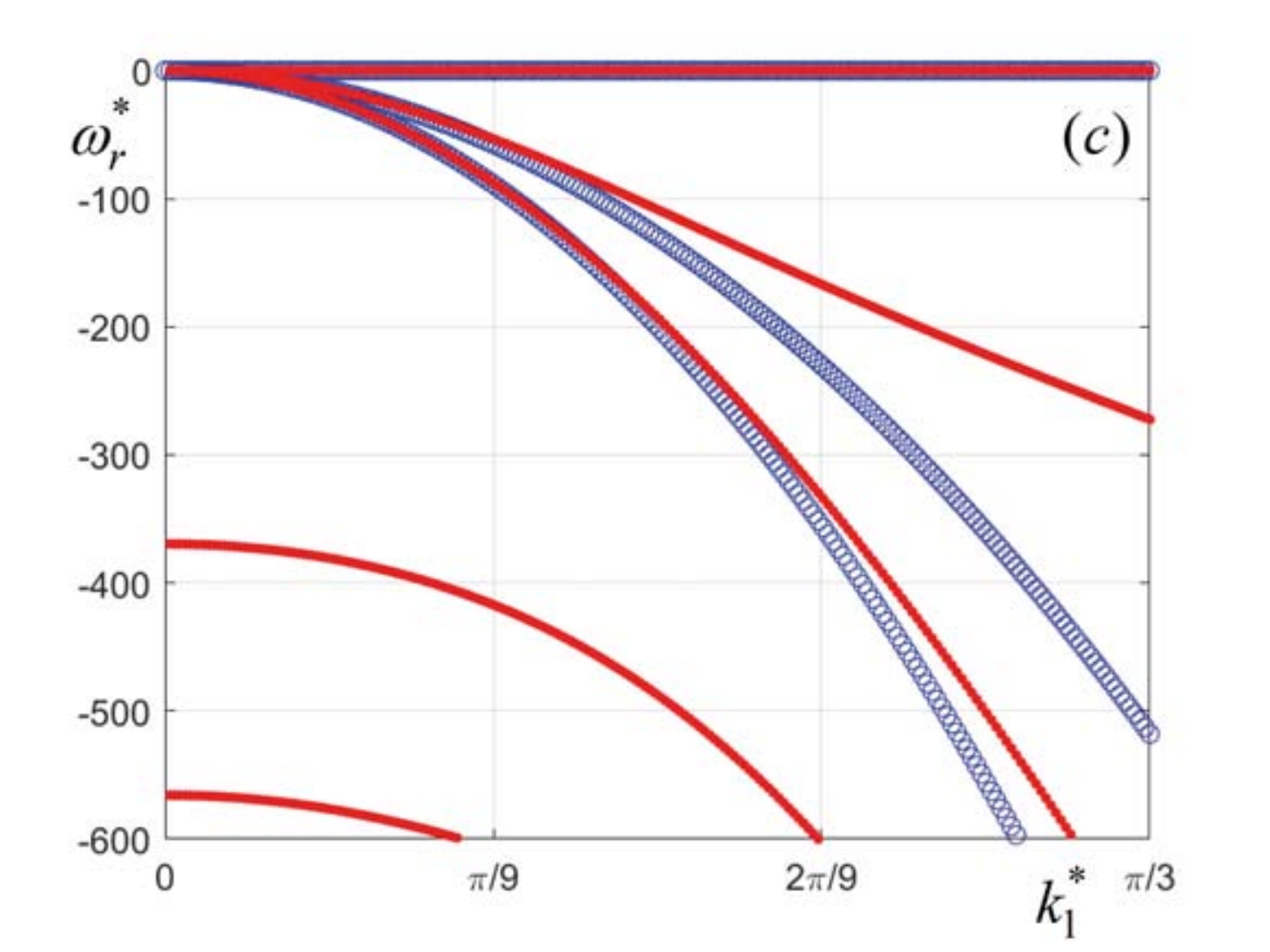}
  &
 \hspace{-1cm}
  \includegraphics[width=8cm]{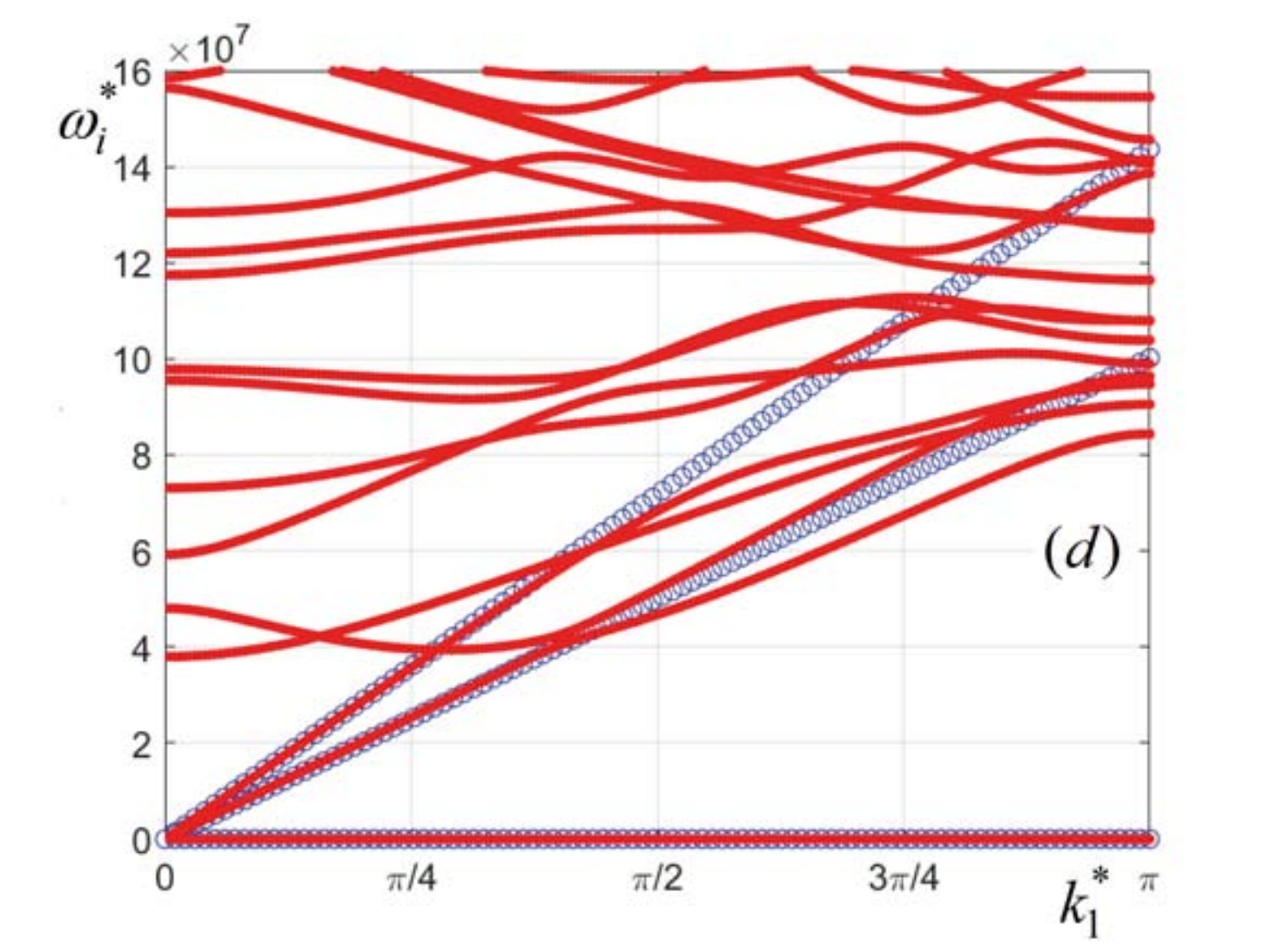}
  \end{tabular}
  \caption{\it Complex frequency spectrum of the heterogeneous SOFC-like material (red curves) and of the first-order equivalent medium (blue curves) along direction $\mathbf{m}_1$ for vanishing coupling coefficients. (a) $\omega^*_r$ and $\omega^*_i$ vs $k^*_1$; (b) zoomed view of the 3D spectrum considering $0\leq k_1^* \leq \pi/3$; (c)damping modes in the plane $\omega^*_r-k^*_1$; (d) propagation modes in the plane $\omega^*_i-k^*_1$.  }
  \label{Fig::spettroInKappa1}
\end{figure}
 As one can notice, a very good agreement between the first branches   of the spectrum of the heterogeneous material  and the ones of homogenized medium, is achieved for $0\leq k^*_i \leq \pi/3 $   $\,(i=1,2)$. 
 A decrease of the accuracy  is generally expected for $k^*_i\geq \pi/3$ 
 as a first-order approximation is adopted to describe the equivalent thermo-diffusive medium behavior and obtained results confirm this fact.
 Furthermore, the obtained approximation of the complex frequency band structure results to be more accurate along the $\mathbf{m}_2$ direction than along $\mathbf{m}_1$ and superior performances attained in the direction perpendicular to the  material layering is confirmed by previous results achieved in the literature \citep{BacigalupoGambarotta2014b}.
 No partial gaps are detected along $\mathbf{m}_1$ and $\mathbf{m}_2$ in the frequency ranges taken into account.
 %
 %
 %
 %
 %
 %
 %
Figure \ref{Fig::semiaccoppiato} represents dispersion curves obtained in the plane $\omega^*_i=0$ along directions $\mathbf{m}_1$ (figure \ref{Fig::semiaccoppiato}-(a)) and $\mathbf{m}_2$ (figure \ref{Fig::semiaccoppiato}-(b)) when thermo-diffusive coupling constant $\psi^m$ is introduced such that $\psi^m=2804.27\,kg/(m^3\,K)$ for phase 1, $\psi^m=14008\,kg/(m^3\,K)$ for phases 2 and $\psi^m=1400.8\,kg/(m^3\,K)$ for phase 3.
Figure \ref{Fig::semiaccoppiato} confirms the capabilities of the proposed first-order asymptotic procedure  to approximate dispertion properties of thermo-diffusive materials in the low frequency regime.
A comparison with the two relative spectra in the case of vanishing $\psi^m$ (figures \ref{Fig::spettroInKappa1}-(c) and \ref{Fig::spettroInKappa2}-(c)) brings to light the qualitative  differences spotted in the two  cases between the spectra relative to the heterogeneous material.
In particular, veering phenomena, meaning the repulsion between two branches, are accentuated in the case of non vanishing $\psi^m$, and, in the correspondence of the same $k_i^*$, the absolute values  
of $\omega^*_r$ increases for each branch of the spectrum.

\begin{figure}[h!]
  \centering
  \begin{tabular}{c c }
 \hspace{-0.75cm}
  \includegraphics[width=8cm]{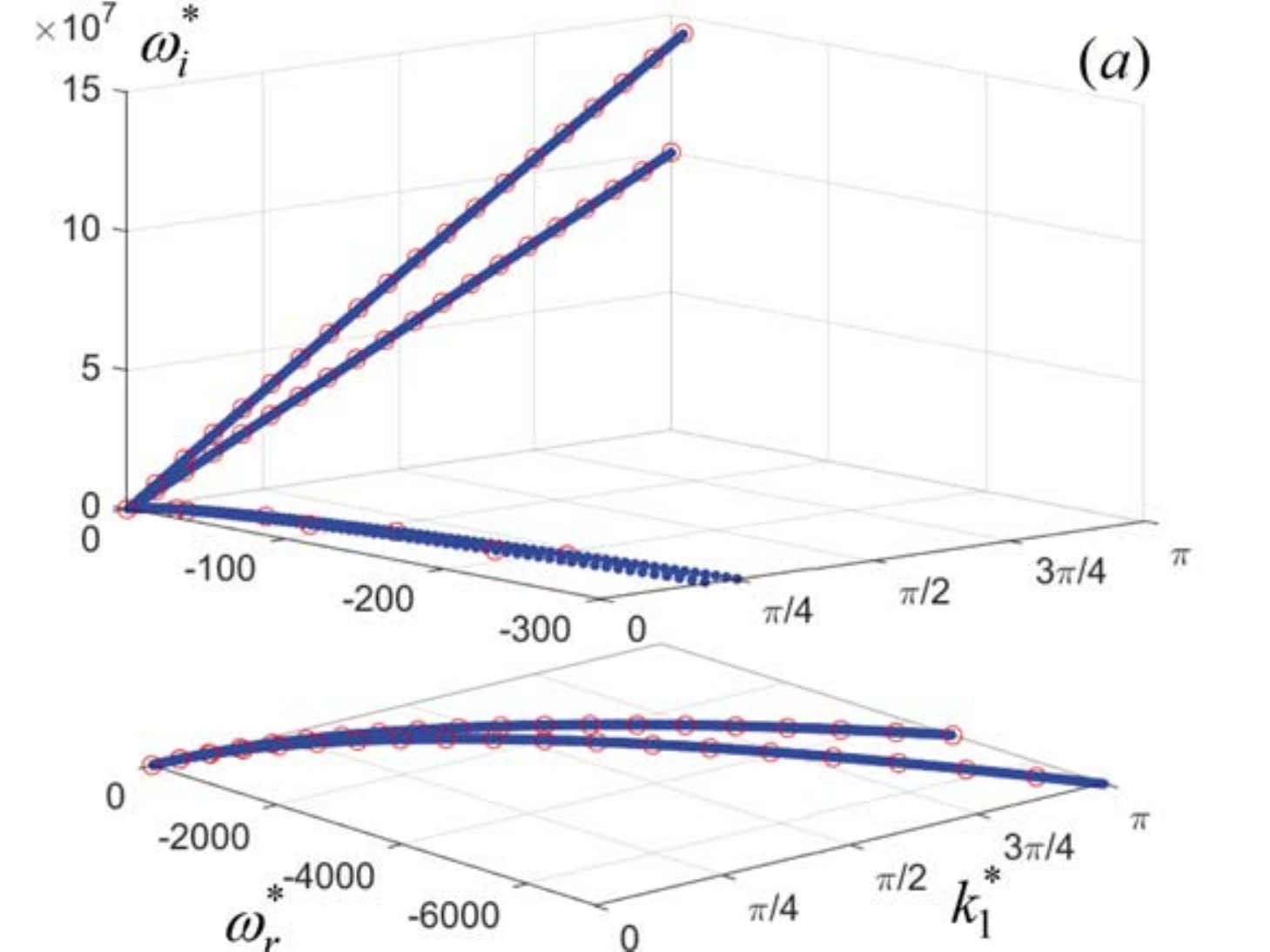}
  &
 \hspace{-1cm}
  \includegraphics[width=8cm]{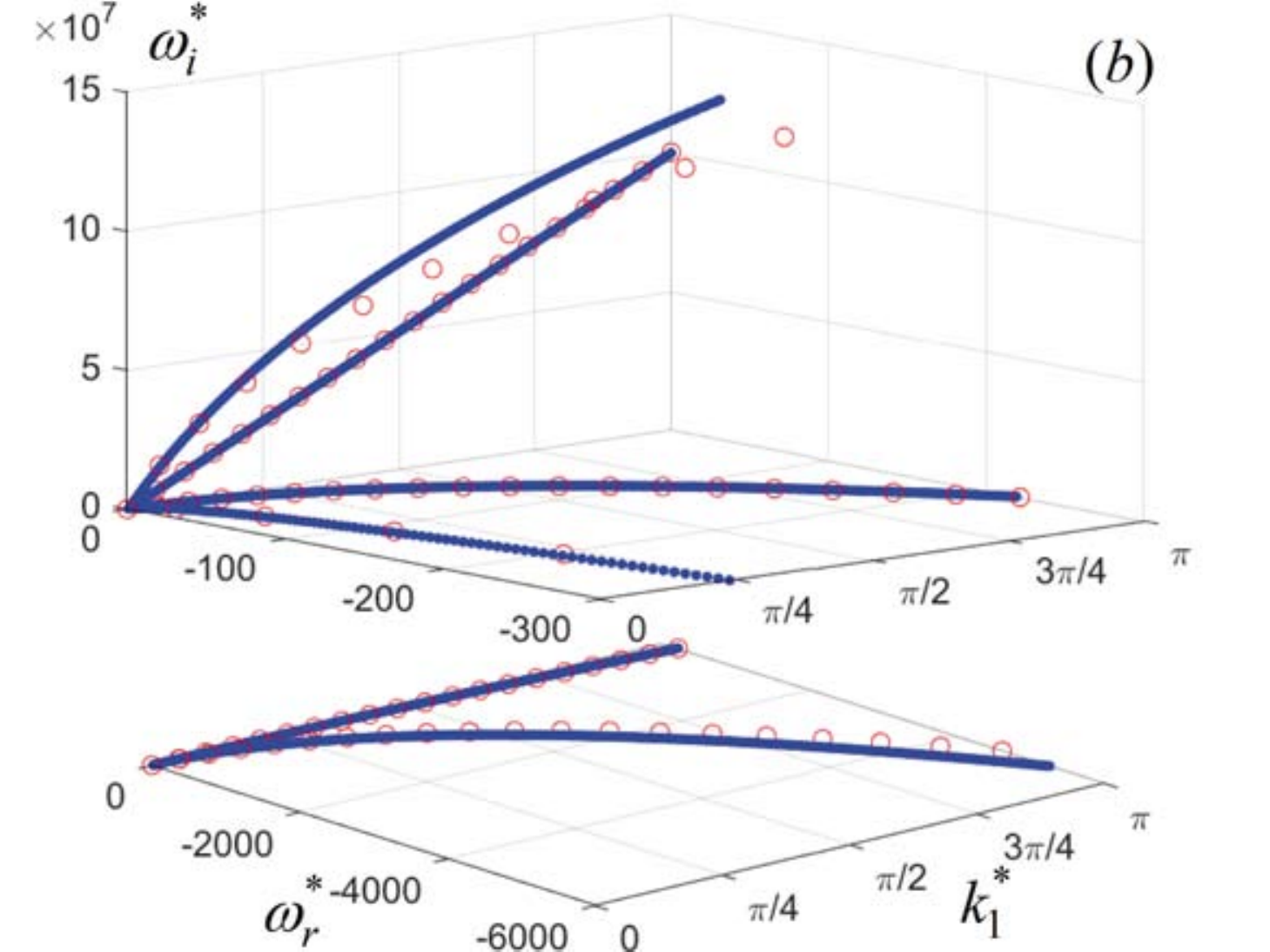}
   \\
   \\
 \hspace{-0.75cm}
  \includegraphics[width=8cm]{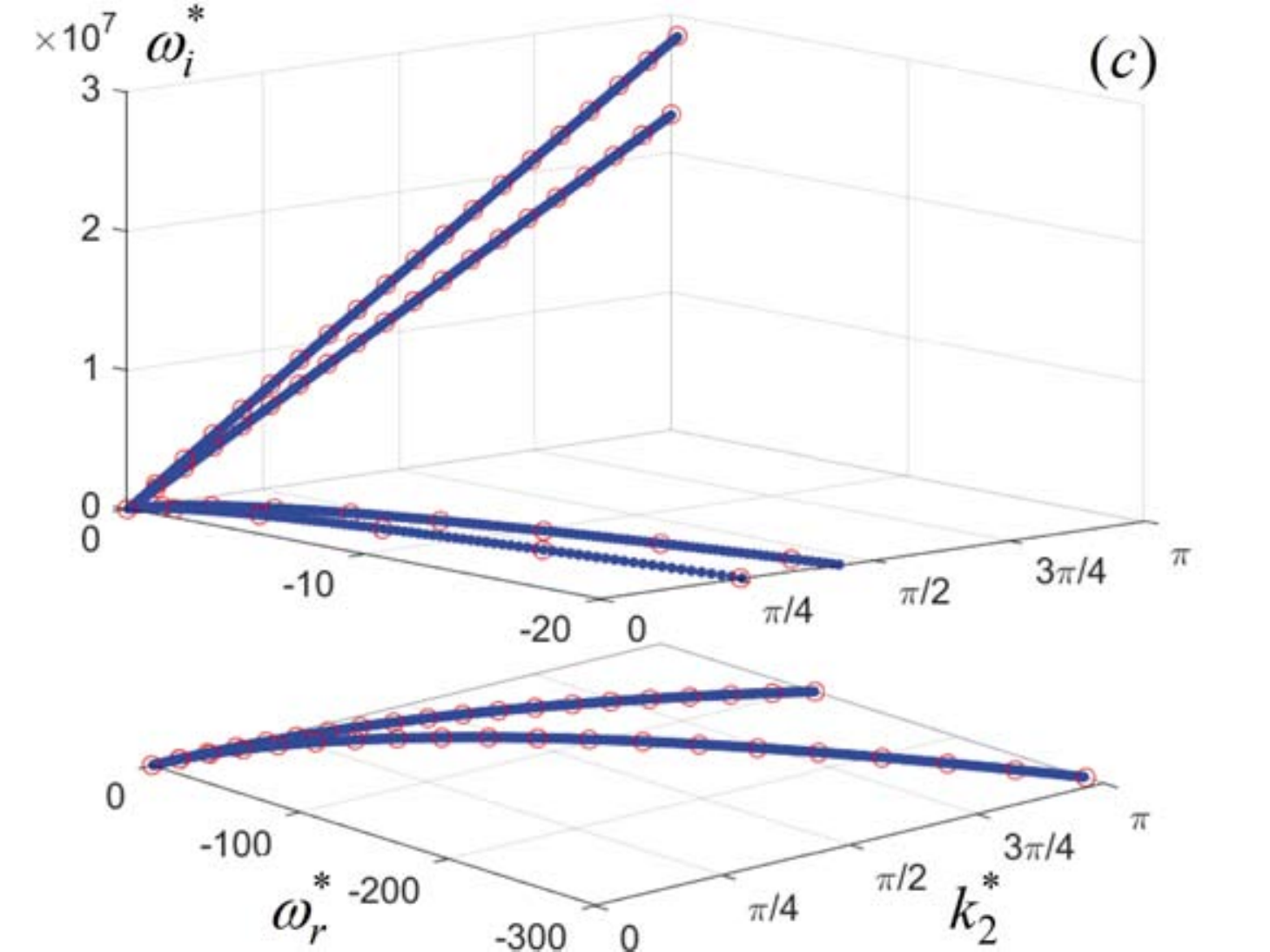}
  &
 \hspace{-1cm}
  \includegraphics[width=8cm]{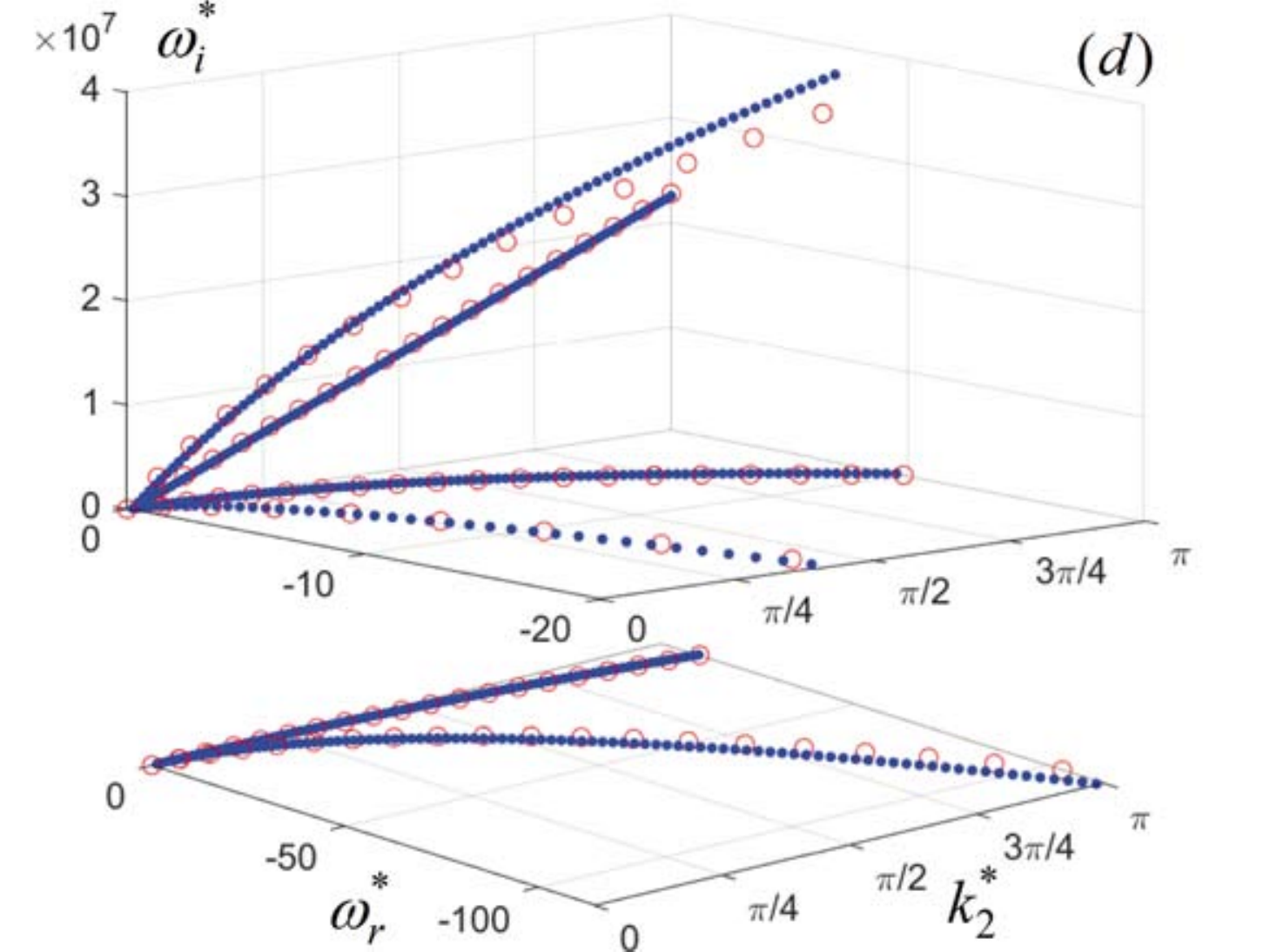}
  \end{tabular}
  \caption{\it Complex frequency spectrum of the first-order equivalent medium (blue curves) and its fourth-order perturbative approximation (red dots) as described in equation (\ref{eq:fourthorderapproximationOfTheEigenvalue}). Each subfigure is enhanced with a perspective view of the plane $\omega^*_i=0$ showing the complete range of values for $\omega^*_r$.  Propagation along direction $\mathbf{m}_1$ and  $\tensor{\alpha}$, $\tensor{\beta}$ and $\psi$ with components as expressed in (\ref{eq:OverallConstitutiveTensorsBenchmark}) (a) and increased by a factor $10^2$ (b).
  Propagation along direction $\mathbf{m}_2$ and  $\tensor{\alpha}$, $\tensor{\beta}$ and $\psi$ with components as expressed in (\ref{eq:OverallConstitutiveTensorsBenchmark}) (c) and increased by a factor $10^2$ (d).    }
  \label{Fig::hom_vs_pert}
\end{figure}

 When all overall coupling tensors $\tensor{\alpha}$ and $\tensor{\beta}$ and overall coupling constant $\psi$ are taken into account with their value as expressed in equation (\ref{eq:OverallConstitutiveTensorsBenchmark}), dispersion curves for the homogenized first-order medium  have the behavior illustrated in figure \ref{Fig::hom_vs_pert}-(a) along $\mathbf{m}_1$ and in  figure \ref{Fig::hom_vs_pert}-(c) along $\mathbf{m}_2$ (blue curves).
 When coupling coefficients are taken into consideration, resolution of quadratic generalized eigenvalue problem (\ref{eq:GeneralizedEigenValueProblemCompactForm}) provides two pure damping branches and four (complex conjugate in twos) mixed mode branches having both components $\omega^*_i$ and $\omega^*_r$ different from zero.
 Red dots in figure \ref{Fig::hom_vs_pert} represent dispersion properties of the equivalent medium obtained by means of the asymptotic approximation  procedure described in Section \ref{SubSec::AsympApproxOfComplexSpectrum}, which allows to achieve a compact and explicit parametric approximation of the eigenvalues in terms of the constitutive coefficients of the homogenized continuum. In particular, a fourth order approximation of type (\ref{eq:fourthorderapproximationOfTheEigenvalue}) is achieved along both $\mathbf{m}_1$ and $\mathbf{m}_2$ by solving recursive perturbation problems $G^{[n]}(r=0)=0$ at the order $k_i^n\,\,(i=1,2)$ in accordance with the solution scheme described in table \ref{Table::SolutionSchemePerturbativeApproach}.
 Figures \ref{Fig::hom_vs_pert}-(b) and \ref{Fig::hom_vs_pert}-(d) represent, respectively, the complex spectrum obtained along the two perpendicular directions  $\mathbf{m}_1$ and $\mathbf{m}_2$ when components of the coupling tensors $\tensor{\alpha}$ and $\tensor{\beta}$ and the value of $\psi$ are multiplied by $10^2$.
 When the absolute values of coupling tensors increases, mixed mode branches bend toward the plane $\omega^*_i=0$ increasing their damping component,  and pure attenuation modes  bend toward  the axis $\omega^*_i=0$, yet remaining in the plane $\omega^*_r=0$.
 As one can notice, the perfect agreement obtained between the eigenvalues of problem (\ref{eq:GeneralizedEigenValueProblemCompactForm}) and their asymptotic approximation (see figures \ref{Fig::hom_vs_pert}-(a) and \ref{Fig::hom_vs_pert}-(c)) deteriorates as the coupling increases as shown in figures \ref{Fig::hom_vs_pert}-(b) and \ref{Fig::hom_vs_pert}-(d), preserving nevertheless the accuracy of the approximation for $0\leq k^*_i\leq \pi/3\,\,(i=1,2)$, as expected.
 %
 %
 %
 
 \section{Conclusions}
 \label{Sec::Conclusions}
 The present work is devoted to the formulation of an asymptotic homogenization technique for periodic microstructured materials characterized by thermo-diffusive phenomena.
 The aim of the proposed technique is twofold: it allows determining the overall constitutive properties of the first-order equivalent medium  and to investigate its complex frequency spectrum by providing its dispersion curves.
 Down-scaling relations are determined, which relate the three microfields, namely displacement, relative temperature and chemical potential to the corresponding macrostructural ones and to their gradients by means of perturbation functions.
 These lasts are regular, $\mathcal{Q}$-periodic functions, which take into account the effects of microstructural heterogeneities. They are solutions of  recursive, non homogeneous differential problems, known as cell problems, obtained inserting asymptotic expansions of the microfields in powers of the microstructural characteristic size $\varepsilon$ into micro governing field equations and reordering at the different orders of $\varepsilon$.
 Substitution of down-scaling relations into local balance equations provides the average field equations of infinite order, whose formal solution can be obtained by inserting an  asymptotic expansion of the macrofields in powers of $\varepsilon$ and reordering at the different orders of $\varepsilon$.
 The attained zeroth order differential problems yield to the governing field equations for the equivalent first-order (Cauchy) thermo-diffusive medium whose overall constitutive tensors are provided in closed form.

 By means of proper integral transforms of such global balance equations  a quadratic generalized eigenvalue problem derives, whose solution provides the complex frequency spectrum of the first-order homogeneous material in the first Brillouin zone.
 %
 %
 In order to assess the capabilities of the presented dynamic asymptotic homogenization technique, a generalization of the Floquet-Bloch theory has been implemented  in order to investigate dispersion properties  of the heterogeneous thermo-diffusive medium.
Thanks to the periodicity of the microstructured material, a quadratic generalized eigenvalue problem is solved over the periodic cell subjected to Floquet-Bloch boundary conditions.
The eigenvalues provide the imaginary and real components of the angular frequency, related respectively to the propagation and attenuation modes of the wave that propagates inside the medium, as functions of the wave vector.
The very good matching obtained between dispersion curves of the first-order homogenized continuum and the lowest frequency ones relative to the heterogeneous   medium, confirms the accuracy of the proposed homogenization technique in predicting the behavior of the acoustic branches of the complex spectrum of the material under consideration, at least in the range of wave number values admissible for a first-order approximation.
%
%

Furthermore, an asymptotic approximation of the complex spectrum is here presented based upon the resolution  of recursive perturbation problems at the different orders of $r$, here intended as the Euclidean norm of the wave vector.
Perturbation problems derive from a Taylor series expansion of the implicit characteristic equation of the equivalent medium in the transformed space and frequency domain and their solutions provide the sensitivities  of the eigenvalues at the different orders of $r$.
Parametric approximation of the complex angular frequency allows obtaining a compact analytical  solution of the characteristic equation, in which the dependence upon the overall constitutive coefficients is made explicit.
A fourth order asymptotic approximation of the spectrum demonstrates to be in good  agreement with dispersion curves of the homogenized material, also in the case of increased coupling coefficients of field equations.

In the context of renewable energy devises, numerical experiments have been conducted referring to a Solid Oxide Fuel Cell (SOFC)-like material, whose typical building block can be modeled as a periodic thermo-diffusive elastic multi-layered material.
SOFC are typically subjected to high operating temperatures and to intensive ions flows, which can increase their vulnerability to damage and undermine their efficiency.
A correct prediction of their behavior is therefore of fundamental importance in order to design high performances batteries.
When scale separation holds, homogenization techniques reveal to be particularly useful to obtain an accurate, but concise at the same time, description of the material, both in static and dynamic regime. 
In this regard, proposed multifield asymptotic homogenization is an efficient and rigorous tool for the investigation of thermo-diffusive materials having periodic microstructure.
%
When non local phenomena connected to the microstructural length scale and/or size effects come into play, first-order homogenization methods result to be inadequate in approximating the behavior of the periodic material.
In these cases more accurate approximations could be obtained by considering higher-order cell problems.
Alternatively, homogenized higher-order materials can be properly modeled by means of non local higher-order homogenization approaches, which allow to consider a characteristic length scale linked to microstructural effects, but the employment of such techniques is out of the scope of the present study.

\bigskip
\bibliographystyle{elsarticle-harv}
\bibliography{Bibliography}

\begin{appendices}
\section*{Appendix A. Proof of equivalence among thermo-diffusive homogeneous tensors  }
\label{Appendix::Dimostrazione uguaglianze overall tensors}
%
In the present Section the following equivalences between the components of overall constitutive tensors  that appear in the average field equations of infinite order (\ref{eq:AverageFieldEquationsInfiniteOrder}) are demonstrated in detail 
\begin{equation}
\tilde{n}_{pq_1}^{(2)}=\tilde{m}_{pq_1}^{(2,1)},
\hspace{0.2 cm}
\hat{n}_{pq_1}^{(2)}=\tilde{w}_{pq_1}^{(2,1)},
\hspace{0.2 cm}
\hat{m}^{(2,1)}=\hat{w}^{(2,1)}.
\end{equation}
This allows relating components of tensors $\tilde{\mathbf{n}}^{(2)}=\tilde{n}^{(2)}_{pq_1}\,\mathbf{e}_{p}\otimes\mathbf{e}_{q_1}$, $\tilde{\mathbf{m}}^{(2,1)}=\tilde{m}^{(2,1)}_{pq_1}\,\mathbf{e}_{p}\otimes\mathbf{e}_{q_1}$, $\hat{\mathbf{n}}^{(2)}=\hat{n}^{(2)}_{pq_1}\,\mathbf{e}_{p}\otimes\mathbf{e}_{q_1}$, $\tilde{\mathbf{w}}^{(2,1)}=\tilde{w}^{(2,1)}_{pq_1}\,\mathbf{e}_{p}\otimes\mathbf{e}_{q_1}$ to the corresponding ones of overall constitutive tensors $\tensor{\alpha}$ and $\tensor{\beta}$ and to relate constants $\hat{m}^{(2,1)}$ and $\hat{w}^{(2,1)}$ to the overall coupling constant $\psi$.
\\
\\
{\itshape Proof of {$\tilde{\mathbf{n}}^{(2)}=\tilde{\mathbf{m}}^{(2,1)}$}}
\\
\\
Components of tensors $\tilde{\mathbf{n}}^{(2)}$ and $\tilde{\mathbf{m}}^{(2,1)}$ come from the known terms of  cell problems (\ref{eq:SecondCellProblemDisplacement+0}) and (\ref{eq:SecondCellProblTherm+0}), respectively, and have the following expressions 

\begin{subequations}
\begin{align}
&\tilde{n}_{pq_1}^{(2)}=
\left\langle \alpha_{pq_1}^{m}-
C_{pq_1kj,j}^{m}\tilde{N}_{k,j}^{(1)}
\right\rangle,
\label{eq:coeff1}\\
&
\tilde{m}_{pq_1}^{(2,1)}=
\left\langle
\alpha_{pq_1}^{m}
+
\alpha_{iq_2}^{m}
N_{ipq_1,q_2}^{(1)}
\right\rangle.
\label{eq:coeff2}
\end{align}
\end{subequations}
The weak form of the first mechanical  cell problem (\ref{eq:FirstCellProblemDisplacement-1}) at the order $\varepsilon^{-1}$ 
\begin{equation}
\left(C_{pjkl}^{m}N_{kiq_1,l}^{(1)}
\right)_{,j}
+
C_{pjiq_1,j}^{m}
=0,
\end{equation}
can be written in the following way, considering as test function the perturbation function $\tilde{N}_p^{(2)}$
\begin{equation}
\left\langle
\left[
\left(C_{pjkl}^{m}N_{kiq_1,l}^{(1)}
\right)_{,j}
+
C_{pjiq_1,j}^{m}
\right]
\tilde{N}_p^{(1)}
\right\rangle=0.
\label{eq:WeakForm1}
\end{equation}
Divergence theorem and $\mathcal{Q}$-periodicity of  micro constitutive tensors and perturbation functions, allow writing equation (\ref{eq:WeakForm1}) as
\begin{equation}
\left\langle
\left[
C_{pjkl}^{m}N_{kiq_1,l}^{(1)}
+
C_{pjiq_1}^{m}
\right]
\tilde{N}_{p,j}^{(1)}
\right\rangle=0.
\label{eq:WeakForm2}
\end{equation}
Adding vanishing term (\ref{eq:WeakForm2}) to expression (\ref{eq:coeff1}), one obtains
\begin{equation}
\tilde{n}_{pq_1}^{(2)}=
\left\langle 
\alpha_{pq_1}^{m}-
C_{pq_1il,l}^{m}\tilde{N}_{i,l}^{(1)}
+
C_{ijkl}^{m}N_{ipq_1,j}^{(1)}\tilde{N}_{k,l}^{(1)}
+
C_{klpq_1}^{m}\tilde{N}_{k,l}^{(1)}
\right\rangle
=
\left\langle 
\alpha_{pq_1}^{m}
+
C_{klik}^{m}\tilde{N}_{k,l}^{(1)}N_{ipq_1,j}^{(1)}
\right\rangle.
\label{eq:coeff3b}
\end{equation}
Analogously, from the second mechanical cell problem (\ref{eq:SecondCellProblemDisplacement-1}) at the order  $\varepsilon^{-1}$ 
\begin{equation}
\left(
C_{ijkl}^{m}
\tilde{N}_{k,l}^{(1)}
\right)_{,j}
-
\alpha_{ij,j}^{m}
=0,
\end{equation}
the following weak form can be written in terms of test function $N_{ipq_1}^{(1)}$
\begin{equation}
\left\langle
\left[
\left(
C_{ijkl}^{m}
\tilde{N}_{k,l}^{(1)}
\right)_{,j}
-
\alpha_{ij,j}^{m}
\right]
N_{ipq_1}^{(1)}
\right\rangle
=0.
\label{eq:WeakForm3}
\end{equation}
Expression (\ref{eq:WeakForm3}) can be transformed into 
\begin{equation}
\left\langle
\left[
C_{ijkl}^{m}
\tilde{N}_{k,l}^{(1)}
-
\alpha_{ij}^{m}
\right]
N_{ipq_1,j}^{(1)}
=0
\right\rangle,
\label{eq:WeakForm4}
\end{equation}
thanks to divergence theorem and $\mathcal{Q}$-periodicity of micro tensors and perturbation functions.
Adding (\ref{eq:WeakForm4}) to (\ref{eq:coeff2}) yields
\begin{equation}
\tilde{m}_{pq_1}^{(2,1)}=
\left\langle
\alpha_{pq_1}^{m}
+
\alpha_{iq_2}^{m}
N_{ipq_1,q_2}^{(1)}
+
C_{ijkl}^{m}
\tilde{N}_{k,l}^{(1)}
N_{ipq_1,j}^{(1)}
-
\alpha_{ij}^{m}
N_{ipq_1,j}^{(1)}
\right\rangle
=
\left\langle
\alpha_{pq_1}^{m}
+
C_{ijkl}^{m}
\tilde{N}_{k,l}^{(1)}
N_{ipq_1,j}^{(1)}
\right\rangle,
\end{equation}
from which identity $\hat{n}^{(2)}_{pq_1}=\tilde{w}_{pq_1}^{(2,1)}$  follows.
\\
\\
\\
{\itshape Proof of {$\hat{\mathbf{n}}^{(2)}=\tilde{\mathbf{w}}^{(2,1)}$}}
\\
\\
Components $n^{(2)}_{pq_1}$ and $\tilde{w}^{(2)}_{pq_1}$ are related to the known terms of cell problems (\ref{eq:ThirdCellProblemDisplacement+0}) and (\ref{eq:ThirdCellProblDiffus+0}), namely
\begin{subequations}
\begin{align}
&\hat{n}_{pq_1}^{(2)}=
\left\langle
\beta_{pq_1}^m
-C_{pq_1kj}^m\hat{N}_{k,j}^{(1)}
\right\rangle,
\label{eq:coeff3}
\\
&
\tilde{w}_{pq_1}^{(2,1)}=
\left\langle
\beta_{pq_1}^m
+
\beta_{iq_2}^m\hat{N}_{ipq_1,q_2}^{(1)}
\right\rangle.
\label{eq:coeff4}
\end{align}
\end{subequations}
Given the first mechanical cell problem (\ref{eq:FirstCellProblemDisplacement-1}) at the order $\varepsilon^{-1}$
\begin{equation}
\left(C_{pjkl}^{m}N_{kiq_1,l}^{(1)}
\right)_{,j}
+
C_{pjiq_1,j}^{m}
=0,
\end{equation}
its weak form has expression
\begin{equation}
\left\langle
\left[
\left(C_{pjkl}^{m}N_{kiq_1,l}^{(1)}
\right)_{,j}
+
C_{pjiq_1,j}^{m}
\right]
\hat{N}_{p}^{(1)}
\right\rangle
=0,
\label{eq:WaekForm5}
\end{equation}
with test function $\hat{N}_{p}^{(1)}$.
Equation (\ref{eq:WaekForm5}) can be written as
\begin{equation}
\left\langle
\left[
C_{pjkl}^{m}N_{kiq_1,l}^{(1)}
+
C_{pjiq_1}^{m}
\right]
\hat{N}_{p,j}^{(1)}
\right\rangle
=0,
\label{eq:WeakForm6}
\end{equation}
for divergence theorem and $\mathcal{Q}$-periodicity of micro constitutive tensors components and weight functions.
By adding vanishing term (\ref{eq:WeakForm6}) to equation (\ref{eq:coeff3}) one obtains
\begin{equation}
\hat{n}_{pq_1}^{(2)}=
\left\langle
\beta_{pq_1}^m
-C_{pq_1il}^m\hat{N}_{i,l}^{(1)}
+
C_{ijkl}^m
{N}_{ipq_1,j}^{(1)}
\hat{N}_{k,l}^{(1)}
+
C_{ilpq_1}^m
\hat{N}_{i,l}^{(1)}
\right\rangle
=
\left\langle
\beta_{pq_1}^m
+
C_{ijkl}^m
{N}_{ipq_1,j}^{(1)}
\hat{N}_{k,l}^{(1)}
\right\rangle.
\end{equation}
From cell problem (\ref{eq:ThirdCellProblemDisplacement-1}) at the order $\varepsilon^{-1}$ 
\begin{equation}
\left(
C_{ijkl}^{m}
\hat{N}_{k,l}^{(1)}
\right)_{,j}
-
\beta_{ij,j}^m=0,
\end{equation}
the following weak form can be derived
\begin{equation}
\left\langle
\left[
\left(
C_{ijkl}^{m}
\hat{N}_{k,l}^{(1)}
\right)_{,j}
-
\beta_{ij,j}^m
\right]
N_{ipq_1}^{(1)}
\right\rangle
=0,
\label{eq:WeakForm7}
\end{equation}
considering $N_{ipq_1}^{(1)}$ as a test function.
Analogously to what done before, equation (\ref{eq:WeakForm7}) can be written in the form
\begin{equation}
\left\langle
\left[
C_{ijkl}^{m}
\hat{N}_{k,l}^{(1)}
-
\beta_{ij}^m
\right]
N_{ipq_1,j}^{(1)}
\right\rangle
=0.
\label{eq:WeakForm8}
\end{equation}
The sum of (\ref{eq:WeakForm8}) and (\ref{eq:coeff4}) leads to
\begin{equation}
\tilde{w}_{pq_1}^{(2,1)}=
\left\langle
\beta_{pq_1}^m
+
\beta_{iq_2}^m\hat{N}_{ipq_1,q_2}^{(1)}
+
C_{ijkl}^{m}
\hat{N}_{k,l}^{(1)}
N_{ipq_1,j}^{(1)}
-
\beta_{ij}^m
N_{ipq_1,j}^{(1)}
\right\rangle
=
\left\langle
\beta_{pq_1}^m
+
C_{ijkl}^{m}
\hat{N}_{k,l}^{(1)}
N_{ipq_1,j}^{(1)}
\right\rangle,
\end{equation}
from which equivalence $\hat{n}_{pq_1}^{(2)}=\tilde{w}_{pq_1}^{(2,1)}$ follows.
\\
\\
\\
{\itshape Proof of {$\hat{m}^{(2,1)}=\hat{w}^{(2,1)}$}}
\\
\\
Constants $\hat{m}^{(2,1)}$ and $\hat{w}^{(2,1)}$ come from the known terms of cell problems (\ref{eq:FourthCellProblemTemperature+0})and (\ref{eq:FourthCellProblDiffus+0}), namely
\begin{subequations}
\begin{align}
&
\hat{m}^{(2,1)}=
\left\langle
\psi^m
+
\alpha_{q_1q_2}^m
\hat{N}_{q_1,q_2}^{(1)}
\right\rangle,
\label{eq:coeff5}
\\
&
\hat{w}^{(2,1)}=
\left\langle
\psi^m
+
\beta_{q_1q_2}^m
\tilde{N}_{q_1,q_2}^{(1)}
\right\rangle.
\label{eq:coeff6}
\end{align}
\end{subequations}
Second mechanical cell problem (\ref{eq:SecondCellProblemDisplacement-1}) at the order $\varepsilon^{-1}$  reads 
\begin{equation}
\left(C_{ijkl}^{m}
\tilde{N}_{k,l}^{(1)}
\right)_{,j}
-
\alpha_{ij,j}^{m}
=0,
\end{equation}
and its weak form, considering $\hat{N}_i^{(1)}$ as test function, is
\begin{equation}
\left\langle
\left[
\left(C_{ijkl}^{m}
\tilde{N}_{k,l}^{(1)}
\right)_{,j}
-
\alpha_{ij,j}^{m}
\right]
\hat{N}_{i}^{(1)}
\right\rangle
=0.
\label{eq:WeakForm9}
\end{equation}
Once again, equation (\ref{eq:WeakForm9}) can be written as
\begin{equation}
\left\langle
\left[
C_{ijkl}^{m}
 \tilde{N}_{k,l}^{(1)}
-
\alpha_{ij}^{m}
\right]
\hat{N}_{i,j}^{(1)}
\right\rangle
=0,
\label{eq:WeakForm10}
\end{equation}
exploiting divergence theorem and $\mathcal{Q}$-periodicity of perturbation functions and micro constitutive tensors.
Adding term (\ref{eq:WeakForm10}) to (\ref{eq:coeff5}), one obtains
\begin{equation}
\hat{m}^{(2,1)}=
\left\langle
\psi^m
+
\alpha_{q_1q_2}^m\hat{N}_{q_1,q_2}^{(1)}
+
C_{ijkl}^m
\tilde{N}_{k,l}^{(1)}
\hat{N}_{i,j}^{(1)}
-
\alpha_{ij}^m
\hat{N}_{i,j}^{(1)}
\right\rangle
=
\left\langle
\psi^m
+
C_{ijkl}^m
\hat{N}_{i,j}^{(1)}
\tilde{N}_{k,l}^{(1)}
\right\rangle.
\label{eq:coeff8}
\end{equation}
The weak form of cell problem (\ref{eq:ThirdCellProblemDisplacement-1}) at the order $\varepsilon^{-1}$, expressed as  %
\begin{equation}
\left(
C_{ijkl}^{m}
\hat{N}_{k,l}^{(1)}
\right)_{,j}
-
\beta_{ij,j}^m=0,
\end{equation}
has the form
\begin{equation}
\left\langle
\left[
\left(
C_{ijkl}^{m}
\hat{N}_{k,l}^{(1)}
\right)_{,j}
-
\beta_{ij,j}^m
\right]
\tilde{N}_i^{(1)}
\right\rangle
=0,
\label{eq:WeakForm11}
\end{equation}
with test function $\tilde{N}_i^{(1)}$.
Equation (\ref{eq:WeakForm11}) turns into
\begin{equation}
\left\langle
\left[
C_{ijkl}^{m}
\hat{N}_{k,l}^{(1)}
-
\beta_{ij}^m
\right]
\hat{N}_{i,j}^{(1)}
\right\rangle
=0,
\label{eq:WeakForm12}
\end{equation}
for divergence theorem and $\mathcal{Q}$-periodicity of terms involved.
The sum of vanishing term (\ref{eq:WeakForm12}) and (\ref{eq:coeff6}) leads to
\begin{equation}
\hat{w}^{(2,1)}=
\left\langle
\psi^m
+
\beta_{q_1q_2}^m
\tilde{N}_{q_1,q_2}^{(1)}
-
\beta_{ij}^{m}
\hat{N}_{i,j}^{(1)}
+
C_{ijkl}^m
\hat{N}_{k,l}^{(1)}
\hat{N}_{i,j}^{(1)}
\right\rangle
=
\left\langle
\psi^m
+
C_{ijkl}^{m}
\hat{N}_{k,l}^{(1)}
\hat{N}_{i,j}^{(1)}
\right\rangle.
\label{eq:coeff7}
\end{equation}
Comparing (\ref{eq:coeff8}) and (\ref{eq:coeff7}), the equivalence between constants $\hat{m}^{(2,1)}$ and $\hat{w}^{(2,1)}$ trivially derives. 
%
\section*{Appendix B. Fourth order approximation of dispersion functions for the equivalent thermo-diffusive medium   }
\label{Appendix::Sensitivita}
Considering an angular coordinate $\phi=0$, sensitivities $\omega_i^{[n]}$, with $n=1,..,4$ and $i=1,...,6$ of table \ref{Table::SolutionSchemePerturbativeApproach} have the closed form detailed below in terms of the overall constitutive tensors components relative to the homogenized thermo-diffusive medium. 
Such sensitivities are the solutions of the chain of $n$-ordered perturbation equations generated by the perturbative approximation described in Section \ref{SubSec::AsympApproxOfComplexSpectrum} of the characteristic equation (\ref{eq:characteristic_equation}).
Being $k_1=r\,cos(\phi)=r$ and $k_2=r\,sin(\phi)=0$, perturbation parameter is represented by the wave number $k_1$ and the fourth order approximated dispersion function $\omega(k_1) $ has the form
\begin{equation}
\omega(k_1)=\omega^{[0]}+\omega^{[1]}\,k_1
+
\omega^{[2]}\, k_1^2
+
\omega^{[3]}\,k_1^3
+
\omega^{[4]}\,k_1^4+ O(k_1^5)
\end{equation}
As expected, generating solutions at the order $k_1^0$ are all vanishing, namely
$
\omega^{[0]}_{1...6}=0
$ for $i=1,...,6$, meaning that dispersion curves are all acoustic branches departing from the origin.
From perturbation problem $G^{[6]}(r=0)=0$ one derives the following sensitivities
\begin{eqnarray}
&&\omega^{[1]}_{1,2}=0,
\nonumber\\
&&\omega^{[1]}_3=\frac{i\,\sqrt{\rho\,C_{1212}}}{\rho},
\nonumber\\
&&\omega^{[1]}_5=\frac{i\,\sqrt{\rho\,
\left(
p\,q
-
\psi^2
\right)\,\left(
C_{1111}\,p\,q
-
C_{1111}\,\psi^2
+
\alpha_{11}^2\,q
+
2\,\alpha_{11}\,\beta_{11}\,\psi
+
\beta_{11}^2\,
p
\right)}}{\rho\,\left(p\,q-\psi^2\right)},
\label{eq:sensitivities_1}
\end{eqnarray}
where $\omega_4^{[1]}$ and $\omega_6^{[1]}$ are not explicitly written being the complex conjugate of sensitivities $\omega_3^{[1]}$ and $\omega_5^{[1]}$, respectively.
Consistently, sensitivities $\omega_i^{[2]}$ have the form
\begin{eqnarray}
&&\omega^{[2]}_1=\frac{
-
p\,C_{1111}\,D_{11}
-
q\,C_{1111}\,K_{11}
-
D_{11}\,\alpha_{11}^2
-
K_{11}\,\beta_{11}^2
+r_1
}{2\,C_{1111}\,p\,q
-
2\,C_{1111}\,\psi^2
+
2\,\alpha_{11}^2\,q
+
4\,\alpha_{11}\,\beta_{11}\,\psi
+
2\,\beta_{11}^2\,p},
\nonumber\\
&&\omega^{[2]}_2=\frac{
-
p\,C_{1111}\,D_{11}
-
q\,C_{1111}\,K_{11}
-
D_{11}\,\alpha_{11}^2
-
K_{11}\,\beta_{11}^2
-r_1
}{2\,C_{1111}\,p\,q
-
2\,C_{1111}\,\psi^2
+
2\,\alpha_{11}^2\,q
+
4\,\alpha_{11}\,\beta_{11}\,\psi
+
2\,\beta_{11}^2\,p},
\nonumber\\
&&\omega^{[2]}_3=0,
\nonumber\\
&&\omega^{[2]}_5=\frac{1}{2}\frac{
\left(-D_{11}\,\alpha_{11}^2
-
K_{11}\,\beta_{11}^2
\right)\,\psi^2
-
2\,\psi\,\beta_{11}\,
\left(D_{11}\,p
+
K_{11}\,q
\right)\,\alpha_{11}
-
D_{11}\,\beta_{11}^2\,p^2
-
K_{11}\,\alpha_{11}^2\,q^2}{
\left(
p\,q-\psi^2
\right)\,
\left(
\left(
C_{1111}\,q
+
\beta_{11}^2
\right)\,p
-
C_{1111}\,\psi^2
+
2\,\beta_{11}\,\alpha_{11}\,\psi
+
\alpha_{11}^2\,q\right)}.
\label{eq:sensitivities_2}
\end{eqnarray}
Sensitivities $\omega_i^{[3]}$ are expressed as
\begin{eqnarray}
&&\omega^{[3]}_1=\omega^{[3]}_{2}=\omega^{[3]}_{3}=0,
\nonumber\\
&&\omega^{[3]}_5=\left(
\alpha_{11}^4\,
a_{40}
+
\alpha_{11}^3\,
\beta_{11}\,
a_{31}+
\alpha_{11}^2\,
\beta_{11}^2\,
a_{22}+
\alpha_{11}\,
\beta_{11}^3\,
a_{13}+
\beta_{11}^4\,
a_{04}+
\alpha_{11}^2\,
a_{20}
+
\alpha_{11}\,
\beta_{11}\,
a_{11}
+
\beta_{11}^2\,
a_{02}
\right)/s_1,
\nonumber\\
\label{eq:sensitivities_3}
\end{eqnarray}
and sensitivities $\omega_i^{[4]}$ result 
\begin{eqnarray}
\omega^{[4]}_1&=&\left(
D_{11}^4\,
b_{40}
+
D_{11}^3\,
K_{11}\,
b_{31}
+
D_{11}^2\,
K_{11}^2\,
b_{22}
+
D_{11}\,
K_{11}^3\,
b_{13}
+
K_{11}^4\,
b_{04}
+
D_{11}^3
\,b_{30}
+
\right.
\nonumber\\
&+&
\left.
D_{11}^2
\,
K_{11}
\,
b_{21}
+
D_{11}
\,
K_{11}^2\,
b_{12}
+
K_{11}^3
\,
b_{03}
\right)/s_2,
\nonumber\\
\omega^{[4]}_{2}&=&
\left(
D_{11}^4\,
c_{40}
+
D_{11}^3\,
K_{11}\,
c_{31}+
D_{11}^2\,
K_{11}^2\,
c_{22}+
D_{11}
\,K11^3\,
c_{13}
+
K_{11}^4\,
c_{04}
+
D_{11}^3\,
c_{30}
+
\right.
\nonumber\\
&+&
\left.
D_{11}^2\,
K_{11}\,
c_{21}
+
D_{11}\,
K_{11}^2\,
c_{12}
+
K_{11}^3\,
c_{30}\right)/s_3,
\nonumber\\
\omega^{[4]}_3&=&0,
\nonumber\\
\omega^{[4]}_5
&=&\left(
\alpha_{11}^6\,
d_{60}
+
\alpha_{11}^5\,
\beta_{11}\,
d_{51}
+
\alpha_{11}^4\,
\beta_{11}^2\,
d_{42}
+
\alpha_{11}^2\,
\beta_{11}^4\,
d_{24}
+
\alpha_{11}\,
\beta_{11}^5\,
d_{15}
+
\beta_{11}^6\,
d_{06}
+
\alpha_{11}^4\,
d_{40}
+
\right.
\nonumber\\
&+&
\left.
\alpha_{11}^3\,
\beta_{11}\,
d_{31}
+
\alpha_{11}^2\,
\beta_{11}^2\,
d_{22}
+
\alpha_{11}\,
\beta_{11}^3\,
d_{13}
+
\beta_{11}^4\,
d_{04}
+
\alpha_{11}^2\,
d_{20}
+
\alpha_{11}\,
\beta_{11}\,
d_{11}
+
\beta_{11}^2\,
d_{02}\right)/s_4.
\label{eq:sensitivities_4}
\end{eqnarray}
Coefficients $r_i, s_i, a_{ij}, b_{ij}, c_{ij}$, and $d_{ij}$ of formulas (\ref{eq:sensitivities_2}), (\ref{eq:sensitivities_3}), and (\ref{eq:sensitivities_4})   are made explicit in Appendix C.
%
\section*{Appendix C. Coefficients involved in the perturbative approximation of dispersion functions   }
\label{Appendix::CoefficientsOfSensitivities}
Coefficient $r_1$ of equation (\ref{eq:sensitivities_2}) reads
\begin{eqnarray}
r_1&=&\left(
\left(
C_{11}\,
p
+
\alpha_{11}^2
\right)^2\,
D_{11}^2
-
2\,K_{11}\,
\left(
\left(
C_{1111}\,q
-
\beta_{11}^2
\right)\,\alpha_{11}^2
+
4\,\psi\,C_{1111}\,\beta_{11}\,\alpha_{11}
+
\left(\beta_{11}^2\,p
+
\right.
\right.
\right.
\nonumber\\
&+&
\left.
\left.
\left.
C_{1111}\,
\left(p\,q
-
2\,\psi^2
\right)
\right)\,C_{1111}
\right)\,D_{11}
+
K_{11}^2\,
\left(C_{1111}\,q
+
\beta_{11}^2
\right)^2\right)^{1/2}.
\end{eqnarray}
Coefficients related to sensitivity $\omega^{[3]}_5$ of equation (\ref{eq:sensitivities_3}) have the form
\begin{eqnarray}
a_{40}&=&
\rho\,
\left(4\,D_{11}^2\,p\,\psi^2\,q
-
3\,D_{11}^2\,\psi^4
+
2\,D_{11}\,K_{11}\,\psi^2\,q^2
+
K_{11}^2\,q^4
\right),
\nonumber\\
a_{31}&=&
4\,\rho\,\psi\,
\left(
2\,D_{11}^2\,p^2\,q
-
D_{11}^2\,p\,\psi^2
-
D_{11}\,K_{11}\,p\,q^2
+
3\,D_{11}\,K_{11}\,\psi^2\,q
+
K_{11}^2\,q^3
\right),
\nonumber\\
a_{22}&=&
4\,
\left(
p\,q^3\,K_{11}^2
+
\left(
-3/2\,p^2\,D_{11}\,K_{11}
+
1/2\,K_{11}^2\,\psi^2
\right)
\,q^2+p\,D_{11}\,
\left(
D_{11}\,p^2
+
2\,K_{11}\,\psi^2
\right)\,
q
+
\right.
\nonumber\\
&+&
\left.
1/2\,D_{11}\,\psi^2\,
\left(
D_{11}\,p^2
+
5\,K_{11}\,\psi^2
\right)
\right)\,\rho,
\nonumber\\
a_{13}&=&
4\,
\left(
\left(
2\,p\,q^2
-\psi^2\,q\right)
\,K_{11}^2
-
p\,K_{11}\,
\left(p\,q-3\,\psi^2
\right)\,D_{11}+p^3\,D_{11}^2\right)
\,\psi\,\rho,
\nonumber\\
a_{04}&=&
\left(-3\,K_{11}^2\,\psi^4
+
\left(
2\,D_{11}\,K_{11}\,p^2
+
4\,K_{11}^2\,p\,q
\right)\,\psi^2
+
p^4\,D_{11}^2
\right)\,\rho,
\nonumber\\
a_{20}&=&
4\,
\left(p\,q-\psi^2
\right)\,
\left(D_{11}^2\,p\,\psi^2
+
2\,D_{11}\,K_{11}\,\psi^2\,q
+
K_{11}^2\,q^3
\right)\,\rho\,C_{1111},
\nonumber\\
a_{11}&=&
8\,
\left(p\,q
-
\psi^2
\right)\,
\left(
D_{11}^2\,p^2
+
K_{11}\,
\left(p\,q
+
\psi^2
\right)\,D_{11}
+
K_{11}^2\,q^2
\right)\,\psi\,\rho\,C_{1111},
\nonumber\\
a_{02}&=&
4\,
\left(p\,q-\psi^2
\right)\,
\left(D_{11}^2\,p^3
+
2\,D_{11}\,K_{11}\,p\,\psi^2
+
K_{11}^2\,\psi^2\,q
\right)\,\rho\,C_{1111},
\nonumber\\
s_1&=&8\, i\,\sqrt{\rho}\,
\left(p\,q-\psi^2
\right)^{3/2}\,
\left(
\left(
C_{1111}\,p
+
\alpha_{11}^2
\right)
\,q
+
\beta_{11}^2\,p
-
C_{1111}\,\psi^2
+
2\,\beta_{11}\,\alpha_{11}\,\psi
\right)^{5/2}.
\end{eqnarray}
Sensitivity $\omega_1^{[4]}$ in equation (\ref{eq:sensitivities_4}) has coefficients
\begin{eqnarray}
b_{40}&=&
4\,
\left(
C_{1111}\,p
+
\alpha_{11}^2
\right)^3
\,
\left(
\alpha_{11}\,\psi
+
\beta_{11}\,p
\right)^2\,C_{1212}\,\rho,
\nonumber\\
b_{31}&=&-4\,
\left(
2\,q\,\beta_{11}\,
\alpha_{11}^4
-
q\,\psi\,C_{1111}\,\alpha_{11}^3
+
\left(
- 2\,p\,\beta_{11}^3
+
C_{1111}\,
\left(
3\,p\,q
+
4\,\psi^2
\right)
\,\beta_{11}\right)\,\alpha_{11}^2
+
\right.
\nonumber\\
&-&
\left.
\left(
-9\,\beta_{11}^2\,p
+
C_{1111}\,
\left(
p\,q
+
3\,\psi^2
\right)
\right)
\,\psi\,C_{1111}\,\alpha_{11}
+
\beta_{11}
\left(
\beta_{11}^2\,p
+
C_{1111}
\left(
p\,q 
-
5\,\psi^2
\right)
\right)
\,p\,C_{1111}
\right)
\nonumber\\
& &
\left(
C_{1111}\,p + \alpha_{11}^2
\right)
\left(
\alpha_{11}\psi
+
\beta_{11}\,p
\right)
\,C_{1212}\,\rho,
\nonumber\\
b_{22}&=&-
4\,
\left(
\alpha_{11}^4\,
q^2
-
2\,
\alpha_{11}^3\,q\,\psi\,\beta_{11}
+
\left(
\left(
-4\,p\,q-2\,\psi^2
\right)
\,\beta_{11}^2
+
q\,C_{1111}\,
\left(
p\,q
+
5\,\psi^2
\right)
\right)
\,\alpha_{11}^2
+
\right.
\nonumber\\
&+&
\left.
8\,\beta_{11}\,
\left(
-1/4\,\beta_{11}^2\,p
+
C_{1111}\,
\left(
p\,q
+1/2\psi^2
\right)
\right)
\psi\,\alpha_{11}
+
\beta_{11}^2\,p\,
\left(
\beta_{11}^2\,p
+
C_{1111}\,
\left(
p\,q
+
5\,\psi^2
\right)
\right)
\right)
\nonumber\\
&&
\left(
C_{1111}\,\psi-\alpha_{11}\beta_{11}
\right)^2
C_{1212}\,\rho,
\nonumber\\
b_{13}&=&-
4\,
\left(
2\,\alpha_{11}\,\beta_{11}^4\,p
-
C_{1111}\,\beta_{11}^3\,p\,\psi
+
\left(\alpha_{11}\,
\left(3\,p\,q
+
4\,\psi^2
\right)\,
C_{1111}-
2\,\alpha_{11}^3\,q
\right)\,\beta_{11}^2
+
\right.
\nonumber\\
&-&
\left.\left(C_{1111}\,
\left(p\,q
+
3\,\psi^2
\right)-
9\,\alpha_{11}^2\,q
\right)\,C_{1111}\,\psi\,\beta_{11}
+
\left(
C_{1111}\,
\left(p\,q-5\,\psi^2
\right)
+\alpha_{11}^2\,q
\right)\,
q\,C_{1111}\,\alpha_{11}
\right)
\nonumber\\
&&
C_{1212}\,
\left(C_{1111}\,q
+
\beta_{11}^2
\right)\,\rho\,
\left(
\alpha_{11}\,q
+
\beta_{11}\,\psi
\right),
\nonumber\\
b_{04}&=&
4\,
\left(C_{1111}\,q
+
\beta_{11}^2
\right)^3\,
\left(\alpha_{11}\,q
+
\beta_{11}\,\psi
\right)^2\,C_{1212}\,\rho,
\nonumber\\
b_{30}&=&-
4\,r_2\,
\left(C_{1111}\,p
+\alpha_{11}^2\right)^2\,
\left(\alpha_{11}\,\psi
+
\beta_{11}\,p
\right)^2\,C_{1212}\,\rho,
\nonumber\\
b_{21}&=&-4\,
\left(\alpha_{11}\,\psi
+
\beta_{11}\,p
\right)\,
\left(2\,C_{1111}\,\alpha_{11}\,p\,q
+
C_{1111}\,\alpha_{11}\,\psi^2
+
3\,C_{1111}\,\beta_{11}\,p\,\psi
+
2\,\alpha_{11}^3\,q
+
\right.
\nonumber\\
&+&
\left.
\alpha_{11}^2\,
\beta_{11}\,\psi
-
\alpha_{11}\,\beta_{11}^2\,p
\right)\,r_2\,\left(C_{1111}\,\psi
-
\alpha_{11}\,\beta_{11}\right)\,C_{1212}\,\rho,
\nonumber\\
b_{12}&=&-4\,
\left(3\,C_{1111}\,\alpha_{11}\,\psi\,q
+
2\,C_{1111}\,\beta_{11}\,p\,q
+
C_{1111}\,\beta_{11}\,\psi^2
-
\alpha_{11}^2\,\beta_{11}\,q
+
\alpha_{11}\,\beta_{11}^2\,\psi
+
2\,\beta_{11}^3\,p
\right)
\nonumber\\
&&
\left(C_{1111}\,\psi
-
\alpha_{11}\,\beta_{11}
\right)\,r_1\,\left(\alpha_{11}\,q
+
\beta_{11}\,\psi
\right)\,C_{1212}\,\rho,
\nonumber\\
b_{03}&=&-4
\left(C_{1111}\,q
+
\beta_{11}^2
\right)^2\,r_1\,
\left(\alpha_{11}\,q
+
\beta_{11}\,\psi
\right)^2\,C_{1212}\,\rho,
\nonumber\\
s_2&=&8\,
C_{1212}\,
\left(
\left(
C_{1111}\,p
+
\alpha_{11}^2
\right)\,q
+
\beta_{11}^2\,p
-
C_{1111}\,\psi^2
+
2\beta_{11}\,\alpha_{11}\,\psi
\right)^4\,r_2,
\nonumber\\
r_2&=&\left(
\left(
C_{1111}\,p
+
\alpha_{11}^2
\right)^2
\,D11^2
-
2\,K_{11}\,
\left(
\left(
C_{1111}\,q
-
\beta_{11}^2
\right)\,\alpha_{11}^2
+
4\,\psi\,C_{1111}\,\beta_{11}\,\alpha_{11}
+
\right.
\right.
\nonumber\\
&+&
\left.
\left.
\left(\beta_{11}^2\,p
+
C_{1111}\,
\left(
p\,q
-
2\,\psi^2
\right)
\right)\,
C_{1111}
\right)\,D_{11}
+
K_{11}^2\,
\left(C_{1111}\,q
+
\beta_{11}^2
\right)^2\right)^{1/2}.
\end{eqnarray}
Coefficients of sensitivity $\omega_2^{[4]}$ in equation (\ref{eq:sensitivities_4}) have the following expression
\begin{eqnarray}
c_{40}&=&-
4\,\left(C_{1111}\,p
+
\alpha_{11}^2\right)^3\,\left(\alpha_{11}\,\psi
+
\beta_{11}\,p
\right)^2\,C_{1212}\,\rho,
\nonumber\\
c_{31}&=&4\,
\left(2\,q\,\beta_{11}\,\alpha_{11}^4
-q\,\psi\,C_{1111}\,\alpha_{11}^3
+
\left(-2\,p\,\beta_{11}^3
+
C_{1111}\left(
3\,p\,q
+
4\,\psi^2
\right)\,\beta_{11}\right)
\,\alpha_{11}^2
+
\right.
\nonumber\\
&-&
\left.
\left(
-
9\,\beta_{11}^2\,p
+
C_{1111}\,
\left(p\,q
+
3\,\psi^2
\right)
\right)\,\psi\,C_{1111}\,\alpha_{11}
+
\beta_{11}\left(\beta_{11}^2\,p
+
C_{1111}\,\left(p\,q-5\,\psi^2\right)\right)\,p\,C_{1111}
\right)
\nonumber\\
&&\left(C_{1111}\,p
+\alpha_{11}^2\right)\,
\left(\alpha_{11}\,\psi
+
\beta_{11}\,p\right)\,C_{1212}\,\rho,
\nonumber\\
c_{22}&=&-
4\,
\left(
\alpha_{11}^4\,q^2
-
2\,\alpha_{11}^3\,q\,\psi\,\beta_{11}
+
\left(
\left(
-4\,p\,q-2\,\psi^2
\right)
\,\beta_{11}^2
+
q\,C_{1111}\,\left(
p\,q+5\,\psi^2
\right)
\right)\,\alpha_{11}^2
+
\right.
\nonumber\\
&+&
\left.
8\,\beta_{11}\,
\left(
-1/4\,\beta_{11}^2\,p
+
C_{1111}\,
\left(
p\,q+1/2\,\psi^2
\right)
\right)
\,\psi\,\alpha_{11}
+
\beta_{11}^2\,p\,
\left(
\beta_{11}^2\,p
+
C_{1111}\,
\left(p\,q+5\,\psi^2
\right)
\right)
\right)
\nonumber\\
&&
\left(
C_{1111}\,\psi-\alpha_{11}\,\beta_{11}
\right)^2\,C_{1212}\,\rho,
\nonumber\\
c_{13}&=&
4\,
\left(
-C_{1111}\,\beta_{11}^3\,p\,\psi
+
2\,\alpha_{11}\,\beta_{11}^4\,p
+\left(
\alpha_{11}\,
\left(3\,p\,q+4\,\psi^2
\right)
C_{1111}
-
2\,\alpha_{11}^3\,q
\right)\,\beta_{11}^2+
\right.
\nonumber\\
&-&
\left.
\left(
C_{1111}\,
\left(p\,q+3\,\psi^2
\right)-9\,\alpha_{11}^2\,q
\right)\,C_{1111}\,\psi\,\beta_{11}
+
\left(
C_{1111}\,
\left(p\,q-5\,\psi^2
\right)+\alpha_{11}^2\,q
\right)
\,q\,C_{1111}\,\alpha_{11}
\right)
\nonumber\\
&&C_{1212}\,
\left(C_{1111}\,q+\beta_{11}^2
\right)\,\rho\,
\left(\alpha_{11}\,q
+
\beta_{11}\,\psi
\right),
\nonumber\\
c_{40}&=&-
4\,
\left(C_{1111}\,q
+\beta_{11}^2\right)^3\,
\left(
\alpha_{11}\,q
+
\beta_{11}\,\psi
\right)^2
\,C_{1212}\,\rho,
\nonumber\\
c_{30}&=&-
4\,r_2\,
\left(
C_{1111}\,p+\alpha_{11}^2
\right)^2\,
\left(
\alpha_{11}\,\psi+\beta_{11}\,p
\right)^2\,C_{1212}\,\rho,
\nonumber\\
c_{21}&=&-
4\,
\left(\alpha_{11}\,
\psi
+\beta_{11}\,p\right)
\left(2\,C_{1111}\,\alpha_{11}\,p\,q
+
C_{1111}\,\alpha_{11}\,\psi^2
+
3\,C_{1111}\,\beta_{11}\,p\,\psi
+
2\,\alpha_{11}^3\,q
+
\right.
\nonumber\\
&+&
\left.
\alpha_{11}^2\,
\beta_{11}\,\psi
-\alpha_{11}\,\beta_{11}^2\,p
\right)\,r_2
\left(
C_{1111}\,\psi
-\alpha_{11}\,\beta_{11}\right)
\,C_{1212}\,\rho,
\nonumber\\
c_{12}&=&-
4\,
\left(
3\,C_{1111}\,\alpha_{11}\,\psi\,q
+
2\,C_{1111}\,\beta_{11}\,p\,q
+
C_{1111}\,\beta_{11}\,\psi^2
-
\alpha_{11}^2\,\beta_{11}\,q
+
\alpha_{11}\,\beta_{11}^2\,\psi
+
2\,\beta_{11}^3\,p
\right)
\nonumber\\
&&
\left(
C_{1111}\,\psi
-
\alpha_{11}\,\beta_{11}
\right)\,r_2\,
\left(
\alpha_{11}\,q
+
\beta_{11}\,\psi
\right)\,C_{1212}\,\rho,
\nonumber\\
c_{03}&=&-
4\,
\left(
C_{1111}\,q
+
\beta_{11}^2
\right)^2\,r_1\,
\left(
\alpha_{11}\,q
+
\beta_{11}\,\psi
\right)^2\,\,\rho,
\nonumber\\
s_3&=&8\,C_{1212}\,
\left(
\left(
C_{1111}\,p
+
\alpha_{11}^2
\right)\,
q
+
\beta_{11}^2\,p
-
C_{1111}\,\psi^2
+
2\,\beta_{11}\,\alpha_{11}\,\psi
\right)^4\,r_2,
\end{eqnarray}
while coefficients of sensitivity $\omega_5^{[4]}$ in equation (\ref{eq:sensitivities_4}) read
\begin{eqnarray}
d_{60}&=&
\rho\,D_{11}^3\,\psi^2,
\nonumber\\
d_{51}&=&
2\,\rho\,D_{11}^2\,\psi\,
\left(
D_{11}\,p
-
K_{11}\,q
\right),
\nonumber\\
d_{42}&=&
\rho\,\left(
D_{11}^2\,p^2
+
\left(
-2\,p\,q-\psi^2
\right)
\,K_{11}\,D_{11}
+
K_{11}^2\,q^2
\right)\,D_{11},
\nonumber\\
d_{24}&=&
\rho\,
\left(
D_{11}^2\,p^2
+
\left(
-2\,p\,q-\psi^2
\right)\,K_{11}\,D_{11}
+
K_{11}^2\,q^2
\right)\,K_{11},
\nonumber\\
d_{15}&=&-
2\,\rho\,K_{11}^2
\left(
D_{11}\,p
-
K_{11}\,q
\right)\,\psi,
\nonumber\\
d_{06}&=&
\rho\,K_{11}^3\,\psi^2,
\nonumber\\
d_{40}&=&
2\rho\,D_{11}^2\,\psi^2\,C_{1111}\,
\left(
D_{11}\,p
+
K_{11}\,q
\right),
\nonumber\\
d_{31}&=&
4\,\rho\,C_{1111}\,D_{11}\,
\psi
\left(
D_{11}^2\,p^2
-
K_{11}^2\,q^2
\right),
\nonumber\\
d_{22}&=&
2\,\rho\,
\left(
D_{11}\,p
+
K_{11}\,q
\right)\,
\left(
D_{11}^2\,p^2
-
2\,K_{11}\,
\left(p\,q
+
\psi^2
\right)\,D_{11}
+
K_{11}^2\,q^2
\right)
C_{1111},
\nonumber\\
d_{13}&=&-
4\,\rho\,C_{1111}\,K_{11}\,
\left(
D_{11}^2\,
p^2
-
K_{11}^2\,q^2
\right)\,\psi,
\nonumber\\
d_{04}&=&
2\,\rho\,K_{11}^2\,\psi^2\,C_{1111}\,\left(D_{11}\,p
+
K_{11}\,q
\right),
\nonumber\\
d_{20}&=&
\rho\,\left(p^2\,D_{11}^3\,\psi^2
+
\left(2\,p\,\psi^2\,q
+
\psi^4
\right)\,K_{11}\,D_{11}^2
+
3\,q^2\,D_{11}\,K_{11}^2\,\psi^2
+
q^4\,K_{11}^3
\right)\,C_{1111}^2,
\nonumber\\
d_{11}&=&
2\,\rho
C_{1111}^2\,
\left(
D_{11}\,p
+
K_{11}\,q
\right)\,
\left(D_{11}^2\,p^2
+
2\,D_{11}\,K_{11}\,\psi^2
+
K_{11}^2\,q^2
\right)\,\psi,
\nonumber\\
d_{02}&=&
\rho\,
\left(p^4\,D_{11}^3
+
3\,p^2\,D_{11}^2\,K_{11}\,\psi^2
+
\left(2\,p\,\psi^2\,q
+
\psi^4
\right)\,K_{11}^2\,D_{11}
+
q^2\,K_{11}^3\,\psi^2
\right)\,C_{1111}^2,
\nonumber\\
s_4&=&2\,
\left(
\left(C_{1111}\,p
+
\alpha_{11}^2
\right)\,q
+
\beta_{11}^2\,p
-
C_{1111}\,\psi^2
+
2\,\beta_{11}\,\alpha_{11}\,\psi
\right)^4.
\end{eqnarray}
%
%
%
\section*{Appendix D. Tensorial fashion for   constitutive equations of thermo-diffusive material }
\label{Appendix::NotazioneTensorialeEquazioni di Campo}
In a 2-D setting, linear constitutive relations (\ref{eq:MicroStress})-(\ref{eq:MicroMassFlux}) for thermo-diffusive materials   can rigorously be written in a tensorial fashion as done in \cite{Mehrabadi1990}. They read
\begin{align}
\label{eq:FEConstitutiveRelationsTensorialFashion}
\left(
\begin{array}{c}
\sigma_{11}\\
\sigma_{22}\\
\sqrt{2}\,\sigma_{12}
\end{array}
\right)=&
\left(
\begin{array}{c c c}
C_{1111}^m & C_{1122}^m & \sqrt{2}\, C_{1112}^m\\
C_{2211}^m & C_{2222}^m & \sqrt{2}\, C_{2212}^m\\
\sqrt{2}\, C_{1211}^m & \sqrt{2}\, C_{1222}^m & 2\, C_{1212}^m
\end{array}
\right)
\left(
\begin{array}{c}
u_{1,1}\\
u_{2,2}\\
\frac{\sqrt{2}}{2}\left(u_{1,2}+u_{2,1}\right)
\end{array}
\right) +\nonumber\\
&
-
\left(
\begin{array}{c }
\alpha_{11}^m \\
\alpha_{22}^m \\
\sqrt{2}\,\alpha_{12}^m
\end{array}
\right)
\theta
-
\left(
\begin{array}{c }
\beta_{11}^m \\
\beta_{22}^m \\
\sqrt{2}\,\beta_{12}^m
\end{array}
\right)
\eta,
\nonumber\\
\left(
\begin{array}{c}
q_{1}\\
q_{2}\\
\end{array}
\right)=&-
\left(
\begin{array}{c c }
K_{11}^m & K_{12}^m\\
K_{21}^m & K_{22}^m 
\end{array}
\right)
\left(
\begin{array}{c}
\theta_{,1}\\
\theta_{,2}
\end{array}
\right),
\nonumber\\
\left(
\begin{array}{c}
j_{1}\\
j_{2}\\
\end{array}
\right)=&-
\left(
\begin{array}{c c }
D_{11}^m & D_{12}^m\\
D_{21}^m & D_{22}^m 
\end{array}
\right)
\left(
\begin{array}{c}
\eta_{,1}\\
\eta_{,2}
\end{array}
\right).
\end{align}
\section*{Appendix E. Frequency band structure of heterogeneous periodic thermo-diffusive material: finite element formulation  }
\label{Appendix::Calcoli condensazione statica spettro eterogeneo}
%
Constitutive relations (\ref{eq:MicroStress})-(\ref{eq:MicroMassFlux}) for thermo-diffusive materials  in indicial form read
\begin{eqnarray}
&&
\sigma_{ij} =C^{m}_{ijkl}\, u_{k,l}-\alpha^m_{ij}\,\theta-\beta^m_{ij}\,\eta,%
\nonumber\\
&&q_i=-K^m_{ij}\,\theta_{,j},
\nonumber\\
&&j_{i}=-D_{ij}^m\,\eta_{,j}.
\label{eq:MicroConstEqsIndexNot}
\end{eqnarray}
Denoting with $\mathbf{b}$ the body force vector, with $r$ the heat source term, and with $s$ the mass source term,  stress tensor $\tensor{\sigma}$,  heat flux vector $\mathbf{q}$, and mass flux vector $\mathbf{j}$ satisfy local balance equations (\ref{eq:MicroEqMotion})-(\ref{eq:MicroDiffusionEquation}), here written in the form
\begin{eqnarray}
&& \left(C^m_{ijkl}\,u_{k,l}\right)_{,j}-
\left(\alpha^m_{ij}\,\theta\right)_{,j}-
\left(\beta^m_{ij}\,\eta\right)_{,j}+b_i
=
\rho^m\,\ddot{u}_{i},\nonumber\\
&&\left(K^m_{ij}\,\theta_{,j}\right)_{,i}-
\alpha^m_{ij}\,\dot{u}_{i,j}-
\psi^m\,\dot{\eta}+
r=p^m\,\dot{\theta},\nonumber\\
&&\left(D^m_{ij}\,\eta_{,j}\right)_{,i}-
\beta^m_{ij}\,\dot{u}_{i,j}-
\psi^m\,\dot{\theta}+
s=q^m\,\dot{\theta}.
\label{eq:MicroBalanceEquationsIndexNot}
\end{eqnarray}
Dirichlet and Neumann part of the boundary $\partial\Omega$ of domain $\Omega$, denoted respectively as $\{\partial\Omega_{\mathbf{u}},\partial\Omega_\theta,\partial\Omega_\eta\}$ and $\{\partial\Omega_{\tensor{\sigma}},\partial\Omega_{\mathbf{q}},\partial\Omega_\mathbf{j}\}$, are such that $\partial\Omega=\partial\Omega_{\mathbf{u}}\cup\partial\Omega_{\tensor{\sigma}}=\partial\Omega_{\theta}\cup\partial\Omega_{\mathbf{q}}=\partial\Omega_{\eta}\cup\partial\Omega_{\mathbf{j}}$
 and $\partial\Omega_{\mathbf{u}}\cap\partial\Omega_{\tensor{\sigma}}=\partial\Omega_{\theta}\cap\partial\Omega_{\mathbf{q}}=\partial\Omega_{\eta}\cap\partial\Omega_{\mathbf{j}}=\emptyset$.
Micro fields satisfy boundary conditions
\begin{eqnarray}
&&\left\{
\begin{array}{l r}
u_i=\bar{u}_i & on\,\partial\Omega_{\mathbf{u}}\\
\sigma_{ij}\,n_j=\bar{t}_i & on \,\partial\Omega_{\tensor{\sigma}}\\
\end{array}
\right.
,
\hspace{0.5cm}
\left\{
\begin{array}{l r}
\theta=\bar{\theta} & on\,\partial\Omega_{\theta}\\
q_{i}\,n_i=\bar{q} & on \,\partial\Omega_{\mathbf{q}}\\
\end{array}
\right.
,
\nonumber\\
&&\left\{
\begin{array}{l r}
\eta=\bar{\eta} & on\,\partial\Omega_{\eta}\\
j_{i}\,n_i=\bar{j} & on \,\partial\Omega_{\mathbf{j}}\\
\end{array}
\right.
.
\label{eq:BoundaryConditionsMicroFields}
\end{eqnarray}  
 where $\bar{t}_i,\bar{q}$ and $\bar{j}$ are the prescribed values of tractions, heat flux, and mass flux, respectively, and $\mathbf{n}$ is the outward normal to the boundary of the domain $\partial\Omega$.
 Taking into account boundary conditions (\ref{eq:BoundaryConditionsMicroFields}), weak form of local balance equations (\ref{eq:MicroBalanceEquationsIndexNot}) reads
 \begin{eqnarray}
 &&
 \int_{\Omega}
 \left(
 C^m_{ijkl}\,u_{k,l}
 -
 \alpha_{ij}^m\,\theta
 -
 \beta_{ij}^m\,\eta\right)\varphi_{u_{i,j}}\,dV
 -
 \int_{\partial\Omega_{\tensor{\sigma}}}
 \bar{t}_i\,\varphi_{u_i}\,dS
 -
 \int_{\Omega} b_i \varphi_{u_i}\,dV
 +\nonumber\\
 &&
 \int_{\Omega}\rho^m\,\ddot{u}_i\,\varphi_{u_i}\,dV=0\hspace{0.2cm}\forall\varphi_{u_i}
 \,s.t.\,\varphi_{u_i}=0\hspace{0.5cm}on\,\,\partial\Omega_{\mathbf{u}},
 \nonumber\\
 \\
 &&
 \int_{\Omega}
 \left(
 K^m_{ij}\,\theta_{,j}\right)\varphi_{\theta_{,i}}\,dV
 +
 \int_{\partial\Omega_{\mathbf{q}}}
 \bar{q}\,\varphi_{\theta}\,dS
 +
 \int_{\Omega} \left(
 \alpha^m_{ij}\,\dot{u}_{i,j}
 +
 \psi^m\dot{\eta}
 -r\right) \varphi_{\theta}\,dV
 +\nonumber\\
 &&
 \int_{\Omega}p^m\,\dot{\theta}\,\varphi_{\theta}\,dV=0\hspace{0.2cm}\forall\varphi_{\theta}
 \,s.t.\,\varphi_{\theta}=0\hspace{0.5cm}
 on\,\,\partial\Omega_{\theta},
 \nonumber\\
 \\
 &&
 \int_{\Omega}
 \left(
 D^m_{ij}\,\eta_{,j}\right)\varphi_{\eta_{,i}}\,dV
 +
 \int_{\partial\Omega_{\mathbf{j}}}
 \bar{j}\,\varphi_{\eta}\,dS
 +
 \int_{\Omega} \left(
 \beta^m_{ij}\,\dot{u}_{i,j}
 +
 \psi^m\dot{\theta}
 -
 s\right) \varphi_{\eta}\,dV
 +\nonumber\\
 &&
 \int_{\Omega}q^m\,\dot{\eta}\,\varphi_{\eta}\,dV = 0\hspace{0.2cm}\forall\varphi_{\eta}
 \,s.t.\,\varphi_{\eta}=0\hspace{0.5cm}on\,\,\partial\Omega_{\eta},
 \label{eq:weakformlocalbalanceequations}
 \end{eqnarray}
 with $\varphi_{u_i},\varphi_{\theta}$ and $\varphi_\eta$ test functions.
 Micro fields $\mathbf{u}(\mathbf{x},t),\theta(\mathbf{x},t)$, and $\eta(\mathbf{x},t)$ are approximated by a linear combination of shape functions $\mathbf{N}(\mathbf{x})$ and nodal unknowns $\mathbf{u}(t),\tensor{\theta}(t)$, and $\tensor{\eta}(t)$, as usual in a finite element discretization, and read
 \begin{eqnarray}
 &&u_i(\mathbf{x},t)=\sum_{j=1}^{N_h} N_j(\mathbf{x})u_{i_j}(t),\hspace{0.2cm}
 \theta(\mathbf{x},t)=\sum_{j=1}^{N_h} N_j(\mathbf{x}) \theta_j(t),
 \hspace{0.2cm}
 \eta(\mathbf{x},t)=\sum_{j=1}^{N_h} N_j(\mathbf{x}) \eta_j(t),
 \label{eq:FiniteElementDiscretizationMicroFields}
 \end{eqnarray}
 and the very same discretization is performed for test functions, with nodal unknowns $\tensor{\delta}\mathbf{u}(t),\tensor{\delta\theta}(t)$, and $\tensor{\delta\eta}(t)$ 
 \begin{eqnarray}
 &&\varphi_
{ u_i}(\mathbf{x},t)=\sum_{j=1}^{N_h} N_j(\mathbf{x})\delta u_{i_j}(t),\hspace{0.2cm}
 \varphi_\theta(\mathbf{x},t)=\sum_{j=1}^{N_h} N_j(\mathbf{x}) \delta\theta_j(t),
 \hspace{0.2 cm}
 \varphi_\eta(\mathbf{x},t)=\sum_{j=1}^{N_h} N_j(\mathbf{x}) \delta\eta_j(t).
 \label{eq:FiniteElementDiscretizationTestFunctions}
 \end{eqnarray}
 In equations (\ref{eq:FiniteElementDiscretizationMicroFields}) and (\ref{eq:FiniteElementDiscretizationTestFunctions}), $N_h$ represents the finite dimension of the space $V_h$ for which $\{N_j|j=1,2,...,N_h\}$ is a basis.
 In a two dimensional setting, denoting with $\mathbf{N}_{\mathbf{u}},\mathbf{N}_\theta$, and $\mathbf{N}_\eta$ matrices collecting shape functions of the single finite element $e$ with $N_{Nnod}$  the number of element nodes, one has
 \begin{eqnarray}
 && \mathbf{N}_{\mathbf{u}}
 =
 \left[
 \begin{array}{c c c c c c c}
 N_1 & 0 & N_2 & 0 & ... & N_{Nnod} & 0 \\
 0 &  N_1 & 0 & N_2 & ... & 0 & N_{Nnod}
 \end{array}
 \right],
 \nonumber\\
 &&
 \mathbf{N}_\theta=\mathbf{N}_\eta=
 \left[ \begin{array}{c c c c}
  N_1 & N_2 & ... & N_{Nnod}
\end{array} \right],
\label{eq:shapefunctionsmatrices}
 \end{eqnarray}
and denoting with $\mathbf{D}_{\mathbf{u}},\mathbf{D}_\theta$, and $\mathbf{D}_\eta$ differential matrices 
\begin{eqnarray}
\mathbf{D}_{\mathbf{u}}=
\left[
\begin{array}{c c}
\partial/\partial x_1 & 0 \\
0 & \partial/\partial x_2 \\
\partial /\partial x_2 & \partial /\partial x_1
\end{array}
\right],
\hspace{0.2cm}
\mathbf{D}_{\theta}=\mathbf{D}_\eta=
\left[
\begin{array}{c}
\partial/\partial x_1\\
\partial/\partial x_2
\end{array}
\right],
\end{eqnarray}
one defines $\mathbf{B}_\mathbf{u}=\mathbf{D}_\mathbf{u}\mathbf{N}_{\mathbf{u}}$, $\mathbf{B}_\theta=\mathbf{D}_\theta\mathbf{N}_\theta$, and $\mathbf{B}_\eta=\mathbf{D}_\eta\mathbf{N}_\eta$.
Weak form (\ref{eq:weakformlocalbalanceequations}) can therefore be written in matrix notation over each element domain $\Omega_e$ as
\begin{eqnarray}
&&
\tensor{\delta}\mathbf{u}^T
\int_{\Omega_e}
\mathbf{B}_{\mathbf{u}}^T\,
\mathbf{C}^m\,
\mathbf{B}_{\mathbf{u}}\,
dV \,
\mathbf{u}
-
\tensor{\delta}\mathbf{u}^T
\int_{\Omega_e}
\mathbf{B}_{\mathbf{u}}^T\,
\tensor{\upalpha}^m\,
\mathbf{N}_{\theta}
\,dV
\,\tensor{\theta}
-
\tensor{\delta}\mathbf{u}^T
\int_{\Omega_e}
\mathbf{B}_{\mathbf{u}}^T\,
\tensor{\upbeta}^m\,
\mathbf{N}_{\eta}
\,dV
\,\tensor{\eta}
+
\nonumber\\
&&
-
\tensor{\delta}\mathbf{u}^T
\int_{\partial\Omega_{e_{\tensor{\sigma}}}}
\mathbf{N}_{\mathbf{u}}^T\,
\bar{\mathbf{t}}\,dS
-
\tensor{\delta}\mathbf{u}^T
\int_{\Omega_e}
\mathbf{N}_{\mathbf{u}}^T\,
\mathbf{b}\,dV
+
\tensor{\delta}\mathbf{u}^T
\int_{\Omega_e}
\mathbf{N}_{\mathbf{u}}^T\,
\rho^m
\mathbf{N}_{\mathbf{u}}\,
dV
\ddot{\mathbf{u}}=0 \hspace{0.5cm}\forall \tensor{\delta}\mathbf{u},
\nonumber\\
\\
&&
\tensor{\delta}\tensor{\theta}^T
\int_{\Omega_e}
\mathbf{B}_{\theta}^T\,
\mathbf{K}^m\,
\mathbf{B}_{\theta}\,
dV \,
\tensor{\theta}
+
\tensor{\delta}\tensor{\theta}^T
\int_{\Omega_e}
\mathbf{N}_{\theta}^T\,
\tensor{\upalpha}^m\,
\mathbf{B}_{\mathbf{u}}
\,dV
\,\dot{\mathbf{u}}
+
\tensor{\delta}\tensor{\theta}^T
\int_{\Omega_e}
\mathbf{N}_{\theta}^T\,
\psi^m\,
\mathbf{N}_{\eta}
\,dV
\,\dot{\tensor{\eta}}
+
\nonumber\\
&&
\tensor{\delta}\tensor{\theta}^T
\int_{\partial\Omega_{e_{\mathbf{q}}}}
\mathbf{N}_{\tensor{\theta}}^T\,
\bar{q}\,dS
-
\tensor{\delta}\tensor{\theta}^T
\int_{\Omega_e}
\mathbf{N}_{\theta}^T\,
r\,dV
+
\tensor{\delta}\tensor{\theta}^T
\int_{\Omega_e}
\mathbf{N}_{\theta}^T\,
p^m
\mathbf{N}_{\theta}\,
dV
\dot{\tensor{\theta}}=0 \hspace{0.5cm}\forall \tensor{\delta}\tensor{\theta},
\label{eq:weakformMatrixNotation0}
\nonumber\\
\\
&&
\tensor{\delta}\tensor{\eta}^T
\int_{\Omega_e}
\mathbf{B}_{\eta}^T\,
\mathbf{D}^m\,
\mathbf{B}_{\eta}\,
dV \,
\tensor{\eta}
+
\tensor{\delta}\tensor{\eta}^T
\int_{\Omega_e}
\mathbf{N}_{\eta}^T\,
\tensor{\upbeta}^m\,
\mathbf{B}_{\mathbf{u}}
\,dV
\,\dot{\mathbf{u}}
+
\tensor{\delta}\tensor{\eta}^T
\int_{\Omega_e}
\mathbf{N}_{\eta}^T\,
\psi^m\,
\mathbf{N}_{\theta}
\,dV
\,\dot{\tensor{\theta}}
+
\nonumber\\
&&
\tensor{\delta}\tensor{\eta}^T
\int_{\partial\Omega_{e_{\mathbf{j}}}}
\mathbf{N}_{\tensor{\eta}}^T\,
\bar{j}\,dS
-
\tensor{\delta}\tensor{\eta}^T
\int_{\Omega_e}
\mathbf{N}_{\eta}^T\,
s\,dV
+
\tensor{\delta}\tensor{\eta}^T
\int_{\Omega_e}
\mathbf{N}_{\eta}^T\,
q^m
\mathbf{N}_{\eta}\,
dV
\dot{\tensor{\eta}}=0\hspace{0.5cm}\forall \tensor{\delta}\tensor{\eta},
\label{eq:weakformMatrixNotation}
\end{eqnarray}
where symbols $\mathbf{C}^m$, $\mathbf{K}^m$, $\mathbf{D}^m$, $\tensor{\upalpha}^m$, and $\tensor{\upbeta}^m$,  denote the matrix form of the corresponding constitutive tensors $\mathfrak{C}^m$, $\tensor{K}^m$, $\tensor{D}^m$, $\tensor{\alpha}^m$, and $\tensor{\beta}^m$.
Elemental stiffness matrices are defined  in the following way
\begin{eqnarray}
&&\mathbf{K}_{\mathbf{u}\mathbf{u}}^e=
\int_{\Omega_e}\mathbf{B}_{\mathbf{u}}^T\,
\mathbf{C}^m\,
\mathbf{B}_{\mathbf{u}}\,dV,
\nonumber\\
&&
\mathbf{K}_{\mathbf{u}\theta}^e=
-\int_{\Omega_e}\mathbf{B}_{\mathbf{u}}^T\,
\tensor{\upalpha}^m\,
\mathbf{N}_{\theta}\,dV,
\nonumber\\
&&
\mathbf{K}_{\mathbf{u}\eta}^e=-
\int_{\Omega_e}\mathbf{B}_{\mathbf{u}}^T\,
\tensor{\upbeta}^m\,
\mathbf{N}_{\eta} \,dV,
\nonumber\\
&&
\mathbf{K}_{\theta\theta}^e=
\int_{\Omega_e}\mathbf{B}_{\theta}^T\,
\mathbf{K}^m\,
\mathbf{B}_{\theta}\,dV,
\nonumber\\
&&
\mathbf{K}_{\eta\eta}^e=
\int_{\Omega_e}\mathbf{B}_{\eta}^T\,
\mathbf{D}^m\,
\mathbf{B}_{\eta}\,dV.
\label{eq:ElementalStiffnessMatrices}
\end{eqnarray}
Analogously, damping matrices relative to each element read
\begin{eqnarray}
&&
\mathbf{C}_{\theta\theta}^e
=
\int_{\Omega_e}
\mathbf{N}_{\theta}^T
p^m
\mathbf{N}_{\theta}\,dV,
\nonumber\\
&&
\mathbf{C}_{\eta\eta}^e
=
\int_{\Omega_e}
\mathbf{N}_{\eta}^T
q^m
\mathbf{N}_{\eta}\,dV,
\nonumber\\
&&
\mathbf{C}_{\theta\mathbf{u}}^e
=
\int_{\Omega_e}
\mathbf{N}_{\theta}^T
\tensor{\upalpha}^m
\mathbf{B}_{\mathbf{u}}\,dV,
\nonumber\\
&&
\mathbf{C}_{\theta\mathbf{u}}^e
=
\int_{\Omega_e}
\mathbf{N}_{\theta}^T
\psi^m
\mathbf{N}_{\eta}\,dV,
\nonumber\\
&&
\mathbf{C}_{\eta\mathbf{u}}^e
=
\int_{\Omega_e}
\mathbf{N}_{\eta}^T
\tensor{\upbeta}^m
\mathbf{B}_{\mathbf{u}}\,dV,
\nonumber\\
&&
\mathbf{C}_{\eta\theta}^e
=
\int_{\Omega_e}
\mathbf{N}_{\eta}^T
\psi^m
\mathbf{N}_{\theta}\,dV,
\end{eqnarray}
and the elemental mass matrix has the form
\begin{equation}
\mathbf{M}_{\mathbf{u}\mathbf{u}}^e=
\int_{\Omega_e}
\mathbf{N}_\mathbf{u}^T\,
\rho^m\,
\mathbf{N}_{\mathbf{u}}\,dV.
\end{equation}
The elemental external force vectors have the following expressions
\begin{eqnarray}
&&
\mathbf{f}_{\mathbf{u}}^e=
\int_{\Omega_e}\mathbf{N}_{\mathbf{u}}^T\,\mathbf{b}\,dV
+
\int_{\partial\Omega_{e_\tensor{\sigma}}}\mathbf{N}_{\mathbf{u}}^t\,\bar{\mathbf{t}}\,dS,
\nonumber\\
&&
\mathbf{f}_\theta^e=
\int_{\Omega_e}\mathbf{N}_{\theta}^T\,r\,dV
-
\int_{\partial\Omega_{e_{\mathbf{q}}}}
\mathbf{N}_{\theta}^T\,\bar{q}\,dS,
\nonumber\\
&&
\mathbf{f}_\eta^e=
\int_{\Omega_e}\mathbf{N}_{\eta}^T\,s\,dV
-
\int_{\partial\Omega_{e_{\mathbf{j}}}}
\mathbf{N}_{\eta}^t\,\bar{j}\,dS.
\label{eq:elemental_force_vector}
\end{eqnarray}
Equations (\ref{eq:weakformMatrixNotation0})-(\ref{eq:weakformMatrixNotation}, therefore,  can be written in the following form, after assembling  elemental contributions (\ref{eq:ElementalStiffnessMatrices})-(\ref{eq:elemental_force_vector}) into the relative global ones
%
%
%
%
\begin{eqnarray}
&&\mathbf{K}_{\mathbf{u}\mathbf{u}}\,\mathbf{u}+
\mathbf{K}_{\mathbf{u}\theta}\,\tensor{\theta}+
\mathbf{K}_{\mathbf{u}\eta}\,\tensor{\eta}=
\mathbf{f}_{\mathbf{u}}-\mathbf{M}_{\mathbf{u}\mathbf{u}}\ddot{\mathbf{u}},
\nonumber\\
&&
\mathbf{K}_{\theta\theta}\,\tensor{\theta}+
\mathbf{C}_{\theta\mathbf{u}}\,\dot{\mathbf{u}}+
\mathbf{C}_{\theta\eta}\,\dot{\tensor{\eta}}=
\mathbf{f}_{\theta}-\mathbf{C}_{\theta\theta}\dot{\tensor{\theta}},
\nonumber\\
&&
\mathbf{K}_{\eta\eta}\,\tensor{\eta}+
\mathbf{C}_{\eta\mathbf{u}}\,\dot{\mathbf{u}}+
\mathbf{C}_{\eta\theta}\,\dot{\tensor{\theta}}=
\mathbf{f}_{\eta}-\mathbf{C}_{\eta\eta}\dot{\tensor{\eta}}.
\nonumber\\
\label{eq:dynamicequationsofmotion}
\end{eqnarray}
After performing bilateral Laplace transform (\ref{eq:LaplaceTrasformata}) on system (\ref{eq:dynamicequationsofmotion}), taking into account derivation rule (\ref{eq:DerivataLaplaceTrasformata}), one obtains the following system expressed in terms of vector $\hat{\mathbf{z}}=(\hat{\mathbf{u}}\hspace{0.1cm} \hat{\tensor{\theta}} \hspace{0.1cm}\hat{\tensor{\eta}})^T$ containing the microfields in the Laplace domain
\begin{eqnarray}
&&(\mathbf{K}_{\mathbf{u}\mathbf{u}}+\omega^2\mathbf{M}_{\mathbf{u}\mathbf{u}})\hat{\mathbf{u}}
+
\mathbf{K}_{\mathbf{u}\theta}\,\hat{\tensor{\theta}}
+
\mathbf{K}_{\mathbf{u}\eta}\,\hat{\tensor{\eta}}=\hat{\mathbf{f}}_{\mathbf{u}},
\nonumber\\
&&
\left(
\mathbf{K}_{\theta\theta}
+
\omega\mathbf{C}_{\theta\theta}
\right)\hat{\tensor{\theta}}
+
\omega\mathbf{C}_{\theta\mathbf{u}}\,\hat{\mathbf{u}}+
\omega\mathbf{C}_{\theta\eta}\,\hat{\tensor{\eta}}
=
\hat{\mathbf{f}}_{\theta},
\nonumber\\
&&
\left(
\mathbf{K}_{\eta\eta}
+
\omega\mathbf{C}_{\eta\eta}
\right)\hat{\tensor{\eta}}
+
\omega\mathbf{C}_{\eta\mathbf{u}}\,\hat{\mathbf{u}}+
\omega\mathbf{C}_{\eta\theta}\,\hat{\tensor{\theta}}
=
\hat{\mathbf{f}}_{\eta}.
\label{eq:GeneralizedchristoffelEquationsEterogeneo}
\end{eqnarray}
Exploiting the periodicity of the medium, generalized Christoffel equations (\ref{eq:GeneralizedchristoffelEquationsEterogeneo})  can be studied in the periodic cell $\mathcal{A}$. By virtue of Bloch's theorem, Floquet-Bloch boundary conditions have to be applied to elementary cell $\mathcal{A}$ in order to obtain its dispersion relations. Following the procedure described in \citep{Langley1993,Phani2006}, degrees of freedom contained in vector $\hat{\mathbf{z}}$ can be reorganized as $\hat{\mathbf{z}}=(\hat{\mathbf{z}}_\ell\hspace{0.1cm}\hat{\mathbf{z}}_r\hspace{0.1cm}\hat{\mathbf{z}}_b\hspace{0.1cm}\hat{\mathbf{z}}_t\hspace{0.1cm}\hat{\mathbf{z}}_{\ell b}\hspace{0.1cm}\hat{\mathbf{z}}_{r b}\hspace{0.1cm}\hat{\mathbf{z}}_{\ell t}\hspace{0.1cm}\hat{\mathbf{z}}_{r t}\hspace{0.1cm}\hat{\mathbf{z}}_i)^T$, where subscripts $\ell,r,b,t$, and $i$ denote, respectively,  the left, right, bottom, top, and internal nodes of a generic cell and double subscripts indicate corner nodes.
Floquet-Bloch boundary conditions are written as
\begin{equation}
\begin{array}{l l}
\hat{\mathbf{z}}_r=e^{ik_1d_1} \hat{\mathbf{z}}_\ell,
 &
  \hat{\mathbf{f}}_r=-e^{ik_1d_1} \hat{\mathbf{f}}_\ell,
  \\
\hat{\mathbf{z}}_t=e^{ik_2d_2}\,\hat{\mathbf{z}}_b,
&
\hat{\mathbf{f}}_t=-e^{ik_2d_2}\,\hat{\mathbf{f}}_b,
\\
\hat{\mathbf{z}}_{r b}=e^{ik_1d_1}\,\hat{\mathbf{z}}_{\ell b},
&
\hat{\mathbf{f}}_{r b}=-e^{ik_1d_1}\,\hat{\mathbf{f}}_{\ell b},
\\
\hat{\mathbf{z}}_{\ell t}=e^{ik_2d_2}\,\hat{\mathbf{z}}_{\ell b},
&
\hat{\mathbf{f}}_{\ell t}=-e^{ik_2d_2}\,\hat{\mathbf{f}}_{\ell b},
\\
\hat{\mathbf{z}}_{r t}=e^{i(k_1 d_1 + k_2d_2)}\,\hat{\mathbf{z}}_{\ell b},
&
\hat{\mathbf{f}}_{r t}=-e^{i(k_1 d_1 + k_2d_2)}\,
\hat{\mathbf{f}}_{\ell b}.
\end{array}
\label{eq:FloquetBlochBoundaryConditions}
\end{equation}
where $i$ is the imaginary unit s.t. $i^2=-1$ and $\mathbf{k}=k_1\mathbf{e}_1 +k_2\mathbf{e}_2\in\mathcal{B}$ is the wave vector with wave numbers $k_1$ and $k_2$, and $\mathcal{B}=[-\pi/d_1,\pi/d_1]\times[-\pi/d_2,\pi/d_2]$ is the first Brillouin zone of cell $\mathcal{A}$ having orthogonal periodicity vectors $\mathbf{v}_1=d_1\mathbf{e}_1$ and $\mathbf{v_2}=d_2\mathbf{e}_2$.
Boundary conditions (\ref{eq:FloquetBlochBoundaryConditions}) allow to define the following transformation
\begin{equation}
\hat{\mathbf{z}}=\mathbf{T}\hat{\mathbf{q}},
\label{eq:LinearTransformation}
\end{equation}
with matrix $\mathbf{T}$ defined as
\begin{equation}
\mathbf{T}=
\left(
\begin{array}{c c c c}
\mathbf{I} & \mathbf{0} & \mathbf{0} & \mathbf{0}\\
\mathbf{I}e^{ik_1d_1} &  \mathbf{0} &  \mathbf{0} &  \mathbf{0}\\
 \mathbf{0} & \mathbf{I} &  \mathbf{0} &  \mathbf{0}\\
  \mathbf{0} & \mathbf{I} e ^{ik_2d_2} &  \mathbf{0} &  \mathbf{0}\\
   \mathbf{0} &  \mathbf{0} & \mathbf{I} &  \mathbf{0}\\
    \mathbf{0} &  \mathbf{0} & \mathbf{I}e^{ik_1d_1} &  \mathbf{0}\\
     \mathbf{0} &  \mathbf{0} & \mathbf{I}e^{ik_2d_2} &  \mathbf{0}\\
      \mathbf{0} &  \mathbf{0} & \mathbf{I}e^{i(k_1d_1+k_2d_2)} &  \mathbf{0}\\
\end{array}
\right),
\end{equation}
and vector $\hat{\mathbf{q}}$ of reduced independent degrees of freedom expressed in the form
\begin{equation}
\hat{\mathbf{q}}=
\left(
\begin{array}{c}
\hat{\mathbf{z}}_\ell\\
\hat{\mathbf{z}}_b\\
\hat{\mathbf{z}}_{\ell b}\\
\hat{\mathbf{z}}_i\\
\end{array}
\right).
\end{equation}
Substitution of equation (\ref{eq:LinearTransformation}) into governing equations of motion (\ref{eq:GeneralizedchristoffelEquationsEterogeneo}) and premultiplication by the Hermitian transpose of $\mathbf{T}$, named $\mathbf{T}^H$, in order to enforce equilibrium, lead to
\begin{equation}
\left(
\omega^2 \mathbf{T}^H\mathbf{M}\mathbf{T}+
\omega \mathbf{T}^H\mathbf{C}\mathbf{T}+
\mathbf{T}^H\mathbf{K}\mathbf{T}
\right)
\hat{\mathbf{q}}=\mathbf{T}^H\hat{\mathbf{f}},
\label{eq:eigenvalueproblemEterogeneo}
\end{equation}
where $\mathbf{M}$, $\mathbf{C}$, and $\mathbf{K}$ represent, respectively, the global mass, damping, and stiffness matrices.
In the case of free wave motion ($\hat{\mathbf{f}}=0$) it results $\mathbf{T}^H\hat{\mathbf{f}}=
\mathbf{0}$,
and equation (\ref{eq:eigenvalueproblemEterogeneo}) defines a quadratic generalized eigenvalue problem whose solution, for each value of wave vector $\mathbf{k}\in\mathbb{R}^2$, gives the complex frequency $\omega$ as the generalized eigenvalue and $\hat{\mathbf{q}}$ as the generalized eigenvector.
Real and imaginary parts of the complex angular frequency $\omega=\omega_r+i\omega_i$, characterize the damping and the propagation mode, respectively, of dispersive Bloch waves propagating inside the heterogeneous material.
Finally, quadratic eigenvalue problem (\ref{eq:eigenvalueproblemEterogeneo}) can be tranformed into an equivalent linear one in the following way
\begin{equation}
\left(
\omega
\left(
\begin{array}{c c}
\mathbf{T}^H\mathbf{M}\mathbf{T} & \mathbf{0}\\
\mathbf{0} & \mathbf{I}
\end{array}
\right)
+
\left(
\begin{array}{c c}
\mathbf{T}^H\mathbf{C}\mathbf{T} & \mathbf{T}^H\mathbf{K}\mathbf{T}\\
-\mathbf{I} & \mathbf{I}
\end{array}
\right)
\right)
\left(
\begin{array}{c}
\omega\hat{\mathbf{q}}\\
\hat{\mathbf{q}}
\end{array}
\right)
=
\left(
\begin{array}{c}
\mathbf{0}\\
\mathbf{0}
\end{array}
\right),
\end{equation}
which admits a non trivial solution $(\omega \hat{\mathbf{q}}\hspace{0.1cm}\hat{\mathbf{q}})^T$ only if the linear operator mutiplying the generalized eigenvector is not invertible.
 
\end{appendices}
\end{document}